\documentclass[letterpaper]{JHEP3}
\pdfoutput=1

\usepackage{amsmath}
\usepackage{amssymb}
\usepackage{amsfonts}
\usepackage{lscape, graphicx}
\usepackage{cite}       
\usepackage{bm}         
\usepackage{verbatim}

\usepackage[all]{xy}
\advance\parskip 1.0pt plus 1.0pt minus 2.0pt   
\def\href#1#2{#2}	

\def\coeff#1#2{{\textstyle {\frac {#1}{#2}}}}
\def\half{\coeff 12}

\def\T{{\cal T}}

\def\R{{\mathbb R}}
\def\S{{\mathbb S}}

\def\tr{{\rm tr}}

\def\Z{{\mathbb Z}}
\def\T{{\mathbb T}}

\def\Dslash{{\rlap{\raise 1pt \hbox{$\>/$}}D}}

\newcommand{\beq}{\begin{equation}}
\newcommand{\eeq}{\end{equation}}
\newcommand{\beqa}{\begin{eqnarray}}
\newcommand{\eeqa}{\end{eqnarray}}

\def\ltap{\ \raise.3ex\hbox{$<$\kern-.75em\lower1ex\hbox{$\sim$}}\ }
\def\gtap{\ \raise.3ex\hbox{$>$\kern-.75em\lower1ex\hbox{$\sim$}}\ }
\def\gl{\ \raise.5ex\hbox{$>$}\kern-.8em\lower.5ex\hbox{$<$}\ }
\def\roughly#1{\raise.3ex\hbox{$#1$\kern-.75em\lower1ex\hbox{$\sim$}}}

\title{2d affine $\mathbf{XY}$-spin model/4d gauge theory duality and 
deconfinement }
  
\author
{
    {
    Mohamed M. Anber,$^1$\footnote{\email{manber@physics.utoronto.ca}}~
    Erich Poppitz,$^2$\footnote{\email{poppitz@physics.utoronto.ca}} ~ and Mithat 
   \" Unsal$^{3,4}$\footnote{\email{unsal@slac.stanford.edu}}
           \\${}^{1,2}${Department of Physics, University of Toronto,
    Toronto, ON M5S 1A7, Canada}
           \\${}^3${SLAC and Physics Department, Stanford University, Stanford, CA 94025/94305, USA}
             \\${}^4${Department of Physics and Astronomy, SFSU, San Francisco, CA 94132, USA}\\
            }
    }%
    
    \abstract{
    
    \smallskip
    
    \smallskip
    
       {\small{We introduce a duality  between two-dimensional  XY-spin models  with  symmetry-breaking perturbations and  certain  four-dimensional   $SU(2)$  and $SU(2)/\Z_2$ gauge theories, compactified on a small spatial circle $\R^{1,2} \times \S^1$, and considered at temperatures near the deconfinement transition. In a Euclidean set up, the theory is defined on $\R^2 \times \T^2$. Similarly, thermal gauge theories of higher rank  are  dual  to  new  families of 
``affine"    XY-spin models with  perturbations. For rank two, these are related to models used to describe the melting of a 2d crystal with a triangular lattice.    
 The connection is made through a multi-component electric-magnetic  Coulomb gas representation for both systems.  Perturbations in the spin system map to topological defects in the gauge theory, such as monopole-instantons or magnetic bions,  
 and the vortices in the spin system map to the  electrically charged $W$-bosons in field theory (or vice versa, depending on the duality frame). 
 The duality permits one to use the two-dimensional technology of spin systems to study the  thermal deconfinement and discrete chiral transitions in four-dimensional $SU(N_c)$ gauge theories with $n_f$$\ge$$1$   adjoint Weyl fermions.  
     }

}}

\begin{document}

\maketitle

\section{Introduction, summary, and outline}

It is well-known since the late 70's that two-dimensional (2d)  XY-spin models  with appropriate 
symmetry-breaking perturbations map to a 2d  Coulomb gas  with electric and magnetic charges  \cite{Jose:1977gm}. This   beautiful  duality  permits analytic calculations of long distance correlation functions, phase diagrams,  and critical indices in a large class of 2d statistical mechanics systems such as  XY-  and ``clock"- (planar Potts) models (for reviews of methods and applications  see, for example,  \cite{Ogilvie:1981mj, Nienhuis:1984wm,Lecheminant:2002va} and references therein). A vectorial version of the XY-model has been used in the study of melting of 
 a 2d crystal with a triangular lattice \cite{Nelson}. The appropriate 
Coulomb gas  is  a system of electric ($e$)  and magnetic ($m$) charges, and involves $e$-$e$ and $m$-$m$ Coulomb interactions, as well as $e$-$m$ Aharonov-Bohm phase interactions \cite{135388}.    

In this paper, we  introduce a long distance  duality (equivalence)  between 2d XY-spin models   and certain four-dimensional (4d)  gauge theories compactified on $\R^2 \times \T^2$, with  boundary conditions as specified below.    The connection is made through an electric-magnetic  Coulomb gas representation for spin  systems and gauge theories.  As stated above, the connection of  the  spin systems with Coulomb gases   is  well understood.  What is new is the realization that some  four-dimensional gauge theories  on $\R^2 \times \T^2$ also admit an electric-magnetic  Coulomb gas representation. This  is the basis of the long-distance duality between  XY-spin models and circle-compactified 4d gauge theories at finite temperature. 

Let us briefly summarize the main progress  which makes the  mapping  of four-dimensional gauge theories to 
electric-magnetic Coulomb gases possible. The gauge theory we study is 4d 
QCD with adjoint fermions, QCD(adj), formulated on $\R^{1,2} \times \S^1_L$. Here,  $\S^1_L$ is a spatial (non-thermal) circle of circumference $L$ and the  fermions obey periodic boundary conditions. This theory does not undergo a center-symmetry changing phase transition  as the radius of the $\S^1_L$ is reduced \cite{Unsal:2006pj}. This implies that the theory ``abelianizes" and  becomes weakly coupled  and semi-classically calculable. 
Non-perturbative properties, which are difficult to study on $\R^{1,3}$, such as the generation of mass gap, confinement, and the realization of chiral symmetry can  be studied analytically at small $\S^1_L$  \cite{Unsal:2007vu, Unsal:2007jx}.  
 The non-perturbative long-distance dynamics of the theory is governed by magnetically charged topological molecules, the ``magnetic bions"---molecular (or correlated) instanton events whose proliferation leads to mass gap and confinement.  The vacuum of the theory is a dilute plasma of magnetic  bions.  The  theory also possesses electrically charged particles, e.g., $W$-bosons, which decouple from the long-distance  dynamics at  $T=0$.
The zero-temperature physics of QCD(adj) on $\R^{1,2} \times \S^1_L$ is reviewed in Section \ref{zerotemperature}. 
  
We now turn to  
finite temperature,  corresponding to   compactifying the theory on  $\R^{2} \times \S^1_L \times \S^1_\beta$. Electrically charged particles, e.g., $W$-bosons, can now be excited according to their  Boltzmann weight  $e^{-\beta m_W}$. It turns out that their effect is non-negligible, as in   studies of deconfinement in the 3d Polyakov model 
 \cite{Agasian:1997wv, Dunne:2000vp} and in deformed Yang-Mills theory on $\R^{1,2} \times \S^1_L$ \cite{Simic:2010sv}. Thus,  near the deconfinement transition we must consider Coulomb gases of both electrically and magnetically charged excitations.  This is the main  rationale under the 
gauge theory/electric-magnetic Coulomb gas mapping on  $\R^{2} \times \S^1_L \times \S^1_\beta$. 
This mapping    reveals a rather rich structure that we have only begun to unravel. 

Below, in Sections \ref{SU2intro}, \ref{su3intro}, and \ref{suNintro}, we summarize the  main results of this paper.

\subsection{Deconfinement in $\mathbf{SU(2)}$ QCD(adj)}  
\label{SU2intro}

We first review our results 
for an $SU(2)/\Z_2$ gauge theory with $n_f$ massless adjoint Weyl fermions on $\R^{1,2} \times \S^1_L$. We show that the physics near the deconfinement temperature    is described by a classical 2d XY-spin model with a $U(1) \rightarrow \Z_4$-breaking perturbation. This is the theory of angular ``spin" variables  $\theta_x \in (0, 2 \pi]$,  		``living" on the sites $x$ of a 2d lattice (of lattice spacing set to unity) with basis vectors $\hat{\mu}$, with nearest-neighbor spin-spin interactions.  The partition function is  $Z = \int{\cal{D}} \theta e^{- \beta H}$ (``$\beta H$" is used to denote the action, i.e. $\beta$ below is not the inverse temperature), where: \begin{equation}
\label{z4model}
- \beta H =   \sum_{x; {\hat\mu = 1,2}} { \kappa \over 2 \pi} \cos ( \theta_{x  + \hat\mu} -\theta_{x}) + \sum_{x} \tilde{y} \cos 4 \theta_{x}~.
\end{equation}
When matching to the gauge theory, the lattice spacing is of the order of the size $L$ of the spatial circle. The spin-spin coupling $\kappa$ is normalized in a way useful for us later. 

We emphasize  that the equivalence of  (\ref{z4model}) to the finite-temperature gauge theory is not simply an effective model for an order parameter 
 based  on  Svetitsky-Yaffe universality \cite{CLNS-82/530}. Instead, the parameters of the lattice spin theory (\ref{z4model}) can be precisely mapped to the  microscopic parameters of the gauge theory, owing to the small-$L$ calculability of the gauge dynamics. This map is  worked out in Section \ref{su2non-perturbativethermal}, where the nature and role of the  perturbative or non-perturbative objects driving  the deconfinement transition is made quite explicit. We note that there have been earlier proposals and discussions of the role of various topological objects in the deconfinement transition, in the continuum and on the lattice, and that some bear  resemblance to our discussion (see, for example,  \cite{hep-ph/0610409,arXiv:0704.3181, arXiv:0806.1736,  Liao:2006ry, Giovannangeli:2001bh, KorthalsAltes:2005ph,D'Alessandro:2007su,D'Alessandro:2010xg} and references therein). However, the  study here stands out by being both analytic and  under complete theoretical control. In addition, we study QCD(adj), a theory with massless fermions, and not pure Yang-Mills theory.

In the lattice theory defined as  in (\ref{z4model}), the  XY-model vortices map to electrically charged $W$-bosons, while the $\Z_4$-preserving perturbation represents  the magnetic bions of the $SU(2)/\Z_2$ QCD(adj) theory.
The spin-spin coupling is expressed via the four-dimensional gauge coupling $g_4(L)$, the size of the spatial circle $L$, and the temperature $T$, as:
\begin{equation}
\label{kappasu2}
\kappa = {g_4^2(L)\over 2 \pi L T},
\end{equation}
and determines the strength of the Coulomb interaction between the $W$-bosons, while the (dual-) Coulomb interaction between magnetic bions  is proportional to $\kappa^{-1}$. 

The global $U(1)$ symmetry of the XY-model, $\theta_x \rightarrow \theta_x + c$, is explicitly broken to $\Z_4$ by the magnetic-bion induced
potential   term in (\ref{z4model}). 
 In terms of the symmetries of the microscopic $SU(2)/\Z_2$ gauge theory, the $\Z_4$ symmetry of the spin model (\ref{z4model}) contains a discrete $\Z_2^{\rm d \chi}$ subgroup of the chiral symmetry of the gauge theory 
and the topological $\Z_2^{\rm t}$ symmetry, which arises due to the nontrivial homotopy $\pi_1(SU(2)/\Z_2)$. The combination of these two symmetries accidentally enhances to give a  $\Z_4^{\rm t/ d \chi}$
symmetry. 

There exists a   lattice formulation different from (\ref{z4model}) and appropriate to an $SU(2)$ gauge theory (which has trivial $\pi_1$, see the discussion in Section \ref{dualsymmetries}), where, instead of the topological $\Z_2^{\rm t}$ symmetry, one finds a $\Z_2^{\rm c}$ center symmetry. In this case, the discrete symmetry group of the theory is $\Z_2^{\rm c} \times \Z_2^{\rm d \chi}$, and does not enhance to a  $\Z_4$.  The realization of the symmetries and the behavior of the correlators of appropriate 't Hooft and Polyakov loops above and below the deconfinement transition  are  discussed  in Section \ref{symabovebelow} and are  summarized in eqns.~(\ref{so3symmetry}) and (\ref{su2symmetry}).

The lattice-spin model (\ref{z4model}) is a member of a class of $\Z_p$-preserving models, defined as in (\ref{z4model}), but  with $\cos 4 \theta_x \rightarrow \cos p \theta_x$ instead; these are sometimes also called the $\Z_p$ ``clock" models, because, in the limit of large $\tilde{y}$, the ``spin" $e^{i \theta}$ is forced to take one of $p$ ``clock" values. The $\Z_4$ model  stands out in this class in that the critical renormalization group trajectory is under theoretical control, at small fugacities, along the entire renormalization group flow to the fixed point describing the theory at $T_c$. This makes it possible to obtain the analytic result for the divergence of the correlation length,  given  below in eqn.~(\ref{corrlength1}), see also Section \ref{rgessection} and Appendix \ref{corrappx}. 
The phase transition in the $\Z_4$ model occurs at $\kappa = 4$, thus, from (\ref{kappasu2}), the critical temperature is given by $T_c  \simeq {g_4^2(L) \over 8 \pi L}$.
We note that $\kappa=4$ is also the critical coupling for the usual Berezinskii-Kosterlitz-Thouless (BKT) transition of the XY-model without the $U(1) \rightarrow \Z_4$-breaking term. The  transition in the $\Z_4$ model is also  continuous, however, as opposed to BKT, it is of finite (but very large, at small fugacities) order. As we show in Section 
\ref{rgessection}, the correlation length   diverges as:
\begin{equation}
\label{corrlength1}
\zeta \sim|T - T_c|^{ -\nu } = |T - T_c|^{ -  {1\over 16 \pi \sqrt{y_0 \tilde{y}_0}} }~,
\end{equation}
as $T\rightarrow T_c$ from both sides of the transition.
Here,  $y_0$ and $\tilde{y}_0$ are exponentially small parameters, essentially determined by the fugacities of the $W$-bosons and magnetic bions at scales of order the lattice cutoff $L^{-1}$. At small values of these fugacities, the    critical   correlators  are essentially governed by a free field theory with  BKT ($\kappa = 4$) exponents. 
The phase transition in the $\Z_4$ ``clock" model at $\kappa = 4$ corresponds to the confinement/deconfinement transition in the $SU(2)$(adj) theory. Both the string tension $\sigma$ and the dual string tension  $\tilde\sigma$ vanish as 
the inverse correlation length $\zeta^{-1}$ as $T\rightarrow T_c$  from below ($\sigma$) and above ($\tilde\sigma$).

An important property of the rank-one case is that the electric-magnetic Coulomb gas dual to the spin model (\ref{z4model}) exhibits electric-magnetic ($e$-$m$) duality. This duality is not manifest in    eqn.~(\ref{z4model}), but is evident from the Coulomb gas representation, see Section \ref{dualsymmetries}. It involves exchanging the fugacities of bions and $W$-bosons and an inversion of the coupling (\ref{kappasu2}):
\begin{equation}
\kappa \Longleftrightarrow {16 \over \kappa}~.
\end{equation}
 Thus, the critical temperature $T_c$ is precisely determined by  the strength of the interaction  $\kappa$
 at the point where the Coulomb gas is self-dual. This $e$-$m$ duality property is shared by all $\Z_p$ ``clock" models. As usual with Kramers-Wannier-type dualities, it helps establish a candidate critical temperature. Using 2d CFT techniques, it has been  shown  \cite{Lecheminant:2002va} that, indeed, at the self-dual point the $\Z_{p=2,3,4}$ models map to known conformal field theories. For   $p=4$, this is a free massless scalar field, even at large fugacities.\footnote{For completeness, we note that the $p>4$ models have an intermediate massless phase, see, e.g.,\cite{Wenbook}.} 

  \begin{FIGURE}[ht]
    {
    \parbox[c]{\textwidth}
        {
        \begin{center}
        \includegraphics[angle=0, scale=0.50]{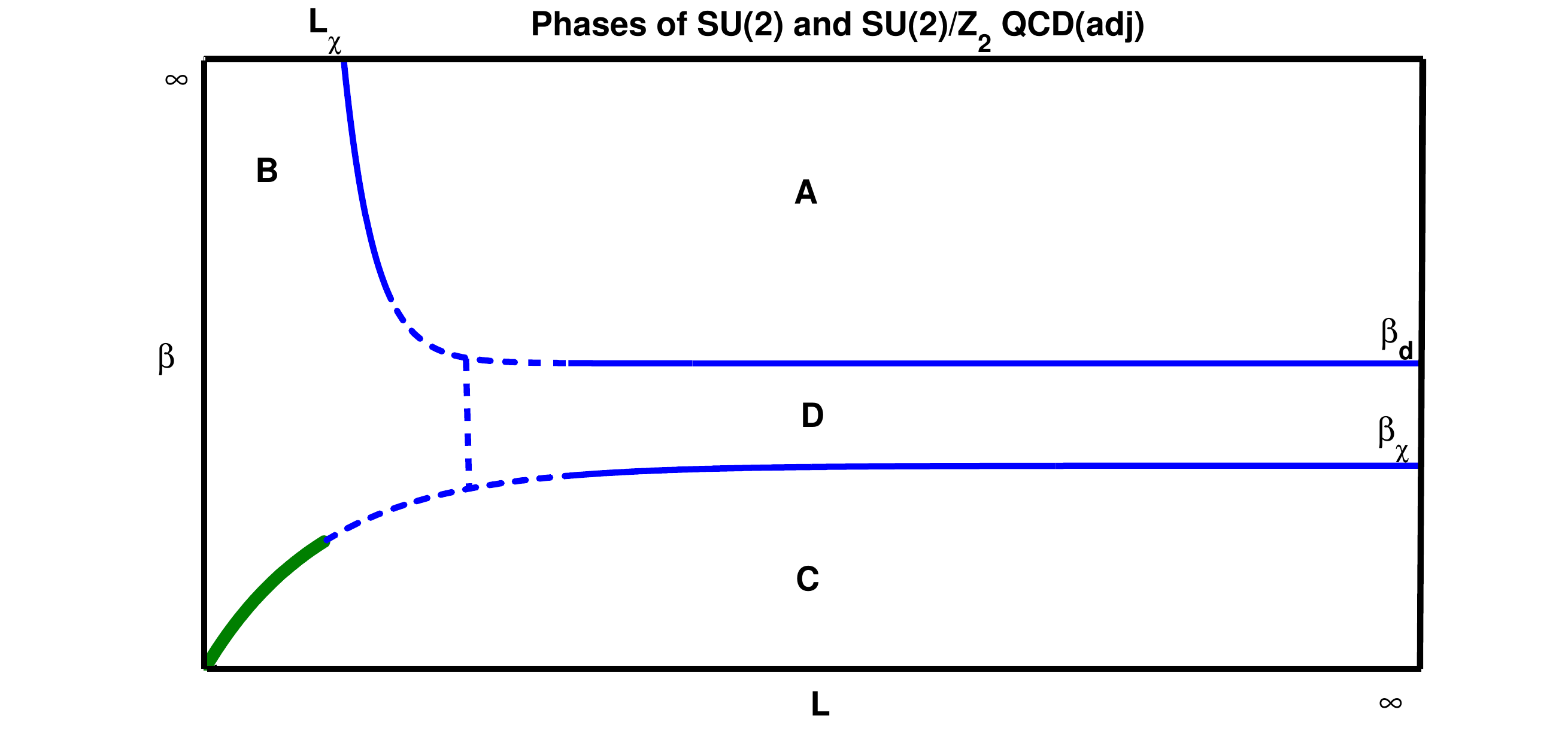}
	\hfil
        \caption 
      {  {\bf Symmetry realizations for  $\bm{ SU(2)}$ theory on $\bm {\R^2 \times \S^1_L \times \S^1_\beta}$:} The  $SU(2)$ theory possesses   $\Z_2^{\rm c} \times \Z_2^{\rm d\chi} \times SU(n_f)$ center, discrete, and continuous chiral symmetries ($\chi$S).       For brevity, in each phase we only list the broken symmetries.    {\bf A}: Discrete and continuous $\chi$S broken.       {\bf B}: Discrete $\chi$S broken. {\bf C}: Center symmetry broken. {\bf D}: All symmetries broken.         
{\bf Symmetry realizations for  $\bm {SU(2)/\Z_2}$ theory:}  
The theory possess  $\Z_4^{\rm t/\chi}  \times SU(n_f)$, where the first factor is the enhanced 
topological times discrete $\chi$S. 
{\bf A}: All symmetries broken. {\bf B}: $\Z_4^{\rm t/\chi}$ broken.  {\bf C}: None of the symmetries is broken. {\bf D}: Discrete and continuous $\chi$S broken.   In this work, we study the transition from {\bf B} to {\bf C} by reliable semi-classical  field-theoretic methods. The thick green line in the lower left corner is the phase boundary at small $L$ where the theory is under complete analytic control. The  field theory in the vicinity of this line can microscopically be mapped, for $SU(2)$ and $SU(2)/\Z_2$, respectively,   to  two types of XY-spin systems with perturbations. The dashed lines represent the conjectured extrapolation where the different phases coexist.
			}
        \label{fig:phase}
        \end{center}
        }
    }
\end{FIGURE}

The phase diagram of the $SU(2)$ theory in the $L-\beta$ plane (here, $\beta={1 \over T}$)  is shown in Fig.~\ref{fig:phase}. According to our current understanding of QCD(adj), the theory should have four phases. Phase-A exhibits confinement with discrete and continuous chiral symmetry breaking ($\chi$SB).
Phase-B has confinement with discrete but without continuous  $\chi$SB. The existence of this phase can be shown analytically \cite{Unsal:2007vu}.  Phase-C is a deconfined, chirally symmetric phase. The existence of this phase  can also be shown  analytically \cite{Gross:1980br}. Phase-D is deconfined with  discrete and continuous 
$\chi$SB. The existence of this phase is understood numerically through  the 
lattice studies \cite{Karsch:1998qj}.  Our current work addresses, 
by reliable continuum field theory techniques, the second-order transition between B and C. 
 
 A few comments on the B-C phase boundary are now due. In different theories, this phase boundary has different interpretation. 
\begin{enumerate} 
 \item In the $SU(2)/\Z_2$  theory as well as in the  $\Z_4$-spin system, the $\Z_4$ is broken in the low temperature phase and restored at high temperature phase. 
   \item In the $SU(2)$  theory as well as in the associated spin system, the symmetry is  $\Z_2^{\rm c}  \times \Z_2^{\rm d \chi}$. It is broken down to $\Z_2^{\rm c}$ in the 
   low-temperature phase, i.e., center symmetry is unbroken. At high temperatures, the symmetry is broken down to   $\Z_2^{\rm d \chi}$.    
\end{enumerate}
This  difference is there because the set of electric and magnetic charges that we are allowed to probe the  two gauge theories and spin systems are different.

\subsection{The theory of melting of 2d-crystals  and $\mathbf{SU(3)}$ QCD(adj)}
\label{su3intro}

Our next result concerns   $SU(3)/\Z_3$ QCD(adj) with $n_f$ massless adjoint Weyl fermions on $\R^{1,2}  \times \S^1_L$.
 The theory near the deconfinement transition is described by a spin model, which is a ``vector" generalization of (\ref{z4model}). This is the theory of two  coupled XY-spins, described by  
 two compact variables whose periodicity is determined by the $SU(3)$ root lattice:
 \begin{equation}
 \vec\theta_x =  (\theta^1_x, \theta^2_x) \equiv \vec\theta_x +  2 \pi  \vec\alpha_1 \equiv  \vec\theta_x +  2 \pi  \vec\alpha_2 ~.
 \end{equation}
 Here $ \vec\alpha_i$ ($i=1,2$) are the simple  roots of $SU(3)$, which can be taken to be $\vec\alpha_1$$=$$(1,0)$, $\vec\alpha_2$$=$$(-{1 \over  {2}} ,{ \sqrt{3}\over 2})$.
The theory is defined, similar to (\ref{z4model}), by a lattice partition function with:
\begin{equation}
\label{z3sqrdmodel}
- \beta H =   \sum_{x; \hat\mu = 1,2}  \sum_{i=1}^3 {\kappa \over 4 \pi} \cos  2  \vec\nu_i \cdot (  \vec\theta_{x  + \hat\mu} - \vec\theta_{x}) + \sum_{x} \sum_{i=1}^3 \tilde{y}   \cos 2 ( \vec\alpha_i - \vec\alpha_{i-1})\cdot \vec\theta_{x} .
\end{equation}
where $\vec\nu_i$ $(i = 1,2,3)$ are the weights of the defining representation (see Section \ref{su3lattice} for their explicit definition) and the sum in the potential term includes also  the affine root. 

The first term in (\ref{z3sqrdmodel}) is essentially\footnote{In disguise: see eqn.~(\ref{z3sqrdmodel1alt}) for another description of (\ref{z3sqrdmodel}), making its relation to \cite{Nelson} more obvious.}  the model used in \cite{Nelson} to describe the melting of a two-dimensional crystal with a triangular lattice. There, the  two-vector $\vec{\theta}_x = (\theta^1_x,\theta^2_x)$ parametrizes fluctuations of the positions of atoms in a 2d crystal around equilibrium (the distortion field) and $\kappa$ is proportional to the Lam\' e coefficients. A vortex of the compact vector field $\vec{\theta}_x$ describes a dislocation in the 2d crystal. The Burgers' vector of the dislocation is the winding number of the vortex, now also a two-component vector. Without going to details of the order parameters and phase diagram, see \cite{Nelson}, we only mention that the melting of the crystal occurs due to the proliferation of dislocations at high temperature, which destroys the algebraic long-range translational order.

To relate (\ref{z3sqrdmodel}) to our theory of interest, $SU(3)$ QCD(adj),  we shall argue in Section \ref{su3lattice}, that the  vortices in (\ref{z3sqrdmodel})   describe the electric excitations, the $W$-bosons, in the thermal theory.  The vortex-vortex coupling $\kappa$ is related to the parameters of the underlying theory exactly as in (\ref{kappasu2}). 
  In the melting applications of the ``vector" XY-model, there is an exact 
$U(1)\times U(1)$ global symmetry $\vec{\theta}_x$$\rightarrow$$\vec{\theta}_x + \vec{c}$. 
This symmetry forbids terms in the action which are not periodic functions of the difference operator. 
In QCD(adj), it is broken to $\Z_3 \times \Z_3$  by the potential term in (\ref{z3sqrdmodel}). This breaking is crucial  and  in the chosen duality frame describes the effect of the magnetically charged particles in the thermal gauge theory---the magnetic bions responsible for confinement. The two $\Z_3$ symmetries of (\ref{z3sqrdmodel}) are the topological $\Z_3$ (associated with the nontrivial $\pi_1(SU(3)/\Z_3)$) and the $\Z_3$ discrete subgroup of the chiral symmetry of the theory.

The nature of the deconfinement transition in the $SU(3)$ QCD(adj) is not yet completely understood.  One obvious feature   that follows from (\ref{z3sqrdmodel}) is that at low temperature (large $\kappa$, see (\ref{kappasu2})), where vortices can be neglected, the  topological/chiral $\Z_3 \times \Z_3$ symmetry is spontaneously broken, as expected in the confining phase (and also follows from the zero-temperature analysis).

The $SU(3)$ theory also exhibits electric magnetic duality,  which interchanges electric and magnetic fugacities and inverts the coupling in (\ref{z3sqrdmodel}):  
\begin{equation}
\label{su3dualitykappa}
\kappa \Longleftrightarrow {12 \over \kappa}~.
\end{equation}
 Thus, a candidate $T_c$ for the deconfinement transition is the self-dual point $T_c = {g^2 \over 4 \sqrt{3} \pi L}$.  
 At present, we do not know if at the self-dual point (\ref{z3sqrdmodel}) is a CFT (which would be the case if the transition was continuous). 
 
In Section \ref{sunrges}, we show that the leading order (in fugacities) renormalization group equation for $\kappa$ does, indeed, exhibit a fixed point at the self-dual point of $e$-$m$ duality (\ref{su3dualitykappa}). However, both $W$-boson and bion fugacities are relevant at that point---indicating that a strong-coupling analysis is necessary,  even if the fugacities are small at the lattice cutoff, unlike the $SU(2)$ case. We note that in $SU(3)$ QCD(adj) it should also be possible to use CFT methods at the self-dual point, as in     \cite{Lecheminant:2002va}, to study the existence and nature of the critical theory.

\subsection{Remarks on the affine XY-spin model and $\mathbf{SU(N_c > 3)}$}
\label{suNintro}

We define the affine XY-spin model by: 
\begin{equation}
\label{affine}
(- \beta H)^{\rm affine}  =   \sum_{x; \hat\mu = 1,2}  \sum_{i=1}^N {\kappa \over 4 \pi} \cos  2 \vec\nu_i \cdot (\vec\theta_{x  + \hat\mu} -\vec\theta_{x})~,   \qquad \theta_x \equiv \vec\theta_x + 2 \pi \vec\alpha_i ~.
\end{equation}
This is a natural Lie-algebraic  generalization of the XY model, and possesses a $U(1)^{N_c-1}$ global symmetry, $\vec\theta_{x}  \rightarrow \vec\theta_{x}  +\vec{c}$, where $\vec\theta$ is now an ($N_c$$-$$1$)-dimensional vector with periodicity defined by the simple roots of $SU(N_c)$. The physics of the phase transition in this model is expected to be a  generalization of BKT. However, as in the 
discussion of $SU(2)$ and $SU(3)$ theories, the spin systems relevant for gauge theories contain  certain symmetry-breaking perturbations.
There are two such  interesting deformations of (\ref{affine}) which relate it to  gauge theories. 
 One preserves a  $\Z_N$ subgroup of the global symmetry and the other preserves a 
 $\Z_N \times \Z_N$ symmetry (which accidentally enhances to $\Z_4$ for  $N=2$). 
 These deformations are:\footnote{As noted earlier, there is a well-known classification of 
 phases for the $U(1) \rightarrow \Z_p$ model, which crucially depend on $p$, see e.g.,\cite{Wenbook}. It is crucial to note that in $U(1)^{N-1} \rightarrow \Z_N$, $N$ does not play the same role as $p$, and the classification is  different.}
  \begin{equation}
\label{deform}
- \beta H = (- \beta H)^{\rm affine} +    \left\{ \begin{array}{ll} 
 \sum_{x} \sum_{i=1}^N \tilde{y}   \cos 2 \vec\alpha_i \cdot \vec\theta_{x} ~, &  \qquad \Z_N^{\rm t}~, 
  \\ \\
  \sum_{x} \sum_{i=1}^N \tilde{y}   \cos 2 (\vec\alpha_i- \vec\alpha_{i-1})\cdot \vec\theta_{x} ~  ,  
  & \qquad \Z_N^{\rm t} \times \Z_N^{\rm d\chi}~,
  \end{array} \right.
\end{equation}
where the sum  now  includes also the affine root (see Section \ref{suncoulombgas}). 

As in the discussion of $SU(2)$ and $SU(3)$, the 
vortices of the affine model can be identified as $W$-bosons in the gauge theory, forming an electric plasma with interactions  dictated by the Cartan matrix $\vec\alpha_i\cdot\vec\alpha_j$. 
The perturbations correspond  to the  proliferation of charges, corresponding to  magnetic monopoles for the $\Z_N^{\rm t}$ case and magnetic bions for   $\Z_N^{\rm t} \times \Z_N^{\rm d\chi}$ case. The interactions in the case of  magnetic monopoles are also governed by the Cartan matrix, up to an overall coupling inversion.\footnote{A brief note on the symmetries of (\ref{deform}) is due.
The $\Z_N^{\rm t}$ symmetry acts on $\vec\theta$ by shifts on the weight lattice, 
\begin{eqnarray}
\label{zntopological}
\Z_N^{\rm t}: && \vec \theta \longrightarrow  \vec \theta - 2 \pi k  \vec \nu_i,   \qquad k=0, \ldots N-1, \qquad   \text{for any } i , \cr\cr 
&& e^{2 i \vec\nu_j \cdot \vec\theta}   \longrightarrow e^{-4\pi k i  \vec \nu_j \cdot \vec \nu_i }   e^{2 i \vec \nu_j. \vec \theta} \;=\;
 e^{i \frac{2 \pi k}{N}}   e^{2 i \vec \nu_j \cdot \vec \theta} ~,
 \end{eqnarray}
where we used identities given in Footnote \ref{conventions} in Section \ref{su3lattice} (it might appear that there are $N-1$ symmetries in (\ref{zntopological}), but note that the difference between  transformations with different $i$'s  is a shift of $\vec\theta$ by a root vector, which by (\ref{affine}) is an identification, not a symmetry). 
Since the  ``monopole" perturbation   in (\ref{deform})   is  
$ e^{2 i \vec \alpha_j\cdot \vec \theta}  \equiv  e^{2 i \vec \nu_j \cdot  \vec \theta}  e^{-2 i \vec \nu_{j+1} \cdot  \vec \theta}$, its invariance under   $\Z_N^{\rm t}$ is manifest. The $\Z_N^{\rm t}$ invariance then trivially holds also for bion perturbations $e^{2 i \vec \alpha_j\cdot  \vec \theta} e^{- 2 i \vec \alpha_{j-1} \cdot \vec\theta}$. 
The bion-induced perturbation (the one on the bottom line in (\ref{deform})) has an extra independent $\Z_N^{\rm d\chi}$ symmetry, which rotates the monopole operator by $N^{\rm th}$ root of unity,  $ e^{2 i \vec \alpha_j\cdot \vec \theta}  \rightarrow 
 e^{i \frac{2 \pi q}{N}} e^{2 i \vec \alpha_j\cdot  \vec \theta} $.  This symmetry is a remnant of the discrete chiral symmetry in the gauge theory. In order to see the action of the symmetry on the  $\vec \theta$ 
 field, we define the Weyl vector, $\vec \rho= \sum_{i=1}^{N-1} \vec \mu_i$, where $\vec \mu_i$ are the fundamental weights of $SU(N)$, defined through the reciprocity relation   $\vec \alpha_i \cdot \vec \mu_j= \half \delta_{ij} $.  Note that $\vec \rho$ satisfies $\vec \rho\cdot  \vec \alpha_i =\half$ for $i=1, \ldots, N-1$, and $\vec \rho\cdot  \vec \alpha_N =\frac{1-N}{2}$. Consequently, 
 \begin{eqnarray}
\Z_N^{\rm d \chi}: && \vec \theta \longrightarrow  \vec \theta + \frac{2 \pi q}{N}  \vec \rho,   \qquad q=0, \ldots N-1, \cr \cr
&& e^{2 i \vec \alpha_j\cdot  \vec \theta}   \longrightarrow e^{i \frac{4\pi  q}{N}   \vec \alpha_j\cdot \vec \rho }   e^{2 i \vec \alpha_j\cdot  \vec \theta} \;=\;
 e^{i \frac{2 \pi q}{N}}   e^{2 i \vec \alpha_j\cdot  \vec \theta} ~.
 \end{eqnarray}
Therefore, the bion induced term   in (\ref{deform}) is also invariant under the $\Z_N^{\rm d \chi}$ symmetry.}

The affine $\Z_N$-model is appropriate for a bosonic gauge theory, such as 
 deformed Yang-Mills theory \cite{Unsal:2008ch}, or equivalently, massive QCD(adj) in a regime of sufficiently small $L$. The 
$\Z_N$ symmetry  in the spin-system  is associated with  the topological $\Z_N^{\rm t}$ in the
bosonic $SU(N)/\Z_N$ theory. The vortex-charge system exhibits  $e$-$m$ duality.
 The   charges (magnetic monopoles)  and vortices ($W$-bosons)  in the plasma have the same interactions, dictated by the Cartan matrix $\vec\alpha_i \cdot \vec\alpha_j$ for both $e$-$e$ and $m$-$m$ interactions.
  
 There are only a few remarks about $N_c>3$ QCD(adj) that we will make in this paper. First, 
the   $\Z_N \times \Z_N$ symmetry in (\ref{deform})  is associated with  the topological  
$\Z_N^{\rm t}$ and discrete $\Z_N^{\chi}$ chiral symmetries of  $SU(N)/\Z_N$ QCD(adj). 
 We show in Section \ref{suncoulombgas} that the  $SU(N_c > 3)$ QCD(adj) theory on $\R^{1,2} \times \S^1_L$ near the deconfinement temperature is described by an  $e$-$m$ Coulomb gas which  does {\it not} exhibit  $e$-$m$ duality. This is    in contrast to the $N_c>3$ case of the   deformed Yang-Mills theory, or  massive QCD(adj).  The absence of duality in the case of QCD(adj)   is due to the fact that, for $N_c>3$, the  magnetic (bions) and electric 
($W$-bosons) particles in the gas have different interactions, governed by:
\begin{eqnarray}
\label{WMproducts}
&&\vec\alpha_i \cdot \vec\alpha_j= \delta_{i,j}-\frac{1}{2} \delta_{i,j+1}-\frac{1}{2} \delta_{i,j-1} \qquad {\rm  for \; W-bosons}~, \cr
&& \vec{Q}_i  \cdot \vec{Q}_j=3\delta_{i,j}-2\delta_{i,j+1}-2\delta_{i,j-1}+\frac{1}{2} \delta_{i,j+2}+ 
\frac{1}{2}  \delta_{i,j-2}\quad \mbox{for magnetic bions}~.
\end{eqnarray}
The magnetic bions (of charges $\vec{Q}_i = \vec\alpha_i - \vec\alpha_{i-1}$) have next-to-nearest-neighbor interactions, in the sense of  their ``location" on the extended Dynkin diagram, while the $W$-bosons only have nearest-neighbor interactions. 
The different interactions are   ultimately due to the particular composite nature of bions, which is itself is due to the presence of massless fermions in the theory.  

In  Section \ref{sunrges} and Appendix \ref{sunrges1}, we derive the renormalization group equations for the $SU(N_c)$ QCD(adj) Coulomb gas to leading order in the fugacities and show that none of them displays a fixed point. 
 Thus, the  picture of the deconfinement transition for $N_c>3$ is, so far, not complete.  
 
  In Section \ref{future}, we list possible avenues for further studies.

\section{The zero temperature dynamics of QCD(adj) on $\mathbf{\R^{1,2} \times \S^1_L}$}
\label{zerotemperature}

\subsection{Review of perturbative dynamics at zero temperature}
\label{perturbative}

We consider four-dimensional (4d) $SU(N_c)$ Yang-Mills theory with $n_f$ massless Weyl fermions in the adjoint representation, a class of  QCD-like (vector) theories usually denoted QCD(adj). The action for $SU(N_c)$ QCD(adj) defined on $\R^{1,2}\times \S^1_L$ is:
\begin{eqnarray}\label{basic lagrangian}
S=\int_{\R^{1,2}\times \S^1_L}\mbox{tr}\left[-\frac{1}{2g^2}F_{MN}F^{MN}+ 2 i\bar \lambda^I \bar\sigma^MD_M\lambda_I \right]\,,
\end{eqnarray}
where we use the signature $\eta_{\mu\nu}=\mbox{diag}(+,-,-,-)$, $F_{MN}$ is the field strength, $D_M$ is the covariant derivative, $I$ is the flavor index, $\lambda_I=\lambda_{I,a}t_a$, $a=1,...,N_c^2-1$,  $\sigma_M=(1,\vec\tau)$, $\bar\sigma_M=(1,-\vec\tau)$, $1$ and $\vec\tau$ are respectively the identity and Pauli matrices, and the hermitean generators $t_a$, taken in the fundamental representation, are normalized as $\mbox{Tr}\,t_at_b=\frac{\delta_{ab}}{2}$. The upper case Latin letters $M, N$ run over $0,1,2,3$, while the Greek letters $\mu,\nu$ run over $0,1,2$. The components $0$ and $3$ denote time and the compact dimension, respectively. Thus $x^3 \equiv x^3 + L$, where $L$ is the circumference of the $\S^1_L$ circle.

 Notice that  we are compactifying a spatial direction and the time direction  $\in \R^{1,2}$   is non-compact.  Thus the fermions and the gauge fields obey periodic boundary conditions around $\S^1_L$. The action (\ref{basic lagrangian}) has a classical  $U(1) \times SU(n_f)$ chiral symmetry acting on the fermions.  
 The quantum theory has a dynamical strong scale $\Lambda_{QCD}$ such that, to one-loop order, we have:
\begin{equation}\label{beta function to two loop order}
g^2 (\mu)=\frac{16 \pi^2}{\beta_0} ~\frac{1}{\log(\mu^2/\Lambda_{QCD}^2)}\,, 
\end{equation}
where $\mu$ is the renormalization scale and $\beta_0=  {11 N_c - 2 n_f N_c\over 3}$; we will only consider a small number of Weyl adjoint fermions, $n_f < 5.5$, to preserve asymptotic freedom. The theory also has BPST instanton solutions with $2 N_c$ zero modes for every Weyl fermion and the corresponding 't Hooft vertex, 
$\sim e^{- {8 \pi^2 \over g^2}} \left({\rm det_{I,J}} \lambda_{I,a} \lambda_{J,a}\right)^{N_c}$, explicitly breaks the classical $U(1)$ chiral symmetry to its anomaly-free $\Z_{2 N_c n_f}$ subgroup.
The classical $SU(n_f) \times U(1)$ chiral symmetry   of the theory is thus reduced to:
\begin{equation}
\label{chiralsymmetry}
 {SU(n_f ) \times \Z_{2 N_c n_f} \over \Z_{n_f}}~ ,
\end{equation} where the $\Z_{n_f}$ factor common to $SU(n_f)$ and $\Z_{2 N_c n_f}$ is factored out to prevent double counting. The $\Z_2$ subgroup of the $\Z_{2 N_c n_f}$ is fermion number modulo 2, $(-1)^F$, which cannot be spontaneously broken so long as Lorentz symmetry is unbroken. Thus, the only genuine discrete chiral symmetry of $SU(N_c)$ QCD(adj) which may potentially be broken is the remaining $\Z_{N_c}$, which we label as  $\Z_{N_c}^{\rm d \chi}$, 
irrespective of the number of flavors. 

An important variable is the Polyakov loop, or holonomy, defined as the path ordered exponent in the $\S^1_L$ direction:
\begin{equation}
\label{omega1}
\Omega_L(x)=Pe^{i\int_{\S^1_L} A^3(x,x^3)}\,,
\end{equation}
where $x\in \R^{1,2}$. This quantity transforms under $x$-dependent gauge transformation as $\Omega_L\rightarrow U^{-1}\Omega_L U$, and hence the set of  eigenvalues of $\Omega_L$ is gauge invariant.  The gauge invariant trace of the Polyakov loop $\tr \Omega_L$ serves as an order parameter for the  $\Z^{(L)}_{N_c}$ global center symmetry  $\tr \Omega_L \rightarrow e^{i {2 \pi k\over N_c}} \tr \Omega_L$, with $k = 1, \ldots N_c$.
If we take $L \Lambda_{QCD} \ll 1 $, a reliable one-loop  analysis  can be performed to study the realization  of the $\Z^{(L)}_{N_c}$ center symmetry. Integrating out the heavy Kaluza-Klein modes along $\S^1_L$, with periodic boundary conditions for the 
 fermions---remembering that our $\S^1_L$ is a spatial, not a thermal circle---results in an effective potential for $\Omega$:
\begin{eqnarray}
\label{abc}
V_{\mbox{\scriptsize eff}}(\Omega_L)=(-1+n_f)\frac{2}{\pi^2L^4}\sum_{n=1}^{\infty}\frac{1}{n^4}|\mbox{tr}\, \Omega_L^n|^2\,.
\end{eqnarray} 
Since $\Omega_L(x)\equiv e^{iL A_3(x)}$, see eqn.~(\ref{omega1}), the action for the $x^3$-independent modes of the gauge field is:\footnote{\label{higgsnote}To avoid confusion, we note that in subsequent sections we study the Euclidean theory, where, for convenience, we relabel the compact direction $x^4$ and the ``Higgs field" $A_4$.}
\begin{eqnarray}
\label{action1}
S_0=\int_{\R^{1,2}}\frac{L}{g^2}\tr\left[ -\frac{1}{2}F_{\mu\nu}F^{\mu\nu}+\left(D_\mu A_3\right)^2  - {g^2\over 2}V_{eff}(A_3)+ 2 i g^2\bar \lambda ^I \left(\bar\sigma^\mu D_\mu - i \bar \sigma_3\left[A_3,\right] \right)\lambda_I\right].
\end{eqnarray} 
The minimum of the potential $V_{\mbox{\scriptsize eff}}$ (\ref{abc}) for $n_f > 1$ is located, up to conjugation by gauge transformations, at  \cite{Unsal:2006pj}: 
\begin{equation}
\label{omegavev}
\langle \Omega_L \rangle= \eta \; {\rm diag} (1, \omega_{N_c}, \omega_{N_c}^2, \ldots \omega_{N_c}^{N_c-1} ) ~,~~  \omega_{N_c} \equiv e^{i {2 \pi \over N_c}}~, 
\end{equation}
where $\eta = e^{\pi i \over N_c}$ for even $N_c$ and $\eta =1 $ otherwise.  
Since $\tr \langle \Omega \rangle=0$, the $\Z^{(L)}_{N_c}$ center symmetry is preserved. This is in contrast with the $n_f = 0$ 
case, where the theory on $\S^1_L$ is equivalent to a finite temperature pure Yang-Mills theory, where the theory deconfines at sufficiently high temperature \cite{Gross:1980br}.
In the supersymmetric $n_f = 1$ case the one-loop  potential vanishes and the center-symmetric vacuum is stabilized due to non-perturbative corrections \cite{Seiberg:1996nz,Aharony:1997bx,Davies:1999uw}. 

In the vacuum (\ref{omegavev}), the $SU(N_c)$ gauge symmetry is broken\footnote{The dynamical abelianization on  $\R^{1,2} \times \S^1_L$ in QCD(adj),   as in  the Coulomb 
branch of the Seiberg-Witten theory \cite{Seiberg:1996nz} and the
 3d Polyakov model \cite{Polyakov:1976fu},  is the ultimate  reason  that the theory admits a semi-classical analysis. It also helps to  continuously connect two  confinement mechanisms, which were previously thought to be unrelated \cite{Poppitz:2011wy}. See  \cite{Nishimura:2011md} for a  bosonic version.    Another interesting recent  work taking advantage of the abelianization, this time on $\R^{1,1} \times \S^1_L$, is \cite{Armoni:2011dw}.}  by the Higgs mechanism down to $U(1)^{N_c-1}$.
 Because the unitary ``Higgs field" $\Omega_L$ is in the adjoint, its off-diagonal components are eaten by the gauge fields which become massive. At the lowest mass level, the $W$-bosons have mass $m_W = {2 \pi \over  L N_c}$; the heavier $W$-bosons have mass $2 \pi k\over L N_c$ with $k=2,\ldots N_c$.  The $N_c-1$ diagonal components of the Higgs field also obtain masses $\sim {2 \pi k\over L N_c} \sqrt {g^2N_c} $,  due to the one-loop potential $V(A_3)$ \cite{Argyres}. In addition, the off-diagonal  fermion components (the color space components of $\lambda$ that do not commute with $\langle A_3 \rangle$ also acquire mass, with masses identical to those of the $W$-bosons). The coupling $g$ ceases to run at $1/L \gg \Lambda N_c$. 

 The conclusion is, then, that for distances greater than $L N_c$ (not $L$), the three-dimensional (3d) low energy theory is described in terms of the diagonal components of the gauge field (the $N_c-1$ ``photons") and  the $(N_c-1) \times n_f$ diagonal components of the adjoint fermions (denoted by $\lambda_{k, I}$, with $I = 1, \ldots n_f$, $k = 1, \ldots N_c -1$). At the perturbative level, up to higher dimensional operators suppressed by powers of $L \sim m_W^{-1}$, the Lagrangian is: 
\begin{equation}
\label{longdistancefree}
S=\int_{\R^{1,2}}-\frac{L}{4g^2} \sum_{i = 1}^{N_c-1} F_{(i) \mu\nu}^2+i L \sum_{k = 1}^{N-1} \sum_{I=1}^{n_f}\bar\lambda_{(k)}^I\bar\sigma^\mu\partial_{\mu}\lambda_{(k) I}\,,
\end{equation}  
i.e.,  that of a free field theory. In the next section, we review how non-perturbative effects change this Lagrangian. 
 
 \subsection{Review of the $\mathbf{SU(2)}$   non-perturbative dynamics at zero temperature}
 \label{non-perturbativesu2review}
 
We will specialize to $N_c=2$ in this section. 
 To further cast the free theory (\ref{longdistancefree}) into a form that will be useful in our further analysis, we note that we can dualize the 3d ``photon" into a periodic scalar field---the ``dual photon." The duality transformation can be performed by first introducing  a Lagrange multiplier field $\sigma$ imposing the Bianchi identity. Thus, we introduce an additional term to the action $S$, i.e., eqn.~(\ref{longdistancefree}) with $N_c=2$:  
\begin{equation}
\label{deltas}
\delta S= { 1 \over 8 \pi} \int_{\R^{1,2}}  \sigma \epsilon_{\mu\nu\lambda} \partial^\mu F^ {\nu\lambda}~.
\end{equation} Note that we are still considering the duality in Minkowskian signature and that in Euclidean space, a factor of $i$ will appear.   Integrating over $F_{\mu\nu}$ in the path integral then amounts to  substituting:
\begin{equation}
\label{duality}
F_{\mu\nu} = - {g^2 \over 4 \pi L} \epsilon_{\mu\nu\lambda} \partial^\lambda \sigma
\end{equation}
 into $S + \delta S$.  The result is the action written in terms of the ``dual photon" field $\sigma$: 
\begin{equation}
\label{perturbativeSeff}
S = \int_{\R^{1,2}}{1\over 2 L} \left( g \over 4 \pi  \right)^2  \left(\partial \sigma\right)^2+i L \sum_{I=1}^{n_f}\bar\lambda^I\bar\sigma^\mu\partial_{\mu}\lambda_{ I}\,. 
\end{equation}  
which captures the long-distance physics of the photons to all orders in perturbation theory and to leading order in the derivative expansion. 

The next step is to consider the modification of (\ref{perturbativeSeff}) by non-perturbative effects.
However, before we study the non-perturbative dynamics, we shall make some comments on 
 $SU(2)$ vs.~$SU(2)/\Z_2$ theory,  discrete fluxes,  and  the topological $\Z_2$ symmetry. Our comments are complementary to those of \cite{hep-ph/0009138}; they also generalize to higher-rank groups and will be presented elsewhere in more detail. The brief  discussion in Section \ref{hamiltonian} below should be useful  to get   a picture of the origin of the topological $\Z_2$ symmetry---which plays an important role in the analysis of the finite-temperature theory and the deconfinement transition in 
 $SU(2)/\Z_2$ theory. For  $SU(2)$, the center symmetry plays an analogous role.

\subsubsection{Hamiltonian quantization of compact $\mathbf{U(1)}$, $\mathbf{SO(3)}$ vs. $\mathbf{SU(2)}$,  discrete fluxes, and the topological 
$\mathbf{\Z_2}$ symmetry.}
\label{hamiltonian}

In this section, we explain the difference between the $SU(2)$ vs.~$SU(2)/\Z_2$ theory in the Hamiltonian formulation. 
Let us first focus on the bosonic part of  (\ref{action1}), without considering  fermions. Such a bosonic theory  arises in the study of QCD(adj) with massive fermions or deformed Yang-Mills theory.  This bosonic theory, in the long-distance regime and to all-orders in perturbation theory, admits a 
description in terms of the  bosonic part of (\ref{perturbativeSeff}):
\begin{equation}
\label{dualSigma1}
S = \int_{\R^1 \times \T^2} {1\over 2 L} \left( {g \over 4 \pi  }\right)^2  \left(\partial \sigma\right)^2, 
\end{equation}  where we are now considering the  theory with space additionally compactified on a large $\T^2$ of area $A_{\T^2}$, i.e.,~on $\R^{1} \times \T^2 $.  
This  is helpful 
 to understand the appearance of discrete fluxes and the associated topological $\Z_2$ symmetry.
 
  It is well known that on $\T^2$, a compact $U(1)$ theory has magnetic flux sectors, see, e.g.,\cite{Wenbook}, labeled by integers  $n$ counting the unit of flux quanta on $\T^2$ and that, for a large $\T^2$, the energies of these flux sectors are almost degenerate.
 To see these sectors in the dual-photon language, consider the dynamics of the zero mode $\sigma_0$ of the $\sigma$-field on $\T^2$. The zero mode is governed by the quantum-mechanical action: 
 \begin{equation}
\label{dualSigma2}
S_0 = {A_{\T^2} \over 2 L} \left( {g \over 4 \pi } \right)^2 \int dt \left({\partial \sigma_0 \over \partial t }\right)^2\equiv {M \over 2} \int d t \; \dot{\sigma}_0^2,\;\;\;{\rm where} \; \; M \equiv {A_{\T^2} \over   L} \left( {g \over 4 \pi } \right)^2~.
\end{equation}  
The action $S_0$ is   obtained upon integration of (\ref{dualSigma1}) over the $\T^2$ by keeping only the zero mode. Clearly, the zero mode dynamics is that of a free ``particle" of ``mass" $M$.  To see that the ``particle" is actually moving on a circle and infer the size of the circle, i.e., its radius $R$, $\sigma_0 \equiv \sigma_0 + 2 \pi R$, consider  the solution  of the equation of motion, $\sigma_0 = t\; V + const$, where $V$ is the constant ``velocity" of the particle. From the duality relation (\ref{duality}), a constant value of $\dot\sigma_0 = V$ implies that there is a constant magnetic field on the $\T^2$: 
 \begin{equation}
 \label{f12}
 F_{12} = -{ g^2 \over 4 \pi L} \; V~.
 \end{equation}  
 
At this point, it is useful to ask whether the nonabelian theory underlying the $U(1)$ is an
$SO(3)$=$SU(2)/\Z_2$ or an $SU(2)$ theory. These are distinguished by the kinds of $U(1)$-charges that are allowed in the spectrum, as we now explain:
 \begin{enumerate}
 \item In the $SO(3)$ theory, a charged particle can only have integer charge $q=\pm1, ...$ under the unbroken $U(1)$. In order that the phase factor $e^{ i q \int_{C=\partial S} A} = e^{i q \int_S B}$   acquired by such a particle upon its motion on $\T^2$ be unambiguously defined,\footnote{Consider an  oriented Wilson loop $C$ on a a compact two dimensional space such as $\T^2$. 
 There is no unique  surface  whose  boundary is $C$.  Both ``inside" and ``outside" surfaces, $S_i$ and $S_o$, are acceptable choices leading to   a two-fold ambiguity, 
 which is resolved provided    $e^{ i q \int_{C} A} = e^{i q \int_{S_i} B} =  e^{- i q \int_{S_o} B}$, i.e., the phase  is independent of the choice. This implies,  $q \int_{S_i} B +  q \int_{S_o} B =  q \int_{\T^2} B =  2 \pi n$.}
  it must be that $ \int_{\T^2} B = F_{12} A_{\T^2} = 2 \pi n$, with $n\in \Z$. Eqn.~(\ref{f12}) then implies that the ``velocity" of   $\sigma_0$ is quantized, $V  = n \;  {8 \pi^2 L \over g^2 A_{\T^2}}$. The  momentum conjugate to $\sigma_0$ is  
  $\Pi_{\sigma_0} = M V =  {n \over 2 } $ and the energy is:
 \begin{equation}
 \label{so3fluxenergy}
 E^{SO(3)}   = {1\over 2 M}\; \Pi_{\sigma_0}^2  = {1 \over2 M} \; \left({n \over 2 } \right)^2~,  \qquad 
  \left\{ \begin{array}{l} 
   \Pi_{\sigma_0}   \in \half\Z, \qquad   {\rm for}  \;\; n\in  \Z \\
  \Pi_{\sigma_0}   \in \Z,  \qquad \;\;\;  {\rm for} \;\;   n\in  2\Z
  \end{array}\,.
  \right.
 \end{equation} 
We refer to the sector with half-integer (integer)  conjugate momenta $\Pi_{\sigma_0}$  as half-integer  (integer) flux sector. 
 The expression of the energy (\ref{so3fluxenergy}) is the same obtained upon quantizing the motion of a particle of mass $M$   on a circle of radius $R=2$. This implies that the coordinate  $\sigma_0$  of the particle from (\ref{dualSigma2}) is a periodic variable of period $4 \pi$. Hence, we conclude that the dual photon $\sigma$  in the $SO(3)$ theory is a compact field with periodicity $2\pi$:
 \begin{equation}
 \label{so3sigma}
\sigma \equiv \sigma + 4\pi ~, ~~ {\rm for \; the } ~ SO(3) ~ {\rm theory}. 
 \end{equation}
Furthermore, since $M \sim A_{\T^2}$, see (\ref{dualSigma2}), the above-mentioned near-degeneracy of the flux sectors at large $\T^2$ is evident from (\ref{so3fluxenergy}).

As functions of the zero mode $\sigma_0$ of the dual photon  on $\T^2$, the flux sectors have Schr\" odinger wave functionals $\Psi_n(\sigma_0) = e^{i n \sigma_0/2}$.  We shall later argue that  shifts of $\sigma$ by $2\pi$ play a special role, namely that such shifts constitute  a symmetry of the $SO(3)$ theory, called the ``topological $\Z_2$ symmetry," associated with the nontrivial $\pi_1(SO(3))$. Under such  shifts of $\sigma$, we have:
\begin{equation}
\label{psin}
\Z_2^{\rm t}: \;~  \sigma_0 \rightarrow \sigma_0 + 2\pi,~ \; \Psi_n(\sigma_0) \rightarrow (-)^n \Psi_n(\sigma_0).
\end{equation}
All states in the Hilbert space (at finite $\T^2$) of the $SO(3)$ theory are either even or odd under the topological $\Z_2$, as in (\ref{psin}).  
 \item In the $SU(2)$ theory, on the other hand,  particles   can have half-integer charge $q = \pm {1 \over 2},...$ under the $U(1)$ (imagine adding a heavy ``electron" in the fundamental representation to QCD(adj)). In order that their propagation on $\T^2$ be well defined (i.e., the phase factor $e^{ i {1\over 2} \int_C A}$ be unambiguous), it must be that ${1 \over 2} F_{12} A_{\T^2} = 2 \pi n$, $n \in \Z$. Then, the ``velocity" of $\sigma_0$ is quantized as $V = 2 n  {8 \pi^2 L \over g^2 A_{\T^2}}$, the conjugate momentum is $\Pi_{\sigma_0} = M V =  n$, and the energy is: 
  \begin{equation}
 \label{su2fluxenergy}
 E^{SU(2)}   = {1\over 2 M} \; \Pi_{\sigma_0}^2  = {1 \over2 M} n^2~,  \qquad  
  \Pi_{\sigma_0}   \in \Z,  \qquad \;\;\;  {\rm for} \;\;   n\in  \Z  ~.
 \end{equation} 
Thus, in the $SU(2)$ theory,  only the integer flux sector exists. Eqn.~(\ref{su2fluxenergy}) is exactly the quantum of energy of a particle of mass $M$ moving on a circle of radius $R=1$. Thus, the coordinate of the particle  $\sigma_0$ in the $SU(2)$ theory has period $2 \pi R = 2\pi$. We conclude that
 the dual photon in the $SU(2)$ theory is a compact field, but with half the periodicity of (\ref{so3sigma}): 
 \begin{equation}
 \label{su2sigma}
\sigma \equiv \sigma +  2 \pi ~, ~~ {\rm for \; the} ~ SU(2) ~ {\rm theory}. 
 \end{equation}
 \end{enumerate}

Next, we recall that the compact $U(1)$ theory (\ref{dualSigma1}) is a low-energy remnant of the broken nonabelian theory and that the latter has monopole-instanton solutions (at sufficiently large $\T^2$ one can use the infinite volume solutions). 
The monopole-instantons  are finite Euclidean action solutions  and play an important role in the non-perturbative dynamics. They affect  the perturbative long-distance Lagrangian (\ref{perturbativeSeff}) in a manner to be
 described in the following section.
 
  These solutions are constructed from the 4d static 't Hooft-Polyakov monopoles as recently reviewed, e.g., in \cite{Anber:2011de}. As explained there,   in a $U(1)$ theory which descends from an $SU(2)$ or  $SO(3)$ theory with a compact adjoint Higgs field, there are two types of monopoles,\footnote{Of lowest action, which are most relevant in the small-$L$ regime we study here; there is an infinite Kaluza-Klein  tower of monopole-instanton solutions of increasing action.} ${\cal M}_1$ and ${\cal M}_2$, and their anti-monopoles, 
 see also Section \ref{nonpert2} (the second type of monopole is present due to the topology of the adjoint Higgs-field).
  However, most of the argument in the bosonic  theory described below is independent of this aspect, because the magnetic charge and action of  ${\cal M}_1$ and  $\overline {\cal M}_2$ (not ${\cal M}_2$) coincide.   These classical solutions are, of course, identical in the $SU(2)$ and $SO(3)$ theories.  
The magnetic charge of a 't Hooft-Polyakov monopole-instanton  is:
\begin{equation}
\label{charge}
 \int_{\S^2_\infty}  \vec{B} \equiv {1 \over 2} \int_{\S^2_\infty}    \epsilon_{\mu\nu\lambda}  F^{\nu\lambda}  = 4 \pi Q_m, ~ {\rm with} ~ |Q_m|=1~,
 \end{equation}
  i.e., we take the abelian long-range field of a monopole to be $\vec{B} = Q_m {\vec{r} \over r^3}$, without a factor of $4\pi$ in the denominator. The Dirac quantization condition is \cite{colemanlectures}:
\begin{equation}
\label{diracquant}
Q_m Q_e = {n \over 2}, \qquad  n \in \Z~,
\end{equation}  and the electric charge of $W$-bosons is $Q_e = \pm 1$. The 't Hooft-Polyakov monopole solutions have $Q_m= \pm 1$. This is  twice the allowed minimal value of magnetic charge in the $SO(3)$ theory with $Q_e^{min}=\pm 1$, and is exactly equal to the minimal value of magnetic charge in the $SU(2)$ theory, where $Q_e^{min}=\pm {1 \over2}$ instead.

 The Hilbert space interpretation of the monopole-instantons is that they facilitate tunneling between  states of different magnetic flux on $\T^2$. From (\ref{charge}), it is clear that a monopole-instanton tunneling event changes the magnetic flux on $\T^2$ by one unit of magnetic flux.\footnote{To see this,  deform  the $\S^2_\infty$ integral (\ref{charge}) to an integral over two infinite planes separated by a large Euclidean time interval $2T$, i.e., $4\pi = \int_{\S^2_\infty} \vec{B} = \int_{\T^2_\infty, T}\vec{B} - \int_{\T^2_\infty, -T}\vec{B}$, and recall that a ``half-unit" flux on $\T^2$ corresponds to $\int_{\T^2} B = 2 \pi$; to avoid confusion, recall that after eqn.~(\ref{so3fluxenergy}) we adopted the convention of calling the minimal flux allowed ``half-unit", corresponding to a half-integer $n$ in eqn.~(\ref{instflux}) below.}  
 We note, for completeness, that 
 the  tunneling events that connect the vacua in different flux sectors are of two-types, call them $I_1$ and $I_2$, these are the descendants of  ${\cal M}_1$ and ${\cal M}_2$ monopoles mentioned above,    and they lead to change in the flux by  one unit, i.e., 
 \begin{eqnarray}
 \label{instflux}
&& I_1 \; : \;|n \rangle  \rightarrow  |n+1 \rangle, \qquad  I_2 \; : \;|n \rangle  \rightarrow  |n-1 \rangle~, ~~ n = 0, \pm {1 \over 2}, \pm 1, \ldots~, \cr 
 && \overline {I}_1 \; : \;|n \rangle  \rightarrow  |n-1 \rangle, \qquad 
  \overline{ I_2} \; : \;|n \rangle  \rightarrow  |n+1 \rangle \,.
 \end{eqnarray}
 For the purpose of our discussion,  $I_1$ and  $\overline{ I_2}$  play the same  role as both increase the flux by one unit.\footnote{In the  Hamiltonian formulation of the Polyakov model, we would only have the   $I_1$-type instanton. This follows from the non-compactness of the adjoint Higgs. 
 In deformed Yang-Mills, or massive QCD(adj), due to the compact nature of the adjoint Higgs, 
 there are  two-types of flux changing instantons. Also, note that the existence of this second type of instanton is the main difference with respect to ref.~\cite{Wenbook}.
 The existence of the two-types of flux-changing instantons has important  physical implications in  (at least)  two  cases: when massless fermions or a non vanishing $\theta$ angle is introduced. These  aspects  will be discussed elsewhere. }
 To study the consequences of the tunneling between the various flux sectors,  we have to, again, make a distinction between the $SO(3)$ and $SU(2)$ theories (as illustrated on Figure 2):

  \begin{FIGURE}[ht]
    {
    \parbox[c]{\textwidth}
        {
        \begin{center}
        \includegraphics[angle=-90, scale=0.55]{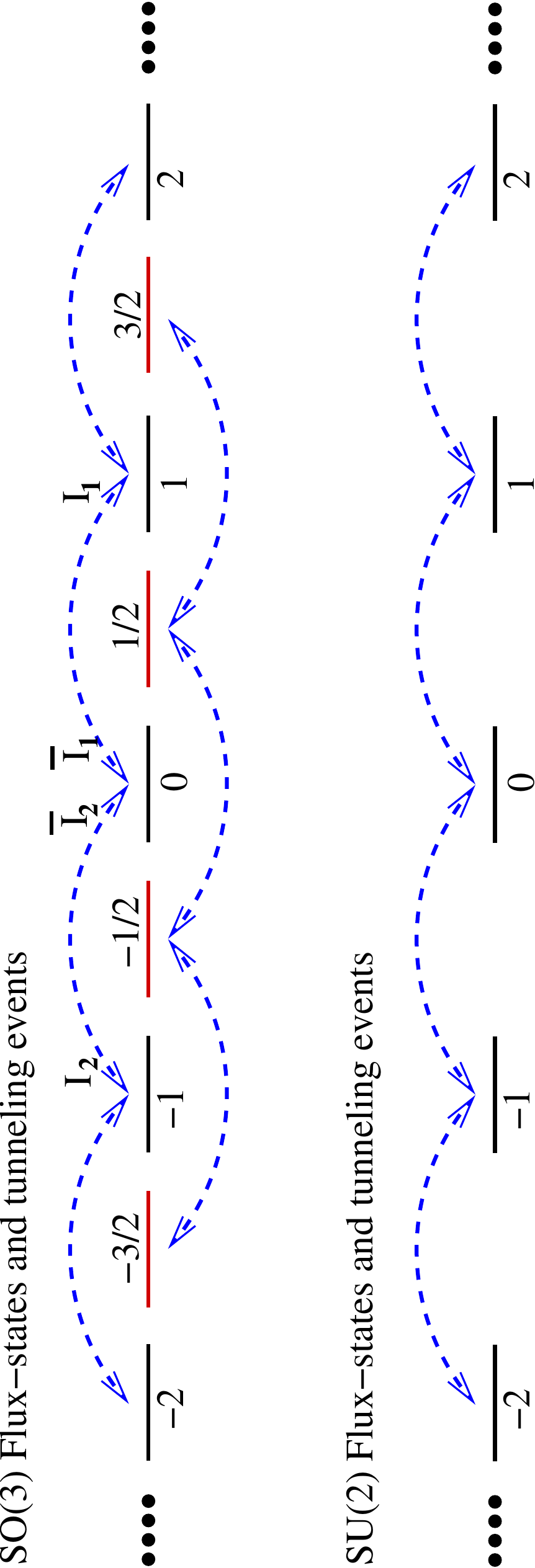}
	\hfil
        \caption 
      {  In an $SO(3)$ theory on $\T^2$, the flux sectors are associated  with integers and half-integers, as per eqn.~(\ref{so3fluxenergy}). In an $SU(2)$ theory, only integer flux sectors exist, as (\ref{su2fluxenergy}) shows. In both, the tunneling events 
      (which are a realization of the 't Hooft-Polyakov monopole-instanton  on  $\T^2$) 
      change flux by one unit.  
      Consequently, the $SU(2)$ theory has a unique vacuum, which is a superposition of all flux sectors, whereas the $SO(3)$ theory has two vacua, distinguished by their $\Z_2$ quantum number (\ref{psin}):  one is the  superposition of half-integer flux sectors and the other is the one of the integer flux sectors. This is the Hamiltonian realization of 
      topological $\Z_2$ symmetry.  
			}
        \label{fig:minima}
        \end{center}
        }}
\end{FIGURE}

 \begin{enumerate}
 \item
  In   the $SU(2)$ theory, the 't Hooft-Polyakov instanton-monopoles facilitate tunneling between sectors of magnetic flux differing by one unit of flux quantum. Since the $SU(2)$ theory only allows an integer number of magnetic flux quanta, as per (\ref{su2fluxenergy}), one expects that at sufficiently large $\T^2$ the true ground state is a unique superposition of all flux sectors.

    \item In the $SO(3)$ theory,   both integer  and half-integer sectors  are  allowed, (\ref{so3fluxenergy}), but,  as in the $SU(2)$ theory, 
  tunneling events can only change the flux on $\T^2$ by one unit of flux---there is only tunneling between sectors of integer  flux   and between sectors of half-integer flux, but not between integer  and half-integer  flux sectors. Thus, the broken $SO(3)$ theory actually has two flux sectors that do not interact with each other. These  can be labeled by a $\Z_2$ quantum number, as in (\ref{psin}): states in Hilbert space have eigenvalue $+1$ under the topological $\Z_2$ symmetry if they are in the integer  flux sector  and $-1$ if they are in the half-integer
   flux sector.

  The integer and half-integer flux sectors in the $U(1)$ theory are the long-distance remnants of the  two 't Hooft magnetic flux sectors\footnote{In the full  theory, the $\Z_2$-odd (``half-integer" in our convention) flux sector is constructed by considering  $SO(3)$  bundles on $\T^2$  twisted  along one of the non contractible loops  on $\T^2$   by gauge transformations in the topologically nontrivial class $\pi_1(SO(3))=\Z_2$. This way to construct  discrete flux sectors on tori is explained in \cite{PRINT-79-0117 (UTRECHT)}; see  \cite{hep-th/0006010} for different perspectives.} \cite{PRINT-79-0117 (UTRECHT)} in $SO(3)$ theories  on $\T^2$. In the limit of large $\T^2$ they become degenerate. The even and odd  magnetic flux\footnote{Also  their $\T^3$ generalization and the related discrete electric flux sectors, which we do not consider here.}  sectors in $SO(3)$ theories on $\T^2$ are distinct superselection sectors of the theory, unless  local operators  on the Hilbert space that can change the value of the $\Z_2$ flux are introduced.

The topological $\Z_2$ symmetry (\ref{psin}) is generated by an infinitely large spacelike Wilson loop \cite{thooft1}.  Let $C$ denote boundary of a surface $S$, which we eventually take to be  $\R^2$, formally, $C= \partial \R^2$.  
In the low-energy theory of the dual photon,  using  (\ref{duality}) and the definition of conjugate momenta, the Wilson loop operator  can be written as:
\begin{equation}
\label{Z2generator}
W=e^{i\int_C A dl }=e^{i\int B d^2x }=e^{i2 \pi \int d^2 x \Pi_\sigma(x)}~,
\end{equation}
where $\Pi_\sigma(x)$ is the momentum canonically conjugate to $\sigma$, i.e., with equal time commutator 
$ \left[\Pi_\sigma(x), \sigma(y)\right] =- i \delta(x-y)$ (the equal time argument is suppressed). The physical interpretation of $W$ is that it measures the amount of $\Z_2$ magnetic flux present at the given time.
The operator  $e^{i \sigma(x)/2}$ and $W$ obey the equal time  commutation relation:
\begin{equation}
\label{thooftalgebra}
W e^{i \sigma(x)/2} = - e^{i \sigma(x)/2} W,
\end{equation}
which is the 3d version of 't Hooft   algebra  for $SO(3)$. The interpretation of the operator $e^{i \sigma(x)/2}$ is that it creates a unit of $\Z_2$ flux at $x$ (a $\Z_2$ vortex). In the context of the underlying $\R^{1,2} \times \S^1_L$ theory,  the commutation relation  (\ref{Z2generator}) may be seen as the dimensional reduction of the 't Hooft commutation relation:
\begin{equation}
W(C) T(C')=  e^{i \pi \ell (C, C')} T(C') W(C) 
\end{equation}
where $\ell (C, C')$ is the linking number of the two closed loops $C$ and $ C'$, and 
 $T(C')$ is the  't Hooft operator.  We consider the limit in which the  inside of $C$ fills $\R^2$. $e^{i \sigma(x)/2}$ is the long-distance ``residue" of a 't Hooft loop of a $Q_m^{min} = {1\over 2}$ monopole in the 4d theory, winding around $\S^1_L$ and located at $x \in \R^2$, hence,  $\ell (C, C')=1$.

 If we allow the ``$\Z_2$-vortex" operator $e^{i \sigma(x)/2}$ in the Hilbert space,  see \cite{Argyres},
  correlation functions like $\langle e^{i \sigma(x)/2} e^{- i \sigma(y)/2} \rangle$ are observable,  and we have to consider the possibility of spontaneous symmetry breaking of the topological $\Z_2$ symmetry.\footnote{If long-distance correlations in the values of $e^{i\sigma/2}$ are present, not allowing for $\Z_2$ symmetry-breaking states leads to violation of cluster decomposition, see, e.g., Ch. 23 in \cite{ZinnJustin:2002ru}.} Thus, in the limit of large $\T^2$, the ground state of the Hamiltonian may not be an eigenstate of $\Z_2$   as in (\ref{psin}), but may be a linear superposition between even and odd sectors corresponding to a fixed value of $\langle e^{i\sigma/2}\rangle = \pm 1$ instead; we shall see in the next section that this is indeed the case in the $SO(3)$ theory.
 \end{enumerate}
The remarks in this section concerning the topological $\Z_2$ symmetry and the difference between bosonic $SU(2)$ and $SO(3)$ theories will be useful when considering the different descriptions of the deconfinement transition in Section \ref{symabovebelow}.

{\flushleft{\bf Fermionic zero modes and  flux sectors:}} The above discussion refers to a theory without fermions.
In the theory with fermions, because of fermion zero modes, tunneling events that change the
magnetic flux $|n\rangle \rightarrow |n+1\rangle $ ($n = 0, \pm 1/2, \pm 1É$)    cannot occur without also changing the fermion number of the state. 
However, tunneling events between states of zero fermion number that change the flux 
sectors as  $|n\rangle \rightarrow |n+2\rangle $,  do occur.  This effect is a descendant  of 
 magnetic bions, a certain type of topological molecule which changes the  magnetic flux by 
 two  units (their magnetic charge is two times the 't Hooft-Polyakov monopole charge or four times 
 the  $\half$-vortex charge  mentioned above) and that  have no fermion zero modes, see Figure \ref{fig:bions}.  These defects on $\R^3 \times \S^1$  will be reviewed in the next Section \ref{nonpert2}. 
  Thus, in the $SU(2)$ theory with fermions we will obtain two  ground states, almost degenerate at sufficiently large 2d volume, while in the SO(3) theory, there will be four  almost degenerate ground states. The detailed structure of the ground states in theories with fermions can be given via the quantum mechanics of zero modes of both the dual photon
and the massless components of the fermions 
and will be discussed elsewhere.

To sum up this section, in the bosonic theory, $SU(2)$ has a unique vacuum, and trivial topological symmetry,  $SU(2)/\Z_2$ has two vacua and a $\Z_2$ topological symmetry. The presence of adjoint fermions doubles the number of vacua in each case, making it two and four, respectively.

\subsubsection{Nonperturbative effects in the $\mathbf{SU(2)}$ theory on $\mathbf{\R^{1,2} \times \S^1_L}$. }
\label{nonpert2}

The perturbative Lagrangian (\ref{longdistancefree}) in terms of the compact dual photons and neutral fermions does not properly  reflect the dynamics of the long-distance theory. There are important non-perturbative effects due to the existence of monopole-instantons in the broken QCD(adj) theory. 
These are not only qualitatively, but also quantitatively  important, and their contributions are under full theoretical control in the small-$L$ weak-coupling regime. In what follows, we will briefly describe the properties of the relevant monopole-instantons and their role in the dynamics. 

In this section, we will, for the most part, have in mind the dual photon appropriate to the $SU(2)$ theory, i.e., $\sigma$ with periodicity (\ref{su2sigma}). We stress that all previous studies of gauge theories on $\R^3 \times \S^1$  (supersymmetric or otherwise) have been also in the context of $SU(2)$
 theories. 

Because the effective 3d theory (\ref{perturbativeSeff}) originates in a 4d theory compactified on $\R^{3} \times \S^1_L$ (we now consider the Euclidean setting appropriate for studying instantons), there is a whole Kaluza-Klein tower of monopole-instantons, in addition to the single fundamental monopole-instanton that  exists in a purely 3d  theory. These solutions have been extensively described in the literature \cite{Lee:1997vp,Kraan:1998sn}.  In this paper, we  work in the small-$L$ regime and to leading order in the semi-classical expansion---so that the contributions of the higher-action Kaluza-Klein monopole-instantons will not be important (unlike  the study of ref.~\cite{Poppitz:2011wy}). Thus, we  will only use the lowest action monopole-instantons. Below, we will describe the   two types of solutions of   lowest action of the self-duality equations $F_{MN} =  \tilde{F}_{MN} =   {1 \over 2} \epsilon_{MNKL} F^{KL}$. 
 In the center-symmetric vacuum (\ref{omegavev}), their action equals $1\over 2$  the BPST instanton action. We denote the action by $S_0$:
\begin{equation}
S_0 ={1 \over 2} \times {8\pi^2 \over g^2 } = {4 \pi^2 \over g^2}.
\end{equation}
Because the solutions are  self-dual, their topological charges,  $Q_T = \frac{1}{32\pi^2}\int_{\R^3\times \S^1} F_{MN}^a\tilde F_{MN}^a$, are all equal to each other and to $1\over 2$ in the center-symmetric vacuum. 
The unbroken-$U(1)$ magnetic field far from the core of the 
two  types of self-dual monopole-instantons is given by (for a review of the explicit solutions, see, e.g., \cite{Anber:2011de}):
  \begin{equation}
  \label{pthinstanton}
  \vec{B}^{\pm} = \pm {\vec{r} \over |r|^3}~.
  \end{equation}
    Thus, according to our normalization (\ref{charge}) the two lowest-action solutions--- the monopole-instanton ($+$) and the KK-monopole-instanton ($-$) (as the extra self-dual solution is often called)---have magnetic charge $\pm 1$. We stress again that  the $+$ and $-$ charge monopole-instanton solutions are both self-dual and that the corresponding anti-self-dual antimonopole solutions  carry opposite magnetic charges.

  In a theory with massless adjoint fermions, the monopole-instantons  have fermionic zero modes (see \cite{Nye:2000eg, Poppitz:2008hr} for the relevant index theorem) and generate, instead of the usual  monopole operator  $e^{-S_0} e^{\pm  i  \sigma}$,  operators of the form:
  \begin{equation}
  \label{monoperator}
{\cal M}_1 =   e^{-S_0} e^{ i  \sigma} \det_{I,J} \lambda_J \lambda_I~, \qquad  
{\cal M}_2 =   e^{-S_0} e^{- i  \sigma} \det_{I,J} \lambda_J \lambda_I~, \qquad 
 \end{equation}
   where the determinant is over the $n_f$ flavor indices. 
  The form of the monopole-instanton induced operators (\ref{monoperator}) has interesting implications for the physics of mass gap generation and confinement:
 \begin{itemize}
 \item{
First, since under the $\Z_{4 n_f}$  discrete chiral symmetry of  (\ref{chiralsymmetry})
$ \det_{I,J} \lambda_J \lambda_I$ $\rightarrow$ $-\det_{I,J} \lambda_J \lambda_I$, 
 the invariance of the monopole-instanton amplitude under the exact discrete chiral symmetry 
 demands that the dual photon transform as well, as shown on the second line below:
\begin{eqnarray}
\Z_2^{{\rm d}\chi}: &&\det_{I,J} \lambda_J \lambda_I \rightarrow -  \det_{I,J} \lambda_J \lambda_I  \nonumber \\
&& e^{ i  \sigma} \rightarrow  e^{i \pi} e^{ i  \sigma} \qquad~.
\label{dcs}
\end{eqnarray}
Here, $\Z_2^{{\rm d}\chi}$ denotes the genuine discrete chiral symmetry  which cannot be rotated away by a discrete nonabelian flavor transformation,  or be  included in $(-1)^F$, as explained just after eqn.~(\ref{chiralsymmetry}). Thus, $e^{ i  \sigma}$ is the topological order parameter (disorder operator) associated with the  $\Z_2^{{\rm d}\chi}$ symmetry  $\sigma \rightarrow \sigma + {\pi }$.}
\item{
Second,     symmetry considerations alone show that the $\Z_2^{{\rm d}\chi}$ symmetry permits a purely bosonic potential term, $\sim$$\cos 2 \sigma$, in the long-distance effective action. This term was shown \cite{Unsal:2007vu,Unsal:2007jx} to be due to a novel type of topological molecule and referred to as ``magnetic bions."  In a Euclidean context, where monopoles are viewed as classical particles, the magnetic bions are  molecular (correlated) instanton events  of self-dual monopole-instantons and anti-self-dual  KK-monopole-instantons that carry twice the charge of a   magnetic monopole.  Due to this reason, we sometimes refer to these defects as topological molecules.  Magnetic bions   arise in  second order in the semi-classical  $e^{-S_0}$ expansion as  $\sim e^{-2 S_0} \cos 2 \sigma$.}
\end{itemize}
Considering the second point in some more detail, we note that the ``magnetic bion" topological molecules  are stable ``bound states," where the magnetic repulsion between the charge $+1$ monopole and the charge $+1$ anti-KK-monopole is balanced by attraction due to fermion exchange, resulting in an interaction potential of the form:
\begin{equation}
V_{\mbox{\scriptsize bion}}(r)=\frac{4\pi L}{g^2 r}+4n_f\ln r\,, 
\end{equation}
with stabilization radius $r_*=\pi L/g^2 n_f$.
The resulting ``magnetic bion" molecule has size $r_*$, larger than the $1/L$ UV-cutoff  of the long distance theory (\ref{perturbativeSeff}), see  \cite{Anber:2011de}  for a detailed study. Thus, confinement in this theory is generated by this novel type of non-self-dual composite topological excitation. Despite being analyzed in the framework of a 3d effective theory valid at distances greater than $L$, the locally 4d nature of the theory is crucial.\footnote{This is because  the ``twisted" KK monopoles do not exist in 3d theories: their action, for general expectation value  $\langle A_4^3\rangle = v $, is ${8\pi^2 \over g_4^2} - {4 \pi L v \over g_4^2}$. A 3d limit $L \rightarrow 0$ requires also taking $g_4^2 \rightarrow 0$, with $g_3^2 = {g_4^2\over L}$ kept fixed. Thus the action of KK monopoles is
$ {8 \pi^2 \over g_4^2}- {4 \pi  v \over g_3^2}$, and, in the 3d limit $L \rightarrow 0$, with fixed $v$ and $g_3^2$, it becomes infinite.} We note that a recent lattice work gave numerical evidence in favor of  topologically neutral topological molecules\cite{Bruckmann:2011cc}.
Although this work is done in Yang-Mills theory, such molecules exist in pure gauge theory as well. This will be discussed in a forthcoming paper.

  \begin{FIGURE}[ht]
    {
    \parbox[c]{\textwidth}
        {
        \begin{center}
        \includegraphics[angle=0, scale=0.80]{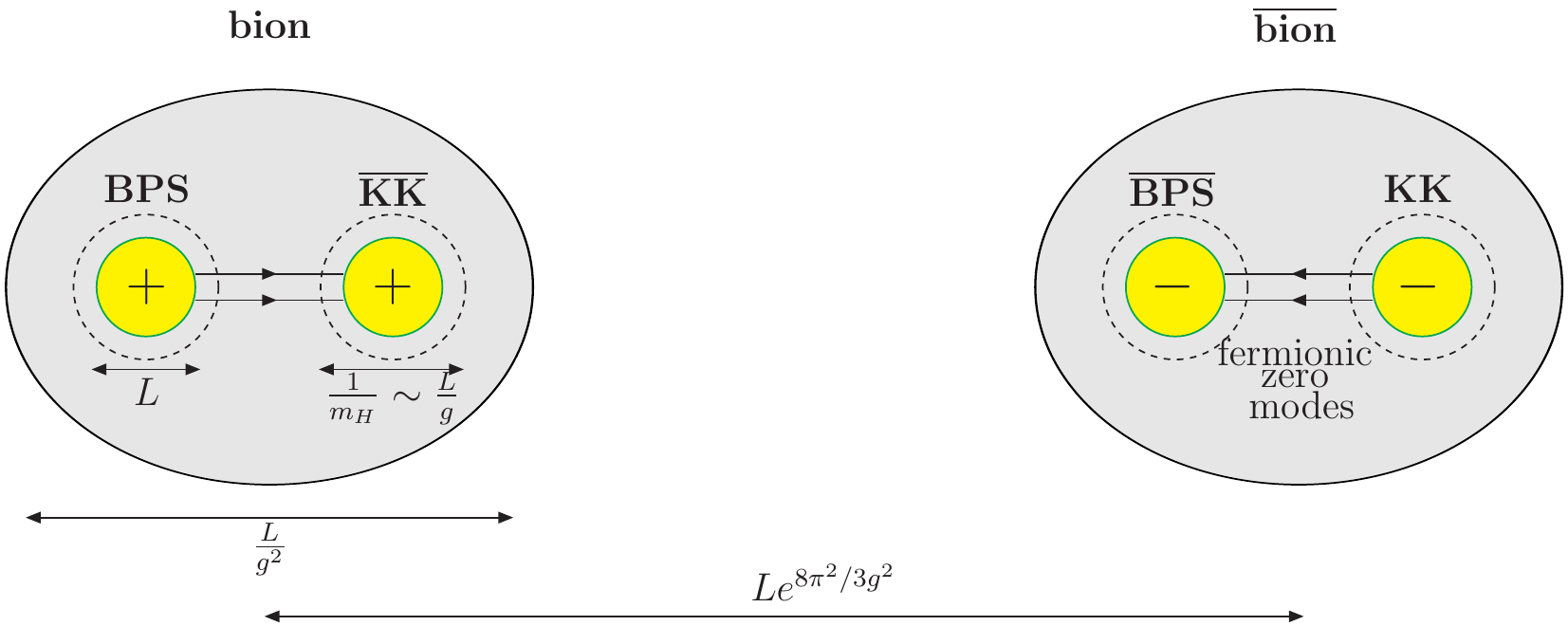}
	\hfil
        \caption 
      { The magnetic bion in the $SU(2)$ theory is a composite of the BPS and $\overline {\rm KK}$ monopoles, which is magnetically charged, but topologically neutral. In Euclidean space, where instantons are viewed as ``particles", bions should  be considered as molecules.  The size of a bion is much larger than the monopole's, but much (exponentially) smaller than inter-monopole separation. This allows a semi-classical dilute gas approximation  for  the magnetic bion plasma.  
			}
        \label{fig:bions}
        \end{center}
        }
    }
\end{FIGURE}

The proliferation of bion-antibion pairs in the vacuum means that the non-perturbative zero-temperature partition function of the $SU(2)$ theory with $n_f$ adjoints can be represented as a dilute gas of magnetic bions (of charge 2) and anti-bions (of charge -2), which interact due to the long range magnetic force:
\begin{equation}
\label{zgas1}
Z_{ bion\; gas}=\sum_{N_{\pm}, q_a=\pm 2 }\frac{Z_{\mbox{\scriptsize bion}}^{N_+ + N_-}}{N_+! N_-!}\prod_j^{N_+ + N_-} \int \frac{d^3R_j}{L^3} \exp\left[-\frac{ { 4} 
 \pi L}{g^2 }\sum_{a >  b}\frac{q_aq_b}{|\vec R_a-\vec R_b|}\right]\,.
\end{equation}
Here, $R_i$ are the locations of the (anti)bions, 
the quantity in the exponent is the interaction ``energy" due to the long-range magnetic field of the bions,
 and $Z_{\mbox{\scriptsize bion}}$ is a single (anti)bion molecular partition function, given by  \cite{Anber:2011de}: \begin{equation}
\label{finalzbion}
Z_{{bion}} \left(g(L)\right)\sim \frac{1}{g_{\scriptsize\mbox{1-Loop}}^{14-8n_f}\left(L\right)} \; e^{-{8\pi^2\over g_{\scriptsize \mbox{2-Loop}}^2\left(L\right)} (1 + c g_{\scriptsize \mbox{2-Loop}} \left(L\right))}~,
\end{equation}  
where  coefficients that play no important role are omitted (the known positive coefficient $c$ is also not relevant for our present studies). Using standard methods  \cite{Polyakov:1976fu}, one then shows that the bion partition function leads to a nonzero mass for the dual photon $\sigma$ at zero temperature. Thus, the    bosonic part of the long-distance effective action (\ref{perturbativeSeff}) for  QCD(adj) on $\R^{1,2} \times \S_L^1$  is modified to:
\begin{equation}
\label{non-perturbativeSeff}
S= \int_{\R^{1,2}} {1 \over 2 L} \left( g \over 4 \pi \right)^2 \left[ \left(\partial_\mu \sigma\right)^2 +{{ m_\sigma}^2 \over { 2}}\cos 2 \sigma \right] ~,  
 \qquad \left\{ \begin{array}{ll} 
 \sigma \equiv \sigma + 2\pi , & {\rm for}  \;\;SU(2) \\
  \sigma \equiv \sigma + 4\pi , & {\rm for}  \;\;SO(3) 
\end{array}\,. \right.
\end{equation}
 In (\ref{non-perturbativeSeff}),  the dual-photon mass is
given by \cite{Anber:2011de}:
\begin{equation}\label{photon mass}
{m_\sigma}^2=  \frac{8 \left(4\pi\right)^2Z_{\mbox{\scriptsize bion}}(g)}{ g^2 L^2} \sim {1 \over L^2 g^{16 - 8 n_f}} e^{-2 S_0}.
\end{equation}
The tension $\gamma$ of the confining string between  test charges is also essentially determined by the mass of the dual photon, 
$\gamma \sim {g^2\over L} {m_\sigma}$.\footnote{It is also useful to give an equivalent  representation for the SO(3) theory  by using  the rescaling $\sigma \rightarrow 2\sigma$ which makes periodicity  into $2\pi$.  Then, $SO(3)$  action becomes: 
\begin{equation}
\label{alternative}
S= \int_{\R^{1,2}} {1 \over 2 L} \left( g \over 2 \pi \right)^2 \left[ \left(\partial_\mu \sigma\right)^2 +{{ m_\sigma}^2 \over { 8}}\cos 4 \sigma \right] ~,   \;\;  \sigma \equiv \sigma + 2\pi ~,
\end{equation}
with four minima within unit-cell. This second representation will also be useful. }
  
Note that   if we were to focus just on the bosonic part of  (\ref{action1}), without considering  fermions, as in the study of QCD(adj) with massive fermions or deformed Yang-Mills theory, 
 we would have obtained: 
 \begin{equation}
\label{non-perturbativeSeffdYM}
S_{\rm b}= \int_{\R^{1,2}} {1 \over 2 L} \left( g \over 4 \pi \right)^2 \left[ \left(\partial_\mu \sigma\right)^2 +{{ m_\sigma}^2 \over { 2}}\cos  \sigma \right] ~,  \qquad \left\{ \begin{array}{ll} 
 \sigma \equiv \sigma + 2\pi , & {\rm for}  \;\;SU(2) \\
  \sigma \equiv \sigma + 4\pi , & {\rm for}  \;\;SO(3)  
\end{array} \right.~.
\end{equation}
In the bosonic $SU(2)$ theory, within the fundamental domain, there is a unique ground state. 
In the $SO(3)$ theory, there are two-ground states related by the topological $\Z_2$ symmetry. 
As discussed in the Hamiltonian framework in Section \ref{hamiltonian}, these numbers are doubled when one considers the theory with massless fermions. 

Going back to QCD(adj), in the $SU(2)$ theory, 
recalling (once again) that    $\sigma\equiv \sigma + 2\pi$  (\ref{su2sigma}), we find that within the fundamental domain, there are two ground states of (\ref{non-perturbativeSeff})  associated with the breaking of the discrete chiral symmetry (\ref{dcs}) by the expectation value of the dual photon:
\begin{equation}
\sigma_{\min}=\{0, \pi \}
\qquad {\rm associated \;  with}  \qquad \frac{1}{2 \pi} \int_{-\infty}^{\infty} \frac{d \sigma}{dz} dz =
\frac{k}{2}, \qquad k=0,1, 
\label{minimasu2}\end{equation} 
where the last equation  shows the ``winding" number of $\sigma$, normalized as appropriate for a variable of period $2\pi$, between the trivial and $k$-th minimum of the potential. The tension of the domain wall between the corresponding vacua is proportional to this winding number. The  domain wall interpolating  between  the vacua with $\langle \sigma \rangle=0$ and $\langle \sigma \rangle={\pi}$ is associated with the spontaneously broken discrete chiral symmetry. 

In the $SO(3)$ theory, we recall that $\sigma \equiv \sigma + 4 \pi$ (\ref{so3sigma}). We now see from (\ref{non-perturbativeSeff}) that, in addition to the  
  non-anomalous  discrete $\Z_2$ chiral symmetry (\ref{dcs}), $\sigma$$\rightarrow$$\sigma+{\pi}$, the effective   bosonic $V(\sigma) \sim -\cos (2 \sigma)$ potential has an (accidental) $\Z_4$ symmetry, which combines the discrete chiral symmetry (\ref{psin}) with the topological $\Z_2$ symmetry $\sigma$$\rightarrow$$\sigma + 2\pi$ (\ref{thooftalgebra}). 
The four minima of the bosonic potential in the fundamental domain are now located at: 
\begin{equation}
\sigma_{\min}=\{0, \pi, 2\pi, 3\pi  \}
\qquad {\rm associated \;  with}  \qquad \frac{1}{4 \pi} \int_{-\infty}^{\infty} \frac{d \sigma}{dz} dz = \frac{k}{4}, \qquad k=0,1,2,3, 
\label{minima}
\end{equation}
where the ``winding" number is now normalized for a field of periodicity $4 \pi$.
The tension of the domain wall interpolating  between $k=0$ and $k=1$ corresponds to the tension of domain wall of the spontaneously broken discrete chiral symmetry. 
The domain wall  tension for the wall interpolating  between $k=0$ and $k=2$ is also equal to the 
string tension $T_1$ for a charge probe in the fundamental representation of $SU(2)$. This is because the winding  of $\sigma$ between  the two sides of the domain wall with  $k=2$ is the one corresponding to the insertion of a Wilson loop of an electric charge $1/2$ used to calculate string tensions \cite{Polyakov:1976fu}.

We end this brief review by noting that the study \cite{Unsal:2007vu,Unsal:2007jx} of this and related $N_c > 2$ theories gives the first example of a locally 4d, continuum, and nonsupersymmetric gauge  theory where confinement can be understood within a controllable analytic framework. In particular, the role of the fermions and the importance of non-self-dual topological molecules  carrying magnetic charge is novel in this regard.

It is then natural to ask what  we can learn   about the finite-temperature behavior of the theory.

\section{The $\mathbf{SU(2)}$(adj) theory at finite temperature and the $\mathbf{Z_p \; \; XY}$ model}
\label{su2non-perturbativethermal} 

Before addressing the finite temperature dynamics of the small-$L$ theory, let us recall the  relevant scales in the zero-temperature problem. The $W$-boson and heavy-fermion Compton wavelengths are $\sim L$, and coincide with the characteristic size of monopole-instantons, 
${r_{\rm m.}}$.
The Compton wavelength of the   adjoint
 ``Higgs" field---the uneaten ``radial" component of $A_4$---is  of order $L
 \over g$. 
 The size of the magnetic bions is of order ${L\over g^2}$ and the typical distance between bions is $\Delta R_{bion} \sim L e^{8 \pi^2 \over 3 g^2 }$, see \cite{Anber:2011de}. Finally, the Compton wavelength (\ref{photon mass}) of the dual photon,
 ${1 \over m_\sigma} \sim L e^{4 \pi^2\over g^2}$, is the largest length scale in the problem:
 \begin{equation}
\left( m_W^{-1}  \sim {r_{\rm m.}}\right) \; \ll  \; m_{A_4}^{-1}   \;  \ll   \;   r_{bion}   \;  \ll   \;   \Delta R_{bion}   \;  \ll   \;  m_{\sigma}^{-1} ~.
 \end{equation}
 Needless to say, it is this hierarchy of increasing length scales at small $L$, 
  illustrated in Fig.~\ref{fig:bions}, that 
makes the problem analytically tractable.

\subsection{The thermal partition function as an electric-magnetic Coulomb gas}

Turning on temperature means that we are considering the Euclidean theory  on $\R^2 \times \S^1_\beta \times \S^1_L$, instead on $\R^3 \times \S_L^1$. The fermions are now antiperiodic around the thermal circle, but, as before, periodic in $\S^1_L$.  We will study the behavior of the system with changing $\beta= {1\over T}$ at fixed $L$.
At temperatures low compared to the confinement scale (the mass gap), $T \ll m_\sigma$, 
we expect the thermodynamics of the theory to be rather trivial, described, for $n_f > 1$, by the thermodynamics  of  free massless fermions (the components of the fermions along the Cartan subalgebra) and 
a  massive $\sigma$ field---the low-$T$  free-field theory limit of the bosonic theory (\ref{non-perturbativeSeff}). The thermodynamic potential is $\Omega = - T \ln Z \sim  A T^3 +  A m_\sigma T^2 e^{- {m_\sigma\over T}}$, where $A$ is the 2d volume of the system. 
As the temperature increases past $m_\sigma$,  at first one can still use the $\sigma$ free-field theory description, leading to $\Omega$ given by the high-temperature limit $\Omega \sim A T^3$ (in the limit of neglect of interactions in the dimensionally reduced (\ref{non-perturbativeSeff})). 

  The most interesting behavior occurs as  the temperature increases further above $m_\sigma$, in the range $m_\sigma \ll T \ll {1\over L} \sim m_W$.\footnote{At even higher temperatures, above $m_W \sim {1\over L}$,  in the small-$L$ regime we expect to find essentially free-field behavior,
   with the thermodynamic potential scaling as $N_c^2$.} 
Now recall that at energies below $1\over L$ the non-perturbative dynamics of the zero-$T$ theory is that of   a decoupled (neutral) free fermion and  a scalar with an exponentially small mass.  
The scalar sector can be described as a  non-relativistic  3d Coulomb gas (\ref{zgas1}) of charges $\pm 2$ (the magnetic bions). 
 This picture essentially remains in the temperature range $m_\sigma \ll T \ll {1\over L} \sim m_W$, albeit  with a few important changes, as we now discuss.

First,   all  fermions---even the massless ones, responsible for the binding of monopoles and KK-anti-monopoles into bions---now have a thermal mass of order $T$. Their Euclidean propagators, and thus the attractive potential between the bion constituents, are affected at distances larger than $1 \over T$. Recalling that the bion size is  $r_* = {\pi L \over g^2 n_f}$ and requiring that the behavior of the fermion propagator is only different at distances  larger than the bion size implies   ${1\over T} \gg r_* = {\pi L \over g^2 n_f}$. Thus, the fermion-induced attraction between the constituents of the bions will be unaffected so long as:
\begin{equation}
\label{bionpersistencecondition}
T \ll {1 \over r_*} =  {g^2 n_f \over \pi L}.
\end{equation}
The  critical behavior that we shall find occurs well within the range (\ref{bionpersistencecondition}). In fact, 
 a more detailed calculation of the  attractive potential between bion constituents is possible and one finds that near $T_c$, which,   as we show later, is equal to  ${g^2 \over 8 \pi L}$, the finite temperature contribution to the bion potential near $r_*$ is  suppressed, relative to the zero temperature potential,  by a factor of order $(r_* T_c)^{3} = (8 n_f)^{-3}$, where $r_* T_c =  1/(8 n_f) \ll 1$. Thus, for the dynamics near criticality, well within the range (\ref{bionpersistencecondition}), we can treat bions as pointlike. The ultimate reason for this is that the bions are much smaller than the size of the compact ``thermal" direction, as illustrated on Figure \ref{fig:scales}.

  \begin{FIGURE}[ht]
    {
    \parbox[c]{\textwidth}
        {
        \begin{center}
        \includegraphics[angle=0, scale=0.33]{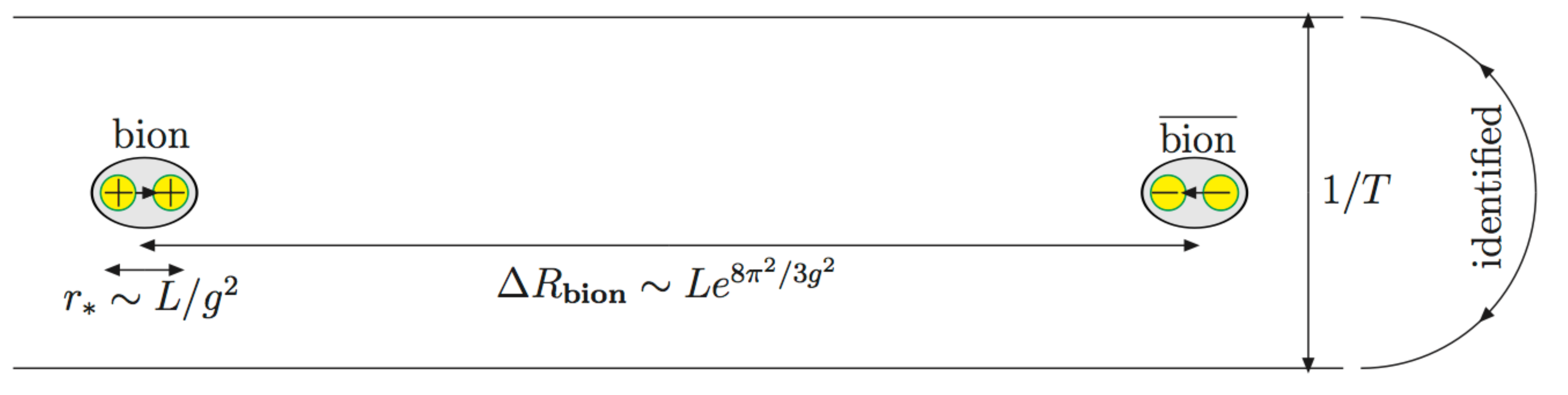}
	\hfil
        \caption 
      {  The scales in the finite-temperature problem.  For the interesting  temperature range (\ref{dimreduction1}), 
      the bion size is much smaller than inverse temperature which in turn is much smaller than inter-bion separation, i.e, $ r_{*} <  \beta \ll {\Delta R}_{\rm bion} $. 
			}
        \label{fig:scales}
        \end{center}
        }
    }
\end{FIGURE}

Second, recall that the inter-bion separation $\Delta R_{bion}$ at $T=0$ is    $\Delta R_{bion} \sim L e^{ 8 \pi^2 \over 3 g^2} \ll {1\over m_\sigma}$. Thus, for $T$ within the range:
\begin{equation}
\label{dimreduction1}
 L e^{ 8 \pi^2 \over 3 g^2}  \sim  \Delta R_{bion}\gg {1 \over T} \gg {r_* } \sim { \pi L \over g ^2 n_f}~,
\end{equation}
the separation between bions is $\gg 1/T$, (and $1/T$, from   (\ref{bionpersistencecondition}), is in turn  larger than the bion size).
This means that in the regime (\ref{dimreduction1}) the bion dynamics, whose $T=0$ partition function is  (\ref{zgas1}), can be described by the dimensional reduction of (\ref{zgas1}).  
Thus, the 3d Coulomb potential between bions can be replaced by the 2d logarithmic one, where the coefficient follows simply from Gauss' law:
\begin{equation}
\label{3d2dCoulomb}
{1 \over |\vec{R}|} \rightarrow - 2 T \log |\vec{R}| T~,
\end{equation}
and $\vec{R}$ denotes now a vector in $\R^2$ (the argument $T$ in the logarithm was inserted for dimensional reasons; it will play no role due to charge neutrality of the 2d gas). Furthermore, the integral over positions of the bions should be 
now over points in $\R^2$, replacing ${d^3 R_j \over L^3 } \rightarrow {d^2 R_j \over T L^3}$. Thus, we obtain for the thermal partition function of the 2d bion gas:
\begin{equation}
\label{zbionthermal1}
Z_{2d\; bion\; gas}=\sum_{N_{\pm}, q_a=\pm 1 }\frac{\xi_{\mbox{\scriptsize bion}}^{N_+ + N_-}}{N_+! N_-!}\prod_j^{N_+ + N_-} \int{d^2 R_j} \exp\left[\frac{32 \pi L T}{g^2 }\sum_{a >  b} {q_aq_b} \log |\vec R_a-\vec R_b| T\right]\,,
\end{equation}
 where we introduced the bion fugacity:
 \begin{equation}
 \label{bionfugacity}
 \xi_{\mbox{\scriptsize bion}} = {Z_{\mbox{\scriptsize bion}} \over L^3 T} \sim {e^{- 2 S_0(1 + c g)} \over L^3 T g^{14 - 8 n_f}}~,
 \end{equation}
 as follows from our comments above (\ref{zbionthermal1}) and from eqn.~(\ref{finalzbion}); we also note that the Coulomb gas partition function in 2d has to obey charge neutrality, $N_+ = N_-$. We took the sum over charges in (\ref{zbionthermal1}) to be over $q_a = \pm 1$, i.e., the factor of $2$ in the bion charge was absorbed in the interaction strength.

Now, recall that the partition function of our theory is $Z = Z_0 Z_{2d \; bion \; gas}$, where $Z_0$ describes the free photon fluctuations, now reduced to 2d and $Z_{2d \; bion \; gas}$ is from (\ref{zbionthermal1}). Recall that the $T=0$ partition function $Z = Z_0 Z_{bion \; gas}$, when expressed in terms of the $\sigma$ field,  equals $Z=\int {\cal{D}} \sigma e^{-S}$, with $S$ given in (\ref{non-perturbativeSeff}). The dimensional reduction of the $T=0$ partition function (appropriate in the range of temperatures (\ref{dimreduction1})) is equal to $Z_0 Z_{2d \; bion \; gas}$ and is precisely
 the partition function of a XY model  in a Villain approximation (see, e.g., \cite{Jose:1977gm}).\footnote{The XY model is  defined by a partition function as in  eqn.~(\ref{z4model}), but with $\tilde{y}=0$ and $\kappa \rightarrow K$, i.e., 
$-\beta H =   \sum_{x; {\hat\mu = 1,2}} { K \over 2 \pi} \cos ( \theta_{x  + \hat\mu} -\theta_{x})$. The contribution of the vortices of the $\theta$ field to the partition function take exactly the form $Z_{2d \; bion \; gas}$ with $K$ defined by (\ref{Kcoupling}); see Section \ref{dualsymmetries} for a derivation in a continuum physicist's language.}
 The ``vortex-vortex" coupling $K$ can be read off from the coefficient of the logarithm in the exponent in (\ref{zbionthermal1}), which describes the interactions of vortices:\footnote{Note that as opposed to usual discussions of the XY-model, for us small-$K$ corresponds to low-$T$.}
 \begin{equation}
 \label{Kcoupling}
 K = {32 \pi L T \over g^2}.
\end{equation} 
It is well known from the 
  physics of the XY model   that at small values of the spin-spin (and vortex-vortex) coupling $K < K_c = 4$  the spontaneous formation of vortices is entropically favored. Note that $K_c=4$ is equivalent, from (\ref{Kcoupling}), to $T_c = {g^2 \over 8 \pi L}$. As we will later see, this value of $T_c$ is the exact (up to exponentially small corrections) value of the critical temperature for the $SU(2)$(adj) theory.
At small $K$, vortices disorder the system and lead to a finite correlation length (mass gap). From the map (\ref{Kcoupling}), we see that in the $SU(2)$(adj) $K < 4$ corresponds to the low-temperature confining region $T<T_c = {g^2 \over 8 \pi L}$. At large values of the spin-spin coupling $K>K_c$, the appearance of vortices is suppressed,  the dynamics is dominated by spin waves (the $Z_0$ partition function), and 
  the XY model exhibits algebraic long-range order with continuously varying critical exponents and no mass gap. From (\ref{Kcoupling}), this   behavior would be attributed to the high-temperature $T > T_c$ 
regime of the $SU(2)$(adj) theory and thus be expected to describe the deconfined phase. If (\ref{zbionthermal1}) was all that was relevant in this temperature range, it   would lead to a   high-temperature phase with an infinite correlation length---while we expect to have finite correlation length above the deconfinement transition due to Debye screening of electric charges.
It has already been noted \cite{Dunne:2000vp}, in the context of the 3d Polyakov model,  that the  BKT  behavior described above is not the one expected of a confinement-deconfinement transition. 

 Thus,  there is essential physics missing from the description  only in terms of the monopole-instanton gas (for the Polyakov model), or bion gas (for $SU(2)$(adj)). The remedy was already suggested   in \cite{Dunne:2000vp}: that effects of the heavy $W$-bosons have to be included in the description of the deconfinement transition. While at $T \ll m_W$ (see (\ref{dimreduction1})) the effects of $W$ bosons are Boltzmann suppressed,  the effects of monopole-instantons (or bions) are also exponentially small, with fugacities $e^{- {8 \pi^2 \over g^2}}$. The Boltzmann suppression of $W$-bosons ($m_W = {\pi \over L}$ in the center-symmetric vacuum) is, at $T= T_c$,  $e^{- {m_W\over T_c} } = e^{- {\pi \over L} {8 \pi L \over g^2} } = e^{ - {8 \pi^2 \over g^2} }$. Thus, near $T_c$ $W$-bosons and bions are equally likely to appear in the plasma and we expect that the deconfinement transition in $SU(2)$(adj) is driven by competition between the interactions of electrically ($W$-bosons) and magnetically (bions) charged particles. This is the analytic realization of the scenario proposed in \cite{Liao:2006ry}.

Including the effects of electric charges on the bion plasma partition function is relatively straightforward.  First, note that the neutral particles, such as the radial mode of the ``Higgs" boson $A_4$ is not expected to play a role in the deconfining dynamics, despite being lighter than the charged $W$-bosons. Recall that  $A_4$ is the Wilson line wrapping the spatial, non-thermal circle. 
 The heavy charged fermions $\lambda^\pm_I$ (of mass equal to $m_W$, $I=1,...,n_f$) contribute similar to the $W$-bosons,  because  at $m_\sigma \ll T\ll m_W$, the $W$-boson,  $\lambda^\pm_I$, and bion gas is (exponentially) dilute: the thermal de Broglie wavelength $(m_W T)^{-{1\over 2}}$ is much smaller than the typical distance between particles $\sim \Delta R_{bion}$, and the  Fermi statistics is irrelevant. Thus, we will further refer to the gas of electrically charged particles as the ``$W$-boson gas" and will account for the $\lambda^\pm_I$ contribution via the multiplicity---see also Footnote \ref{indistinguishability} below.
At  the $T\ll m_W$ temperatures of interest, the  $W$-boson partition function is  that of a 2d gas  of electrically charged non-relativistic particles, which interact via Coulomb forces with themselves (as well as with the magnetically charged objects, the bions, as described below). 
 Thus, the $W$-boson/$\lambda^\pm$ 
 fugacity is:\footnote{The prefactor follows from integrating over the momenta in the non relativistic partition function, equal to the product of factors $\int {d^2 p \; d^2 x \over (2\pi)^2}\; e^{- {m\over T} - {p^2\over 2mT} + V(x) }$ for all particles. The $2 n_f$ factor accounts for the multiplicity of charged fermions. The various terms contributing to the interaction $V(x)$ are given in eqns.~(\ref{Winteraction}) and (\ref{ABinteraction2}). }
 \begin{equation}
 \label{wfugacity}
 \xi_W= (2 n_f + 2)  {m_W T \over 2 \pi} \; e^{-{m_W \over T}}~.
\end{equation}

Another subtlety that we have to discuss is that at the lowest mass level on $\R^3 \times \S^1_L$ in the center-symmetric vacuum there are actually two sets $W^\pm$-bosons (and $\lambda^\pm$ fermions). This is easiest to see in the $D$-brane picture \cite{Lee:1997vp}, which, despite being highly supersymmetric,  greatly helps in studies of the nonsupersymmetric tree-level spectrum. It can also be seen via the usual Kaluza-Klein decomposition: the masses of $W^\pm$ come from their interaction with the Higgs field $A_4$ (recall Footnote \ref{higgsnote}), which always enters the Lagrangian as $D_4= \partial_4 + i A_4$. Replacing $\partial_4$ with $i 2 \pi n \over L$ ($n \in \Z$) and $A_4$ by its vev ${\rm diag }( {\pi\over L}, -{\pi\over L})$,  see eqn.~(\ref{omegavev}), we find that the $W^\pm$ masses are proportional to $|2 \pi n + \pi| \over L$. Thus states of mass $\pi\over L$ appear at the $n=0$ and $n=-1$ Kaluza-Klein levels, explaining the presence of two sets of lowest mass $W^\pm$ bosons. These will be the only $W$-boson states whose contribution we will keep. Similar to accounting for the $\lambda^\pm$ contribution, at $T\ll m_W$, the contribution of these states to the grand partition function can be accounted for by doubling the fugacity (\ref{wfugacity}).\footnote{\label{indistinguishability}Perhaps a comment on this is necessary. We are treating the two kinds of $W^\pm$ bosons  (and the $\lambda^\pm$ fermions), which  have the same charges (and, to the order of our calculation, the same fugacities), as indistinguishable. One can show, via the sine-Gordon representation of  a 2d Coulomb gas partition function, using the fact  that only overall charge-neutral configurations contribute to the 2d partition function, that, indeed, the grand partition function of a gas of two kinds of same-charge  particles with  fugacities $\xi_1$ and $\xi_2$  is equal to that of a gas of one kind of charged particles with  fugacity $\xi_1 + \xi_2$.}

Because the $W$-bosons carry electric charges, two $W$-bosons with electric charges $q_a$ and $q_b$ (equal to $\pm1$) located at $R_a$ and $R_b$ interact via the logarithmic potential:\footnote{A quick way to obtain this formula is to recall that unit electric charges appear as unit winding number vortices of the dual photon field in (\ref{perturbative}); then  (\ref{Winteraction}) is just the interaction energy of two vortices of unit winding.  Equally quickly, since the interaction energy of two static $W$ bosons in $\R^{1,2}$ is ${g^2 \over 2 \pi L} q_a q_b \ln r_{ab}$ and the $W$'s propagate in a (Euclidean) time interval ${1\over T}$, the corresponding action is   precisely (\ref{Winteraction}). }
\begin{eqnarray}
\label{Winteraction}
V_{W\,a b}= - \frac{g^2}{2\pi L T}  \; q_a q_b \; \ln  {|\vec R_a-\vec R_b|} T\,.
\end{eqnarray}

In addition to the 2d Coulomb interaction (\ref{Winteraction}) between $W$-bosons, there is an Aharonov-Bohm phase  due to the presence of magnetic charges (the bions) in the system. The interaction between the $a$-th bion located at the origin in $\R^2$ and the $b$-th static $W$-boson of electric charge $q_b$,  located at $\vec{x}_b$ (in $\R^2$) is given by the integral over the $W$-boson worldline (i.e., along $x_0$): 
\begin{equation}
\label{ABinteraction1}
 q_b \int_{0}^{1/T}dx_0 A^{0\,a}_{bion}(x_0,\vec x_b)~,
\end{equation}
where $A^{0 \, a}_{bion}(x_0, \vec{x_b})$ is the time component of the gauge field of the bion. Note that the exponential of $i$ times  (\ref{ABinteraction1}) contributes a phase factor in the path integral also in Euclidean space and that in our normalization of the fields no factors of $g$ appear.
Next, we note that on $\R^2 \times S^1_\beta$ the field that the $W$-boson experiences is that of an infinite chain of bions (all located at the origin in $\R^2$) and spaced equidistantly  along the $x_0$ axis. Thus, we can replace (\ref{ABinteraction1}) with:
\begin{equation}
\label{ABinteraction2}
  q_b \int_{0}^{1/T} dx_0\sum_{p=-\infty}^{\infty}A^{0\,a}_{bion}(x_0+{p\over T},\vec x_b) = q_b \int_{-\infty}^{\infty}dx_0 A^{0\,a}_{bion}(x_0,\vec x_b)   =  4 q_b \Theta(\vec x_b)~,
\end{equation}
where  $\Theta(\vec{x}_a)$ can be taken to be the angle between the vector $\vec{x}_a$ connecting the monopole and the $W$-boson and the positive $x_2$ axis. To obtain the last equality above, we arranged the Dirac strings of all monopoles to be along the negative $x_2$ axis and used the gauge field of a monopole of charge $Q_a$ (recall $Q_a=2$ for bions). In usual polar coordinates it is given by  $A = Q_a (1 - \cos \theta) d \phi$, when the Dirac string is along the negative $z$-axis. This expression for $A$ is then adapted to our coordinates by replacing $x \rightarrow x_0$, $y \rightarrow x_1$, $z \rightarrow x_2$ (thus the strings are now along the negative-$x_2$ axis). Using it   to calculate the integral  yields (\ref{ABinteraction2}); charge neutrality was also used to drop some irrelevant constant contributions. The heavy (mass $m_W$) fermions $\lambda_\pm$ have the same Aharonov-Bohm interaction with the bions as the $W$ bosons and their spin-orbit interaction with the bion magnetic field is mass-suppressed.

Thus, including the $W$-bosons, in addition to the bions, we arrive at the following partition function (note that, as in (\ref{zbionthermal1}), we absorb the factor of $2$ from the bion charge in the prefactor of the bion-bion interaction):
\begin{eqnarray}
\label{SU2gas}
&&Z_{bion+W} \nonumber \\
&=& \sum_{N_{b\pm}, q_a=\pm 1 } \sum_{N_{W\pm}, q_A = \pm 1} \frac{\xi_{bion}^{N_{b+} + N_{b-}}}{N_{b+}! N_{b-}!}  \frac{(2 \xi_{W})^{N_{W+} + N_{W-}} }{N_{W+}! N_{W-}!} \prod_a^{N_{b+} + N_{b-}} \int {d^2 R_a}   \prod_A^{N_{W+} + N_{W-}} \int {d^2 R_A}  \\
&\times&\exp\left[\frac{32 \pi L T}{g^2 }\sum_{a >  b} {q_aq_b} \ln |\vec R_a-\vec R_b| + {g^2\over 2 \pi LT} \sum_{A >  B} {q_A q_B} \ln |\vec R_A-\vec R_B|  + 4i \sum\limits_{a,B} q_B q_a \Theta(\vec{R}_B - \vec{R_a}) \right].\nonumber
\end{eqnarray}
The meaning of the various terms in the partition function above have already been explained. 
Eqn.~(\ref{SU2gas}) is the nontrivial (i.e., interacting) part of the partition function of the $SU(2)$(adj) theory on $\R^{1,2} \times \S^1_\beta$, for $T$ in the range ${1\over L} e^{-{ 8 \pi^2 \over 3 g^2}} \ll  T  \ll   {g^2 n_f \over \pi L}$, as in(\ref{dimreduction1}). 

An important property of $Z_{bion+W}$ is that it is invariant under electric-magnetic duality. This follows immediately by noticing that exchanging:
\begin{eqnarray}
\label{dualityz4}
 \xi_{bion} &\Longleftrightarrow& 2 \xi_W,  \nonumber \\
 {32 \pi L T \over g^2}  &\Longleftrightarrow& {g^2 \over 2 \pi L T}
\end{eqnarray}
in (\ref{SU2gas}) gives rise to an equivalent partition function. The self-dual point $T_* = {g^2 \over 8 \pi L}$ occurs exactly at the critical temperature of the bion-only gas. This temperature also happens to be the BKT temperature of the $W$-only gas (where the only interactions would be given by (\ref{Winteraction})).  This strongly suggests that the deconfinement transition in $SU(2)$(adj) indeed occurs at $T_c = {g^2 \over 8 \pi L}$.

The partition function $Z_{bion+W}$, when defined on the lattice,  also has a Krammers-Wannier type duality, analogous to (\ref{dualityz4}) (exchanging electric and magnetic charges living on dual lattices), and is known   \cite{135388}  
 to be equivalent to the lattice XY-model with a $\Z_4$ preserving perturbation, defined in (\ref{z4model}).
 We will refer the reader to the quoted literature for the lattice duality; instead, will present a (somewhat shorter but helpful) continuum version later in Section \ref{dualsymmetries}. There, we will also establish that  $\kappa = {g^2 \over  2 \pi L T}$, as    claimed in Section \ref{SU2intro}.
 
Before we continue, we note that if one studies the thermal physics not of QCD(adj) but of deformed Yang-Mills theory on 
 $\R^{1,2} \times \S^1_L$ \cite{Simic:2010sv}, one finds, instead, a lattice XY-model with a $\Z_2$ preserving perturbation.  If the 3d minimally supersymmetric Georgi-Glashow model is studied at finite temperature,   a partition function similar to (\ref{SU2gas}) is obtained \cite{Antonov:2003rj}. The nature of the composite topological excitations in the ${\cal{N}}=1$ 3d theory, analogous to magnetic bions,
  was only elucidated recently in \cite{Poppitz:2009kz}.

\subsection{The renormalization group equations for the magnetic bion/W-boson plasma and the approach to ${\mathbf T_c}$}
\label{rgessection}

The behavior of the magnetic-bion/W-boson Coulomb gas (\ref{SU2gas}) can be studied by various means. One way would be to do Monte-Carlo simulations, using its representation as a lattice XY-model with a $\Z_4$-preserving perturbation, discussed in Section \ref{dualsymmetries}.  
Another way is to study the perturbative renormalization group equations and look for a fixed point where the correlation length diverges. It is interpreted as the critical point corresponding to a continuous phase transition.  
Showing that this point exists for the Coulomb gas (\ref{SU2gas}) is the subject of this section. 

To lowest order in the fugacities $\xi_{bion}$ and $2 \xi_W$, these equations are derived in a physically intuitive way in Appendix \ref{rgeappendix}. There exists much literature on the subject, see, e.g. \cite{Jose:1977gm, Amit:1979ab, Nienhuis:1984wm, Boyanovsky:1988ge,Boyanovsky:1989mc,Boyanovsky:1990iw}, but we find that to the order we are working the derivation given in the Appendix is the simplest we are aware of. 
To describe the renormalization group equations (RGEs), we introduce the dimensionless variables:
\begin{equation}
\label{variables}
y =  2 \xi_W \; a^2, ~~  \tilde{y} = \xi_{bion}\; a^2, ~~ \kappa = {g^2 \over 2 \pi L T}~,
\end{equation}
 where $a$ is the ``lattice spacing" (inverse UV cutoff, $\sim  L$) used to define dimensionless fugacities. In terms of these variables, the duality relations (\ref{dualityz4}) become:
 \begin{equation}
 \label{dualityz42}
  y \leftrightarrow \tilde{y}, ~ \kappa \leftrightarrow {16 \over \kappa}.
  \end{equation}
 To leading order in the fugacities, the RGEs (\ref{ymagneticrge},\ref{yelectricrge},\ref{kappaeff7}), given here for the $\Z_4$ model  are:
 \begin{eqnarray}
 \label{rges}
 \dot{\kappa}  &=& 2 \pi^2 \left(16\; \tilde{y}^2 - \kappa^2\; y^2\right) ~,\nonumber \\
\dot{y}  &=& \left(2 - {\kappa \over 2}\right) y ~,\\
 \dot{\tilde{y}} &=& \left(2 - {8 \over  \kappa}\right) \tilde{y} ~,\nonumber
 \end{eqnarray}
 where $\dot{y} \equiv {d y\over d b}$, and the course graining parameter is $a \rightarrow e^b a$ with $b>0$; thus increasing $b$ corresponds to flow to the IR. 
 The first equation above describes the screening of the $W$-boson Coulomb interaction of strength $\kappa$ by electric charges (the $ - \kappa^2 y^2$ term on the r.h.s.) and its anti-screening by magnetic bions (the $\tilde{y}^2$ term). The origin of the antiscreening term is in the Aharonov-Bohm interaction between $W$-bosons and bions, see discussion above  eqn.~({\ref{kappaeff7}) in the Appendix. The last two equations in (\ref{rges}) describe the change of the fugacities of magnetic bions (\ref{ymagneticrge}) and $W$-bosons (\ref{yelectricrge})  upon coarse-graining. 
 
 The RGEs (\ref{rges}) are invariant under the electric-magnetic duality (\ref{dualityz42}). Thus, the two terms responsible for the ``running" of the magnetic coupling $16 \over \kappa$:
  \begin{equation}
  \label{dualkapparge}
  {d \over d b}\left({16 \over \kappa}\right)= 2 \pi^2  \left( 16 y^2 -  \tilde{y}^2 \left({16 \over \kappa}\right)^2\right),
  \end{equation}
written  in the dual frame, can be  understood as being due to its screening  by bions and  anti-screening by  $W$-bosons. Since the duality is an exact symmetry of the partition function (\ref{SU2gas}), it is also a symmetry of the exact RGEs.

  \begin{FIGURE}[ht]
    {
    \parbox[c]{\textwidth}
        {
        \begin{center}
        \includegraphics[angle=0, scale=0.30]{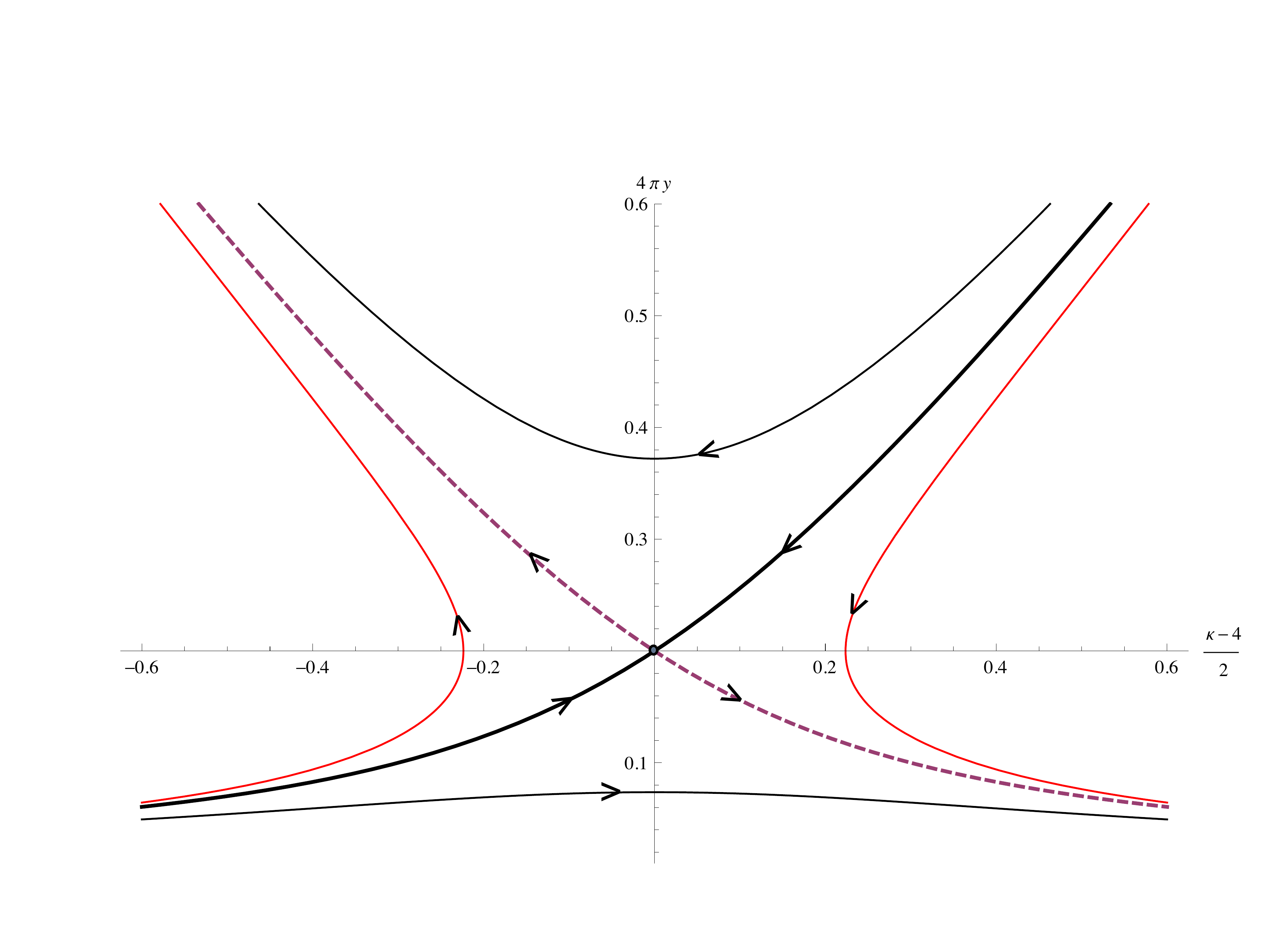}
	\hfil
        \caption 
      { A two-dimensional  (in the $y-\kappa$ plane) cross-section of the RG flow of the $\Z_4$ clock model for small  fugacities, near the critical coupling $\kappa_c = 4$. A two-dimensional flow is appropriate, since $y \tilde{y}$ is approximately RG invariant near $\kappa_c$, see (\ref{rges3}). When the temperature is tuned to $T_c$  the theory flows along one of the two critical trajectories (shown in thick lines) to a  position on the fixed line of (\ref{rges}): $y = \tilde{y} = \sqrt{c^\prime}/4\pi$, $\kappa = 4$, determined by the UV values of the fugacities  ($c^\prime = 16 \pi^2 y_0  \tilde{y}_0$; in the plot, we have chosen $\sqrt{c^\prime} = 0.2$). 
The equation determining $T_c$ and the critical trajectory corresponding to $4 \pi y_0 > \sqrt{c^\prime}$ (the one in the upper right quadrangle) are found in Appendix \ref{z4rgflow}.
 Note that  when one fugacity becomes small the other one grows and that both fugacities remain small only on the critical trajectory. The amount of RG ``time" that trajectories stay close to the critical one diverges as $T \rightarrow T_c$ and is used to find the divergence of the correlation length $\zeta$, given in (\ref{corrlength}), similar to the analysis of the BKT transition.
			}
    \label {fig:RGFLOW} 
        \end{center}
        }
    }
\end{FIGURE}

  At the self-dual point of (\ref{dualityz42}) we have $\kappa = 4$, $y = \tilde{y}$. Thus, for the $\Z_{p=4}$ case of interest (\ref{SU2gas}), this point  corresponds to a line of fixed points of the leading-order renormalization group equations (\ref{rges}). The duality (\ref{dualityz42}) maps high to low temperatures and, as usual, suggests that if there is a unique transition, it should occur at the self dual point. 
 Thus, in the $\Z_4$ theory, the deconfinement transition is expected to 
 occur at the self dual point $T_c \approx {g^2 \over 8 \pi L}$, up to small corrections (the precise equation determining $T_c$ is given in (\ref{tcequation})). The exponential accuracy of this determination of $T_c$ as well as the flow to the critical point are described in detail in Appendix \ref{z4rgflow}.  The important point to make here is that the perturbative expansion in small fugacities in the $SU(2)$(adj) theory is sufficient   to reliably study the approach to the critical point, because the fixed line extends to zero fugacities. This is in contrast with the $\Z_2$ model studied in  \cite{Dunne:2000vp, Simic:2010sv} where the small-fugacities approximation breaks down, but is similar to \cite{Antonov:2003rj}, where it is mentioned but an analysis of the RG flow is not discussed in detail. 

The study of the RGEs near the fixed line is considered in detail in  Appendix \ref{z4rgflow}, where  we   show that the correlation length diverges as:\footnote{As far as we can tell, this result is new; however, it may exist somewhere in the condensed-matter literature.}
 \begin{equation}
\label{corrlength}
\zeta \sim|T - T_c|^{ -\nu } = |T - T_c|^{ -  {1\over 16 \pi \sqrt{y_0 \tilde{y}_0}} }~, 
\end{equation} 
where $y_0$ and $\tilde{y}_0$ are the $W$-boson and bion fugacities taken at the UV-cutoff scale. Hence, the free energy $F\sim \zeta^{-2} \sim  |T - T_c|^{{1\over 8 \pi \sqrt{y_0 \tilde{y}_0}} }$ is continuous at the transition along with its derivatives to a large (but finite, unlike the BKT transition) order. 

The critical exponent  $\eta$, measuring the decay of the correlations at $T=T_c$, also depends on the values of the fugacities on the line of fixed points. 
It has been shown  \cite{Lecheminant:2002va}, using the dual-sine-Gordon description of the $p=4$ Coulomb gas, that the theory at the self-dual point is equivalent to that of a free scalar field (2d $c=1$ conformal field theory), with critical exponents that depend on the fixed-line value of $y=\tilde{y}$ (an earlier perturbative calculation of the conformal anomaly, also indicating $c=1$, is given in \cite{Boyanovsky:1989mc}). We will not need to make use of this result, as the fugacities in the small-$L$ theory are already small; hence at $T=T_c$ the decay of correlations is governed by the BKT exponents with $\kappa=4$, with negligible corrections due to the nonzero $y$, $\tilde{y}$.

\subsection{Dual descriptions and symmetry realizations above and below ${\mathbf T_c}$} 
\label{dualsymmetries}
  
  In this section, we study the duality between the Coulomb gas (\ref{SU2gas}) and XY models with symmetry-breaking perturbations.
  
  \subsubsection{$\mathbf{\Z_p}$ XY models as Coulomb gases}
  
   The lattice XY-model with a $\Z_p$ preserving perturbation is defined by a partition function similar to (\ref{z4model}), but with an ``external field," breaking the continuous $U(1)$ symmetry $\theta_x \rightarrow \theta_x + c$   to a $\Z_p$ subgroup, $\theta_x \rightarrow \theta_x + {2 \pi \over p}$:
 \begin{equation}
\label{XYZ41}
Z_{XY Z_p} =  \int\limits_0^{2 \pi} \prod_i d \theta_i \;\; e^{{K \over 2 \pi} \sum\limits_{\langle i,j\rangle} \cos (\theta_i - \theta_j) + g \sum\limits_i \cos p \theta_i}~,
\end{equation}
where $g$ is the strength of the $U(1) \rightarrow \Z_p$ perturbation (the lattice spacing $a$ is set to unity). We have denoted the couplings by $K$ and $g$ (instead of $\kappa$ and $\tilde{y}$ as in (\ref{z4model})), since we are going to use the duality of a Coulomb gas to (\ref{XYZ41}) in several different ways. 
 
 We will not derive the duality of (\ref{XYZ41}) to (\ref{SU2gas}) in the most rigorous way, as a 
  detailed derivation on the lattice can be found in the literature \cite{135388,Nienhuis:1984wm}.
    Instead, we will give a more qualitative continuum 
 presentation, see \cite{Wenbook,WittenLectures}. While less rigorous than \cite{135388,Nienhuis:1984wm}, it is perhaps more intuitive  to the continuum physicist. 
 
To begin, note that the naive continuum limit of (\ref{XYZ41}), obtained by expanding the cosine and replacing it with the $\theta$ kinetic term (shown below), is the Euclidean theory of a compact scalar $\theta \equiv \theta + 2 \pi$ with Euclidean action and partition function: 
 \begin{equation}
 \label{XYZ42}
 S[K; G, p] =  \int_{\R^2}  {K \over 4 \pi}(\partial_i \theta)^2 - 2 G \cos p \theta ~, ~~ Z[K; G, p; H, w] = \int {\cal D} \theta \; e^{- S[K; G, p]}~, 
 \end{equation}
 where $2 G = {g \over a^2}$ and $a$ is the lattice spacing. The meaning  of most the arguments of $Z$ is clear, except for $w$ and $H$. These will be explained in detail below (but let us  mention that  $w$ is the (minimum) winding number of vortices that appear with nonzero fugacity $H$).
 To argue that $Z[K; G, p; H, w]$ maps to an $e$-$m$ Coulomb gas, begin by expanding the interaction term as follows:
 \begin{equation}
 \label{XYZ43}
 e^{\int d^2 x 2G \cos p \theta}
  = \sum\limits_{k\ge 0} {(2G)^{k} \over k!} \left( \int d^2 x { e^{i p \theta(x)} + e^{-i p \theta(x)} \over 2}\right)^k~. 
  \end{equation}
  Then, noting that  
 performing a Gaussian integration over $\theta$ will make only equal numbers of $e^{i \theta}$ and $e^{- i\theta}$ contribute (this imposes charge neutrality on the resulting Coulomb gas), rewrite (\ref{XYZ43}) as:
 \begin{equation}
 \label{XYZ431}
 e^{\int d^2 x G \cos p \theta} = \sum\limits_{n\ge 0} \sum\limits_{q_A = \pm 1} { G^{2n} \over (n!)^2}
  \prod\limits_{A=1}^{2n} \int d^2 x_A e^{i p   q_A \theta(x_A)}\bigg\vert_{\sum_A q_A =0}~. 
 \end{equation}
 We then insert (\ref{XYZ431}) in (\ref{XYZ42}), interchange the orders of the sum over $n$ and the path integral and perform the  Gaussian integral over $\theta$. 
 Clearly, every term on the r.h.s. of (\ref{XYZ431}) is now a source of $2n$ ``electric" charges $p q_A$ at positions $x_A$. However, 
 the most general solution for $\theta(x)$  also includes a set of an arbitrary number of vortices, allowed due to the periodicity of $\theta$. Take these to have  
 winding numbers $q_a$ and be located at positions $x_a$.\footnote{Including vortices requires a UV definition, provided, e.g., by the lattice model.} Note that the gas of vortices  must also obey a neutrality condition $\sum_a q_a = 0$, otherwise the action will diverge. Thus, the general solution for $\theta(x)$ in a sector with a given number of vortices and charges is:
 \begin{equation}
 \label{XYZ432}
\theta(x)_{class} = - {i p \over K} \sum\limits_A q_A \ln |x - x_A| + \sum_a q_a \Theta(x - x_a) + \theta_0(x)~,
 \end{equation}
 where $\theta_0(x)$ represents periodic spin-wave fluctuations and $\Theta(x)$ is the polar angle between $\vec{x}$ and, say, the positive $\hat{x}$ axis.
 We note that as far as vortices are concerned, we are free to consider vortices of arbitrary winding number, with   fugacities that can depend on the winding. In what follows  we will  only sum  over vortices with $|q_a|=w$.  Note also that while the vortex fugacities do not explicitly appear in   (\ref{XYZ42}), they are part of the definition of the theory, and are required by the 2d compact scalar duality \cite{WittenLectures}.
 
 The final step is to substitute the classical solution in (\ref{XYZ42}) and calculate the action (omitting all self energies which are, of course, absorbed in the fugacity after renormalization). The spin-wave fluctuations decouple from the vortices and the charges, while the latter only interact via an Aharonov-Bohm type interaction, which arises due to the phase factors in the expansion of the $\cos p \theta$ term.
  The final result, including also a sum over the arbitrary number of vortices of fugacity $H$, gives  the partition function of the compact scalar theory (\ref{XYZ42}) in the form:
\begin{eqnarray}
\label{XYZ433}
 &&Z[K; G, p; H, w]  = Z_0 \times \\
&&~~  \sum\limits_{m, q_a=\pm w} \sum\limits_{n, q_A=\pm1} { G^{2n} \over (n!)^2}{ H^{2m} \over (m!)^2} \int [d^{4(n+m)} r] \; e^{ \sum\limits_{A>B} {p^2 \over K}q_A q_B  \ln r_{AB} + \sum\limits_{a>b}K q_a q_b \ln r_{ab} + i p \sum\limits_{A,b} q_A q_b \Theta( \vec{r}_{Ab})}~, \nonumber
 \end{eqnarray}
 where $Z_0$ is the partition function of the massless scalar $\theta_0$ representing the spin waves. The electric and magnetic (winding number) neutrality is understood in (\ref{XYZ433}) and the sums over $n,m$ are from $0$ to $\infty$. The integral over the position of the charges and vortices is denoted by $\int [d^{4(n+m)} r]$.  Finally, as written, the partition function only involves a sum over one set of magnetic vortices, but the generalization to many is trivial. 
 
 {\flushleft{T}}he summary of this section is that we found  that the compact-$\theta$ theory (\ref{XYZ42}),  related to the XY-model (\ref{XYZ41}), has a representation in terms of an $e$-$m$ Coulomb gas (\ref{XYZ433}). In what follows, we shall call (whenever convenient) the charge-$p$ particles labeled by $q_A, x_A$ ``electric" and the winding-number $w$ vortices labeled by $q_a, r_a$---"magnetic".
 
 \subsubsection{$\mathbf{\Z_p}$ XY-models as duals of  the $\mathbf{W}$-boson/bion gas}
 
 Now we   can map the bion-/$W$-boson gas  partition function of the $SU(2)$ QCD(adj) theory to the Coulomb gas  (\ref{XYZ433}) of the $\Z_4$ XY model. To facilitate the comparison, we reproduce here the interaction energy (``$-\beta H$")  of the $W$/bion gas from (\ref{SU2gas}):
 \begin{equation}
 \label{311again}
 {16 \over \kappa} \sum_{a >  b (bions)} {q_aq_b} \ln r_{ab} + \kappa \sum_{A >  B (W \; bosons)} {q_A q_B} \ln r_{AB}  + 4i \sum\limits_{a,B} q_B q_a \Theta(\vec{r}_{Ba}) ~, ~~~q_{a,A} = \pm 1.
 \end{equation}
 There are two   ways to map (\ref{XYZ433}) to the $W$/bion Coulomb gas (\ref{SU2gas}) that are inequivalent  under $e$-$m$ (vortex-charge) duality. The  first  matches to the  
 $SO(3)$$=$$SU(2)/\Z_2$ gauge theory and the other matches to the $SU(2)$ theory. We discuss both in turn:
 \begin{enumerate}
 \item
First,  we can identify the charge-1 
 particles  with the $W$-bosons of (\ref{SU2gas}), and   the   charge-$p$ 
 particles with the magnetic bions. Upon comparing to the bion partition function (\ref{SU2gas}), or the ``energy" displayed  in (\ref{311again}), we find the map between (\ref{XYZ42}) and the QCD(adj) theory:
\begin{equation}
\label{map2}
p=4,~w=1,~ K = \kappa \equiv {g^2 \over 2 \pi L T},~ G= \zeta_{bion},~ H = 2 \zeta_W,
\end{equation} 
where $\kappa$ is as defined in (\ref{variables}). Thus the dual description is the theory defined by (\ref{XYZ42}) with  $Z[\kappa; \zeta_{bion}, 4; 2 \zeta_W,1]$:
 \begin{equation}
 \label{map22}
 S_1=  \int_{\R^2}  {\kappa \over 4 \pi}(\partial_i \theta)^2 - 2 \zeta_{bion} \cos 4 \theta, ~~Z[\kappa; \zeta_{bion}, 4; 2 \zeta_W, 1] = \int {\cal D} \theta \; e^{- S_1}~. 
  \end{equation}
  Note that the bion-induced potential  given in  
(\ref{non-perturbativeSeff})  for  $SO(3)$ is  $4\pi$ periodic. On the other hand,  our angular spin variable is $2\pi$ periodic.  It is, therefore,  more convenient to compare to the equivalent alternative form  for  $SO(3)$ given in (\ref{alternative}), which  agrees with (\ref{map22}).

Since unit-winding vortices are identified with $W$-bosons and 
charge-$4$ particles---with magnetic bions,  which carry magnetic charge $Q_m=2$, the description by means of $S_1$ allows us to also introduce particles of integer  
charges $<4$ into the system. 
 In the language of the original gauge theory, this means that we could introduce monopoles of charges as low as $Q_m= {1\over 2}$---the minimal charge allowed by Dirac quantization in the $SO(3)$ theory, equal to one-quarter of the bion magnetic charge. On the other hand, since the $W$-bosons are the unit-winding vortices and since there are no $1\over 2$-winding vortices---as these would violate Dirac quantization---we can not introduce dynamical\footnote{However, the theory can still be probed by external charge-$1/2$ ``electrons" (despite the fact that they would violate Dirac quantization if they were dynamical). In the 2d Coulomb gas picture, this  corresponds to the insertion of, say, two {\it external} charge-$1/2$ ``magnetic" particles in the plasma and studying how their interaction is affected by the dynamics of the particles in the plasma as the temperature  changes.} particles of $Q_e = {1\over 2}$ into the system. 
Thus, the description  (\ref{map2}, \ref{map22}) is appropriate for the $SO(3)$ theory. 

 The $e$-$m$ dual of (\ref{map2}, \ref{map22}) is obtained by interchanging $p$ and $w$, i.e., by   letting $p=1$, $w=4$, and taking the $W$-bosons to be the ``electric" charge-$p=1$ particles in (\ref{XYZ433}), and the bions---the $w=4$-winding ``magnetic" particles. To get the interactions in (\ref{SU2gas}) reproduced correctly, this requires taking:
\begin{equation}
\label{map2emdual}
p=1, ~w=4, ~ K = {1 \over \kappa} = {2 \pi L T \over g^2}, ~G= 2 \zeta_W, ~ H =\zeta_{bion}~. 
\end{equation}
We note that (\ref{map2emdual}) is simply the transformation of (\ref{map2}) under the $e$-$m$ duality (\ref{dualityz4}) and is also the usual $T$-duality of $\S^1$ sigma models (compact-$\theta$ theory) in 2d \cite{WittenLectures}. $T$-duality interchanges $K \leftrightarrow 1/K$, i.e., inverts the ``radius" of the compact field $\theta$, and also interchanges vortices and charges (or ``electric" and ``magnetic" particles, in the Coulomb gas language). Overall, the electric-magnetic duality amounts to: 
\begin{equation}
p \leftrightarrow w,  \qquad \kappa  \leftrightarrow  \frac{1}{\kappa}, \qquad  H \leftrightarrow  G\,.
\label{emd}
\end{equation}
This dual description is thus the one defined by (\ref{XYZ42}) with $Z[{1 \over \kappa}; 2 \zeta_W, 1; \zeta_{bion}, 4]$:
 \begin{equation}
 \label{map2emdual2}
 \tilde{S}_1 =  \int_{\R^2}  {1 \over  4 \pi \kappa}(\partial_i \tilde\theta)^2 - 4 \zeta_W \cos  \tilde\theta,~ ~ Z[{1\over \kappa}; 2 \zeta_W, 1; \zeta_{bion}, 4] = \int {\cal D} \tilde\theta \; e^{- \tilde{S}_1}~.
 \end{equation}
Both (\ref{map22}) and (\ref{map2emdual2}), the model with action $S_1$ and its dual  $\tilde{S}_1$-model,    describe the same $SO(3)$ theory, but are weakly coupled in different regimes. 
 
 For later use, note that we use $\theta$ to denote the spin variable  in (\ref{map22}), such that $e^{i q \theta}$ creates magnetic charge in the language of the original gauge theory---minimal charge monopoles for $q=1$, 't Hooft-Polyakov monopoles for $q=2$, and magnetic bions for $q=4$.  In contrast, we used $\tilde\theta$ when describing the $T$-dual spin model in (\ref{map2emdual2}), such that 
 $e^{i q \tilde\theta}$ creates electric charges in the original gauge theory, which are $W$-bosons for $q=1$ (as this is the $SO(3)$ theory, ``electrons" are not allowed as dynamical objects). 

  \item The second interesting possibility, which is not related by 2d $e$-$m$ duality to the preceding one and relevant to our later discussion, is to take the $p=2$ 
  charges in (\ref{XYZ433}) to correspond to the magnetic bions. 
  To preserve the Aharonov-Bohm interaction, we must now take the unit winding $(w=1)$ charges  in (\ref{XYZ433}) to have zero fugacity, and instead sum over 
  charges with $w = 2$ (assigning them  fugacity $H$). This restriction due to the Aharonov-Bohm interaction is protected by a $\Z_2$ symmetry of the theory, and consequently, $w=1$ vortex configurations are forbidden.  
The ``magnetic" charge-2 particles of (\ref{XYZ433}) will be now identified with the $W$-bosons.  
  We then have another representation of the bion/$W$ gas (\ref{SU2gas}) in terms of:
\begin{equation}
\label{map3}
p=2, ~w=2,~ K  = {\kappa \over 4} = {g^2 \over 8 \pi L T} , ~G= \xi_{bion}~, H= 2\xi_W ~.
\end{equation}
This dual to the bion/$W$ gas is  the theory  (\ref{XYZ42}) with $Z[{  \kappa \over 4}; \zeta_{bion}, 2; 2 \zeta_W, 2]$:
 \begin{equation}
 \label{map32}
 S_2 =  \int_{\R^2}  {\kappa \over 16 \pi}(\partial_i \theta)^2 - 2 \xi_{bion} \cos 2 \theta ~, ~~ Z[{\kappa \over 4}; \xi_{bion}, 2; 2 \xi_W, 2] = \int {\cal D} \theta \; e^{- \tilde{S}_2}~.
 \end{equation}
In the description (\ref{map3}, \ref{map32}), the $W$-bosons are winding number-2 vortices and the magnetic bions are the charge-2 
particles. The theory permits the introduction of $Q_e = {1\over 2}$  probe ``electrons" (which would be the winding number-1 vortices) as well as of $Q_m =1$ 't Hooft-Polyakov monopoles 
as dynamical objects (the latter can be introduced by adding Majorana masses for the gauginos to lift the fermion zero modes). Thus, the description in terms of $S_2$ is
 appropriate to the $SU(2)$ theory.  Indeed,   (\ref{map32}) is the dimensional reduction of  the bion induced potential (\ref{non-perturbativeSeff})  for  $SU(2)$ down to $\R^2$.

  To construct the $e$-$m$ dual, we can 
take the $p=2$ 
charges in (\ref{XYZ433}) to correspond to  the $W$-bosons and identify the 
 $w=2$ vortices of (\ref{XYZ433}) with the magnetic bions.  Then we have another representation of the bion/$W$ gas (\ref{SU2gas}) in terms of:
 \begin{equation}
\label{map4}
p=2, ~w=2, ~K = {4 \over \kappa} = {8 \pi L T \over g^2} , ~G= 2 \xi_W, ~H= \xi_{bion}~.
\end{equation}
Thus, now   the  theory dual to the bion/$W$ gas is (\ref{XYZ42}) with $Z[{  4\over \kappa  }; 2 \zeta_{W}, 2; \zeta_{bion}, 2]$:
 \begin{equation}
 \label{map42}
 \tilde{S}_2 =  \int_{\R^2}  {1 \over  \pi \kappa}(\partial_i \tilde\theta)^2 - 4 \xi_W \cos 2 \tilde\theta, ~~ Z[{4 \over \kappa}; 2 \xi_W, 2; \xi_{bion}, 2] = \int {\cal D} \tilde\theta \; e^{- S_2}~.
 \end{equation}

This theory allows the introduction of dynamical    ``electric" charge-$1$ particles (the ``electrons" of the original gauge theory with $Q_e = {1\over 2}$) as well as   ``magnetic" charge-$1$ particles (which in the gauge theory correspond to 't Hooft-Polyakov monopoles with $Q_m=1$) in the plasma. However, in (\ref{map42}), we can not introduce dynamical magnetic monopoles  corresponding to $Q_m={1 \over 2}$. 
Thus,  the description (\ref{map4}, \ref{map42}) is appropriate for the $SU(2)$ gauge theory.
   \end{enumerate}
We stress again that the  difference between the two descriptions above lies in the allowed magnetic and electric charges of  dynamical objects that can be added to the theory without violating Dirac quantization. 
We caution the reader against thinking that $S_1$/$\tilde{S}_1$  and $S_2$/$\tilde{S}_2$  map  to different---from each other and from (\ref{SU2gas})---lattice models. This is not so, {\it provided} the unit-winding fugacity in (\ref{map4}) is strictly zero (the models  
differ in the charges of allowed dynamical objects  {\it not} present in the $W$-/bion-gas (\ref{SU2gas})). Their equivalence as Coulomb gases follows from our discussion and, e.g., the lattice representation \cite{135388}.\footnote{One can show that a Coulomb gas of electric particles of charges quantized in units of of a positive integer $k_e$ and magnetic particles with charges quantized in units of a positive integer $k_m$, with Coulomb interaction strength governed by $\kappa$,  is equivalent to a Coulomb gas of electric particles quantized in units of $k_e^\prime = 1$ and magnetic particles quantized in units of $k_m^\prime = k_e k_m$ and an interaction strength $\kappa^\prime = k_e^2 \kappa$, by showing that they map to equivalent lattice models; that this is so should also be intuitively clear from the fact that they have identical values of the ``energy" $-\beta H$.} 

In each case---either  $SO(3)$ or $SU(2)$---we used $e^{i q \theta}$ to describe the creation of magnetically charged objects in the gauge theory, and $e^{i q \tilde\theta}$---to create electrically-charged gauge-theory objects.
Various correlators of these exponentials   probe the phase structure of the theory and will be discussed in the next section. 
   
  {\flushleft{To}} summarize this section, we have established  two  (up to $e$-$m$ duality)  ways, appropriate to $SO(3)$ or $SU(2)$ gauge theories, to map  the  $SU(2)$(adj) partition function (\ref{SU2gas}) to  the lattice XY-model with a $\Z_p$-preserving perturbation (\ref{XYZ41}). In each case, one side  of the duality is a thermal 4d nonabelian gauge theory with massless fermions, compactified on a small circle, while the other side is  a well-studied lattice model.
 The dynamics of the 4d QCD(adj) gauge theories is complicated and not  well-understood. Numerical Monte-Carlo simulations of theories with massless fermions are expensive and time consuming. On the other hand, in the simplifying regime of a small-$L$ compactification, theories with adjoint Weyl fermions have yielded to controlled theoretical analysis. 
We have taken this analysis one step further by having shown    that the  thermal dynamics of the theory in this regime can be studied via a lattice spin model.  
 
 \subsubsection{Topological, center, and chiral symmetry realization}
 \label{symabovebelow}

The  dual representations described in the previous section can be used to study the realization of the  symmetries as a function of the temperature. 
{\flushleft{\bf The $SO(3)$ theory:}} The dual descriptions are given by (\ref{map22}) and its $T$-dual 
 (\ref{map2emdual2}).
  In the  description by $S_1$ of (\ref{map22}), fluctuations of $\theta$ are small at large values of $\kappa$, which from (\ref{map2}) corresponds to low temperature  $T < T_c = {g^2 \over 8 \pi L}$. Then, the state of the theory is determined by minimizing the classical potential $\sim \xi_{bion} \cos 4 \theta$ and corresponds to the spontaneous breaking of the $\Z_4$ symmetry (recall from the $T=0$ discussion near eqn.~(\ref{minima}) that this corresponds to spontaneously broken topological $\Z_2^{\rm t}$ and chiral $\Z_2^{{\rm d}\chi}$ symmetries). 
  
  On the other hand, the $e$-$m$ (or $T$-) dual model described by $\tilde{S}_1$ (\ref{map2emdual2}) becomes weakly coupled at large values of $\kappa$, i.e.~at high temperature $T > T_c$. The classical potential is $\cos \tilde \theta$, with  a unique ground state and no broken symmetry in the high-temperature regime of the $SO(3)$ theory. 
  
  The symmetry realization can also be inferred from  studying the correlators of the order parameter $e^{i \theta }$ for $\Z_2^{\rm t}$  and the  $\Z_2^{{\rm d}\chi}$ chiral symmetry; recall that $e^{i \theta}$ is an operator creating a minimal charge monopole in the plasma (which is really a 4d minimal charge 't Hooft loop winding around the $\S_L^1$). Below, we show the   $e^{i q \theta}$,  $q = 1,2,3,4$, correlators, as well as those of the $Q_e=1$ Polyakov loop ($e^{i \tilde\theta}$, representing a $W$-boson worldline winding around the $\S_\beta^1$). The   
  $Q_e=1$ Polyakov loop  is not associated with any symmetry, and its correlator is shown 
 for completeness.\footnote{For the $Q_e =1$ Polyakov loop correlator the ``$|x| \rightarrow \infty$" limit should be understood in the sense that while $|x|$ is large, it is still sufficiently small  so that  the Polyakov loop is not screened by $W^\pm$ pairs ``popping out"  of the vacuum and leading to string breaking.}
  Recall that  in the  $SO(3)$  theory, there are no center symmetry,   
 $Q_e=\half$ charge, and fundamental Polyakov loop.    
  \begin{equation}
  \label{so3symmetry}
  \begin{array}{|l|c|cc|cc|c|}
\hline
SO(3)\; {\rm theory} & T<T_c &  & T= T_c& & T>T_c \cr
      &  & & &   & \cr
   \hline
 Q_m = {q\over 2}\; \;\; {\rm 't \;Hooft\; loop:}  \;\;\langle e^{i q \theta(x)} e^{- i q \theta(0)} \rangle\big\vert_{|x| \rightarrow \infty} &     1   &  & 
   {1  \over |x|^{{q^2 \over 4}(1 + {\cal{O}}(y))}   } & & e^{ - {\tilde\sigma_q(T)\over T} |x| } \cr
   \hline
  Q_e = 1\;\;  {\rm Polyakov\; loop:}  \;\;     \langle e^{i   \tilde\theta(x)} e^{- i  \tilde\theta(0)} \rangle\big\vert_{``|x| \rightarrow \infty"} &   e^{ - {\sigma_1 (T)  \over T} |x|} &  &  {1  \over |x|^{{4}(1 + {\cal{O}}(y))}   }& &  1  \cr  
  \hline
 \Z_4^{\rm d\chi/t} \;\; \text {chiral and topological symmetry}  & {\rm broken}& & {\rm c=1 \;   CFT} & & {\rm unbroken}   \cr
 \hline
\text{magnetic charges}   & m{\rm-free} & & m{\rm-free}&  &   m{\rm-confined}  \cr
\hline
 \text{electric charges}     & e{\rm-confined} & & e{\rm-free}&  &   e{\rm-free}   \cr 
      \hline
      \end{array}
  \end{equation}
  
 The table in (\ref{so3symmetry}) shows the spontaneous breaking of $\Z_2^{\rm t}$ (and $\Z_2$ chiral) at low temperatures, the almost-BKT scaling at $T_c$, and the confinement of monopoles in the deconfined phase. Both the string tension $\sigma(T)$ and the dual string tension  $\tilde\sigma_q(T)$  are proportional to the spin-model mass gap (the inverse correlation length $\xi^{-1}$), and both vanish   as $T \rightarrow T_c$:
\begin{equation}
\label{dualstringtension}
{\sigma_q(T)\over T}\bigg\vert_{T \rightarrow T_c^-} \sim {\tilde\sigma_q(T)\over T}\bigg\vert_{T \rightarrow T_c^+} \sim \xi^{-1} \sim|T - T_c|^{  \nu } = |T - T_c|^{ {1\over 16 \pi \sqrt{y_0 \tilde{y}_0}} }~.
\end{equation}
 The value of the exponent $\nu$  follows from our  RG analysis, see (\ref{corrlength}) and Appendix \ref{corrappx}; recall that  the RG analysis is valid on both sides of $T_c$, owing to the high-$T$/low-$T$ duality.   
{\flushleft{\bf The $SU(2)$ theory:}} We begin with the $S_2$ theory (\ref{map32}), and observe that it is weakly-coupled at large $\kappa$, i.e., at low temperature $T<T_c$. Thus, semi-classical analysis is appropriate. It shows that the potential $\cos 2 \theta$ has two minima, indicating that the $\Z_2$ chiral symmetry is broken in the low-temperature phase. The order parameter is $e^{i\theta}$, creating a 't Hooft-Polyakov monopole (a 't Hooft loop of  a $Q_m=1$ monopole winding around $\S_L^1$; it is charged under the chiral symmetry due to the intertwining of the topological shift symmetry with chiral symmetry, eqn.~(\ref{dcs})).\footnote{We warn against identifying $e^{i \theta}$-correlators in the $SU(2)$ and $SO(3)$ dual models: a $Q_m=1$ 't Hooft-Polyakov monopole is created by $e^{i \theta}$ in the $SU(2)$ dual and by $e^{i 2 \theta}$ in the $SO(3)$ dual; similar care should be exercised in comparing the $e^{i \tilde\theta}$ correlators.} 
 On the other hand,   the  description by means of $\tilde{S}_2$,  (\ref{map42}), $T$-dual to (\ref{map32}), is semi-classical at small $\kappa$, high temperature ($T>T_c$), and shows that there are two vacua of the $\cos 2 \tilde\theta$ potential. The corresponding order parameter is the ``electron" creation operator $e^{i \tilde\theta(x)}$---the Polyakov loop of a $Q_e = {1\over 2}$ ``electron" with a worldline winding around the thermal circle $\S_\beta^1$---whose expectation value indicates that the 
  $\Z_2^{\rm c}$ center symmetry is broken at high temperature. We can now summarize the relevant correlators---of 't Hooft-Polyakov monopole operators, $e^{i \theta}$, and of ``electrons", $e^{i \tilde\theta}$---in the $SU(2)$ theory as follows:
    \begin{equation}
  \label{su2symmetry}
  \begin{array}{|l|c|cc|cc|c|}
\hline
SU(2) \; {\rm theory} & T<T_c &  & T= T_c& & T>T_c \cr
      &  & & &   & \cr
   \hline
 Q_m = {1}\; \;\; {\rm 't \;Hooft\; loop:}  \;\;\langle e^{i  \theta(x)} e^{- i  \theta(0)} \rangle\big\vert_{|x| \rightarrow \infty} &     1   &  & 
   {1  \over |x|^{ 1 + {\cal{O}}(y) }   } & & e^{ - {\tilde\sigma_{1} (T)\over T} |x| } \cr
   \hline
  Q_e = {1 \over 2} \;\;  {\rm Polyakov\; loop:}  \;\;     \langle e^{i   \tilde\theta(x)} e^{- i  \tilde\theta(0)} \rangle\big\vert_{ |x| \rightarrow \infty } &   e^{ - {\sigma_{1/2} (T)  \over T} |x|} &  &  {1  \over |x|^{ 1 + {\cal{O}}(y)}   }& &  1  \cr  
  \hline
 \Z_2^{\rm c} \;\; \text {center symmetry}  & {\rm  \; unbroken}& & c=1 \;{\rm   CFT} & & {\rm   \; broken}   \cr
 \hline
 \Z_2^{\rm d\chi}  \;\; \text {discrete chiral symmetry}  & {\rm  broken}& &  & & {\rm   unbroken}   \cr
 \hline
\text{magnetic charges}   & m{\rm-free} & & m{\rm-free}&  &   m{\rm-confined}  \cr
\hline
 \text{electric charges}     & e{\rm-confined} & & e{\rm-free}&  &   e{\rm-free}   \cr 
      \hline
      \end{array}~,
  \end{equation}
      which is the $SU(2)$ equivalent of the $SO(3)$ table (\ref{so3symmetry}).
 
\section{On the deconfinement transition in $\mathbf{N_c>2}$ QCD(adj)  }
\label{sunsection}

In this section, we study the deconfinement phase transition for $SU(N_c)$ QCD like theories with $n_f$ adjoint fermions compactified on $\R^{1,2}\times \S_L^1$. The $SU(N_c)$ gauge symmetry is broken by the Higgs mechanism down to $U(1)^{N_c-1}$, as described in Section \ref{perturbative}.

\subsection{The Coulomb gas   for $\mathbf{N_c>2}$ and $\mathbf{N_c=3}$ electric-magnetic duality}
\label{suncoulombgas}

The non-perturbative dynamics is also similar to the one described in Section \ref{non-perturbativesu2review} for $SU(2)$, but one has to take into account the fact that there are now different kinds of non-perturbative monopole-instanton configurations. The monopole-instantons are now labeled by the affine roots of the $SU(N_c)$ Lie algebra. Let the simple roots be $\vec{\alpha}_i$, $i = 1,\ldots N_c -1$. A given root $\vec{\alpha}_i$ can be described as  an $N_c-1$ dimensional vector.
 The affine root $\vec{\alpha}_{N_c}$ is defined as:
\begin{eqnarray}
\vec\alpha_{N_c}=- \vec\alpha_1 - \vec\alpha_2 - \vec\alpha_3 - \ldots - \vec\alpha_{N_c-1}~.
\end{eqnarray}
The extended (or affine) root system  $\Delta_{\rm aff}^0 = \{ \vec{\alpha}^1,  \vec{\alpha}^2, \ldots  \vec{\alpha}^{N_c} \}$, using normalization $\mbox{Tr}\,t_at_b=\frac{\delta_{ab}}{2}$ for the generator of the Lie algebra as stated in Section~\ref{perturbative}, obeys:
\begin{eqnarray}
\label{alphaproducts}
\vec \alpha_i\cdot\vec \alpha_j=\delta_{i,j}- \half \delta_{i,j+1}-\half \delta_{i,j-1}, \; {\rm with}\; \; i,j = 1 \ldots N_c \; (N_c +1 \equiv 1, 0 \equiv N_c)~.
\end{eqnarray}
The self-dual BPS and KK monopoles  in $SU(N_c)$  are labeled by the extended roots---or equivalently, their magnetic charges under the unbroken $U(1)^{N_c -1}$. The  magnetic charge of each monopole  is found by  integrating its $U(1)^{N_c-1}$ magnetic field over spatial infinity in $\R^3$:
\begin{equation}
 \int_{S^2} d{\bm\Sigma} \cdot \mathbf F = \frac {4 \pi}{g} \,\vec{\alpha}^i \equiv  4 \pi \vec{Q}_{M_i}~, 
 \qquad
 \mbox{for the type-$i$ $(= 1 \ldots N_c$) monopole}~.
\label{eq:mag-charge}
\end{equation}
In the center-symmetric vacuum, the action of each monopole (\ref{eq:mag-charge}) is $e^{- S_0} = e^{ - {8 \pi^2 \over g^2 N_c}}$.

Recall that the charges of the $W$-bosons are $\vec{Q}_{W_i} = g \vec{\alpha}^i$, thus, 
\begin{eqnarray}
\vec{Q}_{M_i} \cdot \vec{Q}_{W_j} =  \left\{ \begin{array}  {cl}
\;\; 1 & \qquad i=j \\
-\half  & \qquad i = j \pm 1 \\
\;\; 0 & \qquad  {\rm otherwise} 
\end{array}\,. \right.
\end{eqnarray}
The Dirac quantization condition is saturated for the nearest neighbor charges on the Dynkin diagram, and is twice (as in $SU(2)$) the minimal bound for charges on the same site on 
Dynkin diagram.

Magnetic bions, the topological excitations responsible for the confinement, also exist for $SU(N_c)$ with massless adjoint fermions. Bions are composed of a monopole with charges $\alpha_i$ and anti-monopole with charges $-\alpha_{i-1}$. The monopole constituents of the bions   repel because of Coulomb interaction and attract because of fermion zero-mode exchange; this ensures the stability of the bions, as in the $SU(2)$ case. 
 Clearly, there are also $N_c$ bions, labeled by the position of  (say) the monopole constituent on the extended Dynkin diagram.
 Every bion carries a charge under the unbroken $U(1)^{N_c-1}$ gauge group, given by: 
 \begin{equation}
 \label{bioncharge}
 \vec Q_i=\vec \alpha_i-\vec \alpha_{i-1}~, ~\; i = 1 \ldots N_c~.
 \end{equation}
 The (magnetic) Coulomb potential energy between two bions (or antibions) carrying charges $\vec Q_i
 $, and $\vec Q_j$, and located at $\vec R_i$, and $\vec R_j$, is given by:
\begin{eqnarray}
V_{\mbox{\scriptsize bion}\,i,j}= \pm {4 \pi L \over g^2}   \frac{ \vec Q_i\cdot \vec Q_j}{ |\vec R_i-\vec R_j|}\,~.
\end{eqnarray}
The products of the bion charge vectors can be evaluated, using (\ref{bioncharge}) and (\ref{alphaproducts}), yielding: 
\begin{eqnarray}
\label{qproducts}
\nonumber
\vec Q_i\cdot\vec Q_j&=&3\delta_{i,j}-2\delta_{i,j+1}-2\delta_{i,j-1}+\frac{1}{2} \delta_{i,j+2}+ 
\frac{1}{2}  \delta_{i,j-2}\quad \mbox{for}\, N_c\geq 5~,\\
\vec Q_i\cdot \vec Q_j&=&3 \delta_{i,j}-2\delta_{i,j+1}-2 \delta_{i,j-1}+    \delta_{i,j+2}\quad \mbox{for}\, N_c= 4~,\\
\nonumber
\vec Q_i\cdot\vec Q_j&=&3 \delta_{i,j}-\frac{3}{2} \delta_{i,j+1}-\frac{3}{2} \delta_{i,j-1}\quad \mbox{for}\, N_c=3 \, ~,\\
 \vec Q_i\cdot\vec Q_j&=&4\delta_{i,j}-4\delta_{i,j+1}\quad \mbox {for}\; N_c=2\,.\nonumber
\end{eqnarray}
Note the fact (of later importance) that, for $SU(3)$, we have:
\begin{equation}
\label{su3bionvectors}
\vec Q_i\cdot\vec Q_j = 3 \; \vec\alpha_i \cdot \vec\alpha_j~.
\end{equation}

At temperatures ${m_\sigma\ll T \ll m_W}$, where ${m_\sigma}$ is the dual photon mass, the $1/|\vec R|$ inter-bion potential should be replaced with a logarithmic one, as was done for the $SU(2)$ case, see (\ref{3d2dCoulomb}). Hence, the pairwise interaction between a collection of $n_{i}$ bions (with charge $\vec Q _i$) and $\bar n_{i}$ antibions (with charge $-\vec Q _i$) is given by:\footnote{Note that for $N_c=2$, there is only one bion, hence there is no sum over $N_c$ different kinds of bions.} 
\begin{eqnarray}
\nonumber
S_{\mbox{\scriptsize bions}}&=&-\frac{{ 8}\pi LT}{g^2}\sum_{i\geq j}^{N_c}\vec Q_i\cdot \vec Q_j\left[\sum_{a>b}^{n_i,n_j}\ln\left(T|\vec R^i_a-\vec R_b^j|\right)+ \sum_{\bar a>\bar b}^{\bar n_i,\bar n_j}\ln\left(T|\vec R^i_{\bar a}-\vec R_{\bar b}^j|\right)\right.\\
&&\left.\quad\quad\quad\quad\quad\quad\quad\quad-\sum_{ a,\bar b}^{n_i,\bar n_j}\ln\left(T|\vec R^i_{ a}-\vec R^j_{\bar b}|\right)\right]\,.
\end{eqnarray}
 Next,
  we turn to the $W$ bosons. In Dynkin space we have $N_c$ $W$-bosons at the lowest Kaluza-Klein level, for reasons similar to the ones explained for the $SU(2)$ case. Each $W$-boson carries a charge  given by $g\vec \alpha_i$\,, and each has  mass $M_W={2\pi \over N_c L}$. The would-be 3d interaction between two $W$-bosons is:
\begin{eqnarray}
V_{W\,ij}=\pm \frac{g^2}{4\pi}\frac{\vec\alpha_i\cdot\vec\alpha_j}{|\vec R_i-\vec R_j|}\,.
\end{eqnarray}
 At temperature $T$, the interaction between  $m_{i}$ $W^+$ (with charge $\vec\alpha_i$) and $\bar m_{i}$ $W^-$ (with charge $-\vec\alpha_i$) bosons is:
\begin{eqnarray}
\nonumber
S_{W}&=&- \frac{ g^2}{2\pi LT}\sum_{i\geq j}^{N_c}\vec \alpha_i\cdot \vec\alpha_j\left[\sum_{A>B}^{m_i,m_j}\ln\left(T|\vec R^i_A-\vec R_B^j|\right)+ \sum_{\bar A>\bar B}^{\bar m_i,\bar m_j}\ln\left(T|\vec R^i_{\bar A}-\vec R_{\bar B}^j|\right)\right.\\
&&\left.\quad\quad\quad\quad\quad\quad\quad\quad-\sum_{ A,\bar B}^{m_i,\bar m_j}\ln\left(T|\vec R^i_{A}-\vec R_{\bar B}^j|\right)\right]\,.
\end{eqnarray}
 The Aharonov-Bohm phase interaction between a $W$-boson (with charge $\vec\alpha_j$ at $\vec{y}_j$), and a bion (with charge $\vec{Q}_i$ at $\vec{x}_i$) reads, similar to (\ref{ABinteraction2}):
\begin{eqnarray}
{2} \vec Q_i\cdot\vec\alpha_{j}\Theta(\vec x_i-\vec y_j),  
\end{eqnarray}
where $\Theta(\vec{x}_i-\vec{y}_j)$ is the angle between the vector connecting the monopole and the $W$-boson and a chosen spacial direction. Hence, the interaction between $m_j$ and $\bar m_j$ $W$-bosons (with charge $\vec \alpha_i$ and $-\vec \alpha_i$ respectively), and $n_i$ and $\bar n_i$ bions (with charge $\vec Q_i$ and  $-\vec Q_i$ respectively) is given by:
\begin{eqnarray}
\nonumber
S_{W,\mbox{\scriptsize bions}}&=&- { 2} i \sum_{i,j}^{N_c}\vec Q_i\cdot\vec \alpha_j\left[\sum_{a,B}^{n_i,m_j}\Theta\left(\vec R^i_a-\vec R_B^j\right)+ \sum_{\bar a,\bar B}^{\bar n_i,\bar m_j}\Theta\left(\vec R^i_{\bar a}-\vec R_{\bar B}^j\right)\right.\\
&&\left.\quad\quad\quad\quad\quad\quad\quad\quad-\sum_{ a,\bar B}^{n_i,\bar m_j}\Theta\left(\vec R^i_{a}-\vec R_{\bar B}^j\right)-\sum_{ B,\bar a}^{\bar n_i,m_j}\Theta\left(\vec R^i_{\bar a}-\vec R_{B}^j\right)\right]\,.
\end{eqnarray}
Collecting everything, we arrive to  the 2d Coulomb gas representation of the  partition function of $SU(N_c)$ QCD(adj):
\begin{eqnarray}
\nonumber
Z_{\mbox{\scriptsize bion}+W}&=&\sum_{N^i_{b\pm},\pm \vec Q_i}\sum_{N^j_{W\pm},\pm \vec \alpha_j}\frac{\xi_{\mbox{\scriptsize bion}}^{N_{b+}^i+N_{b-}^i}}{N_{b+}^i!N_{b-}^i!}\frac{\xi_W^{N_{W+}^j+N_{W-}^j}}{N_{w+}^j!N_{w-}^j!}\prod_{a,i}^{N_{b+}^iN_{b-}^i}\int d^2R_a^i \prod_{A,j}^{N_{W+}^jN_{W-}^j}\int d^2R_A^j\\
\label{partition function for SUn}
&&\times \exp\left[-S_{\mbox{\scriptsize bion}}-S_W-S_{W,\mbox{\scriptsize bions}}\right]\,,
\end{eqnarray}
where $\xi_{\mbox{\scriptsize bion}}$ and $\xi_W$ are the bion and $W$ fugacities 
$\xi_{\mbox{\scriptsize bion}} \sim  \frac{e^{-{16\pi^2\over N_c g^2}}}{L^3T g^{14-8n_f}}$, $
\xi_{W}  \sim m_w Te^{-\frac{m_w}{T}}$. 

For further use, let us rewrite (\ref{partition function for SUn}) as follows:
\begin{eqnarray}
\label{sunpartition11}
\nonumber
Z &=&\sum\limits_{(N^i_{e \pm}\geq0,\; i\geq0,\; q_a=\pm1)} \sum\limits_{(N^i_{m \pm}\geq0,\; j\geq0,\; q_A=\pm1)}\frac{\left(\frac{y_m}{a^2}\right)^{  \sum_{i}(N^i_{m +} + N^i_{m -})} \left(\frac{y_e}{a^2}\right)^{  \sum_{i}(N^i_{e +} + N^i_{e -})} }{ \prod\limits_i  N^i_{m+}!N^i_{m-}!N^i_{e+}!N^i_{m-}! } \nonumber \\
&&\times \int \prod_{a,i} d^2R_a^{i}\int \prod_{A,i}d^2R_A^{j}
\nonumber\\
\nonumber
&&\times\exp\left[\kappa_e\sum_{i\geq j}^{N_c}\sum_{A>B}^{N_e}q_Aq_B\vec \alpha_i\cdot \vec\alpha_j\ln\frac{|\vec R^i_A-\vec R_B^j|}{a}+\frac{{4}}{\kappa_m}\sum_{i\geq j}^{N_c}\sum_{a>b}^{N_m}q_aq_b\vec Q_i\cdot \vec Q_j\ln\frac{|\vec R^i_a-\vec R_b^j|}{a}\right.\\
&&\qquad~~ \left.+ {2} i\sum_{i,j}^{N_c}\sum_{a,B}^{N_m,N_e}q_aq_B\vec \alpha_j\cdot \vec Q_i\ln\frac{|\vec R^i_a-\vec R_B^j|}{a}   \right]\,. 
\end{eqnarray}
We have slightly changed notation, compared to (\ref{partition function for SUn}), notably the subscript $m$ refers now to bions and $e$---to $W$'s.
 The interaction strengths are denoted by $\kappa_e$ for the electric charges' Coulomb interaction and $\frac{1}{\kappa_m}$ for the magnetic charges' interaction.  The map between the UV-cutoff ($a$) values of $\kappa_e$, $\kappa_m$, $y_e$, $y_m$, and $g$, $\xi_{\mbox{\scriptsize bion}}$, $\xi_W$, $L$, and $T$, is:
\begin{eqnarray}
\label{maps3}
y_e  \leftrightarrow \xi_W a^2 ~, ~ y_m  \leftrightarrow  \xi_{\mbox{\scriptsize bion}} a^2~, ~ 
\kappa_e(a) =  \kappa_m(a)    \leftrightarrow \frac{g^2}{ { 2} \pi LT}\,.
\end{eqnarray}
 The electric charges ($W$'s) of type $i$ (of charges $q_A \vec\alpha^i$, $q_A = \pm 1$) are located at $\vec{R}^i_A$ and the magnetic charges (bions) of type $i$ (of charges $q_a \vec{Q}^i$, $q_a = \pm 1$) are located at $\vec{R}^i_a$. Note that only $e$- and $m$- charge-neutral, with respect to $U(1)^{N_c-1}$, configurations will give a finite contribution to the partition function. 
 There are arbitrary numbers of electric ($N_{e \pm}^i$) and magnetic ($N_{m \pm}^i$) charges of each type in the plasma, which are summed over in the grand canonical  partition function. 
 Since we are working at a center-symmetric point along the spatial circle,   the fugacities of all $N_c$ bions are equal (as well as those of the $N_c$ $W$-bosons). The dimensionless fugacities are defined as in (\ref{variables}) in terms of the bion and $W$ fugacities. 
 Similar to the fugacities of the different bions, the gauge couplings of all the unbroken $U(1)$'s are equal. The  ``clock" symmetry of the extended Dynkin diagram  responsible for the equal fugacities is preserved by renormalization and greatly simplifies the resulting RGEs.
 
An important observation distinguishing  $N_c=3$ from the $N_c > 3$ theories is now due. From   eqn.~(\ref{su3bionvectors}), it follows that for $N_c=3$ the partition function  (\ref{sunpartition11}) is invariant under the duality transformation:
\begin{equation}
\label{su3dualitymain}
y_m \Longleftrightarrow y_e , ~~~ \kappa_e  \Longleftrightarrow {12\over \kappa_e},  ~~{\rm for } \; SU(3)~{\rm only}, 
\end{equation}
exchanging the fugacities of bions and $W$'s and appropriately inverting the coupling (note that for $SU(3)$ $\kappa_e = \kappa_m$ at any scale, see the next Section \ref{sunrges}). As was already mentioned in the Introduction, this is because the interpretation of (\ref{su3bionvectors}) is that different kinds of bions and $W$'s have identical interactions (which are only nearest-neighbor in the Dynkin diagram).  We expect that the duality relation (\ref{su3dualitymain}) is preserved by the RGE flow (and show that indeed, it is) and that its use should greatly simplify the study of the critical theory.
 
  On the other hand, since for $N_c>3$ there is no relation between $\vec\alpha_i \cdot \vec\alpha_j$ and $\vec{Q}_i \cdot \vec{Q}_j$ (see (\ref{alphaproducts}), (\ref{qproducts})), the partition function is 
  not invariant under electric-magnetic duality. 
   Thus, we expect that the UV cutoff relation $\kappa_e = \kappa_m$ is going to be violated by the renormalization flow. Thus, in addition to the RGEs for the two fugacities $y_e$ and $y_m$, we will have separate RGEs for $\kappa_e$ and $\kappa_m$. This expectation is, indeed, borne out by their explicit form given in the next section.

\subsection{The RGEs for the $\mathbf{SU(N_c)}$ magnetic bion/W-boson plasma}
\label{sunrges}

 Although there are different methods to derive the RGEs, we choose to follow the more intuitive approach introduced for the $SU(2)$ case in Appendix A, and worked out for the general $SU(N_c)$ model in Appendix \ref{sunrges1}. 
 
 Unlike the $SU(2)$ self-dual model, the self-duality is lost in the $SU(N_c)$ group with $N_c > 3$. 
Therefore, one does not expect the magnetic and electric coupling strengths to scale the same under the RG flow. This, in turn, forces us to define different strengths $\kappa_e$ and ${1\over \kappa_m} (\ne {1\over \kappa_e})$ for the Coulomb interaction of electric and magnetic charges. The details of the procedure are given in Appendix \ref{sunrges1}. The final equations (\ref{RG for kappae}), (\ref{RG for kappam}), and (\ref{rgefory}), respectively, read:
\begin{eqnarray}
\nonumber
\frac{d\kappa_e}{db}&=&2 \pi^2 \left(- \kappa^2_ey_e^2\sum_{i=0,\pm1}(\vec\alpha_{p}\cdot\vec\alpha_{p+i})^2+ 4 y_m^2\sum_{i=-1,0,1,2}(\vec\alpha_{p}\cdot\vec Q_{p+i})^2  \right)\,, \nonumber 
\\
\nonumber
\frac{d}{db}\left(\frac{1}{\kappa_m}\right)&=&\frac{2\pi^2}{3}\left(-\frac{4 y_m^2}{\kappa_m^2}\sum_{i=0,\pm1,\pm2}(\vec Q_{p}\cdot\vec Q_{p+i})^2+y_e^2\sum_{i=-2,-1,0,1}(\vec Q_{p}\cdot\vec \alpha_{p+i})^2  \right)\,, \nonumber 
\\
\nonumber
\frac{dy_e}{db}&=&\left(2-\frac{\kappa_e}{2} \right)y_e\,,\\
\label{complete RGES in SUn}
\frac{dy_m}{db}&=&\left(2-\frac{6}{\kappa_m} \right)y_m\,.
\end{eqnarray}
Notice the different number of terms in the sums over bion-pair and $W$-boson pair contributions to the (anti-) screening. This is because a given bion can interact with its next-to-nearest neighbors on the Dynkin diagram (i.e., it interacts with similar kinds bions as well as with four  other kinds), while a given $W$-boson can only interact with the same kind of $W$'s and with its two nearest-neighbors. 
 
  As already mentioned, $N_c=3$  is an exceptional case. There, the sums over $i$  in all terms in (\ref{complete RGES in SUn}) are over $0, \pm 1$ only.  We can evaluate the r.h.s., using (\ref{su3bionvectors}), (\ref{alphaproducts}), and (\ref{bioncharge}), 
  and cast the $SU(3)$ RGEs in the form:
 \begin{eqnarray}
 \label{SU3RGES}
\nonumber
\frac{d\kappa_e}{db}&=&  3 \pi^2 \left(-   \; \kappa^2_e\; y_e^2 + 12 \; y_m^2 \right) ~,\nonumber \\
\nonumber
\frac{d}{db}\left(\frac{1}{\kappa_m}\right)&=& 3 \pi^2  \left(- 12\; \frac{y_m^2}{\kappa_m^2} + \; y_e^2  \right)\,, \nonumber \\
\frac{dy_e}{db}&=&\left(2-\frac{\kappa_e}{2} \right)y_e\,,\\
\frac{dy_m}{db}&=&\left(2-\frac{6}{\kappa_m} \right)y_m\,. \nonumber
\end{eqnarray}
The UV relation $\kappa_e = \kappa_m$ is preserved by the RGEs (one can explicitly check that the second equation follows from the first, third and fourth).
The  RGEs are invariant under the $e$-$m$ duality (\ref{su3dualitymain}) ($\kappa_e \rightarrow 12/\kappa_e$, $y_e \leftrightarrow y_m$). Further, we also observe that there is a fixed point for the $\kappa_e$ (or $1/\kappa_e$) equation at the $e$-$m$ self-dual point $\kappa_e = \kappa_m=  2 \sqrt{3}$, $y_e = y_m$. However, this is not a fixed point of the $y_{e,m}$ RGEs. Instead, both fugacities are relevant at the $\kappa_{e,m} = 2 \sqrt{3}$ fixed point, at least according to the leading-order RGEs. While the existence of a self-dual point implies, as discussed in the Introduction, that $T_c$ is the one corresponding to $\kappa_{e,m} = 2 \sqrt{3}$, further study of the approach to criticality and the critical theory clearly requires extensions of the methods used here. 
  
 We shall make very few comments on the $SU(N_c>3)$ case. The deconfinement transition there also  occurs in the region where both fugacities are relevant, in the window $ {g^2 \over 8 \pi L} < T_c < {g^2 \over 6 \pi L}$ (equivalently ${3 <\kappa_e = \kappa_m< 4}$ ).  
It is also clear that studying the nature of the transition is more challenging than in the $N_c = 2$ case. This is because in the $N_c>2$ case the RGEs  do not display a fixed point, at least to lowest order in fugacities, and because $e$-$m$ duality, which is  helpful to find the value of $T_c$ even in models where small-fugacity RGEs break down (and even study the theory at $T_c$) is absent in QCD(adj).

 In the next section, we describe the construction of a lattice model dual to the $N_c = 3$ Coulomb gas. The model is related to vectorial Coulomb gas models used to study the melting of 2d crystals with a triangular lattice \cite{Nelson} (not surprisingly, in view of the root lattice of $SU(3)$), by the addition of appropriate symmetry breaking fields. In this paper, we will only present the lattice model and will defer the study of its phase diagram for future work.

\subsection{The $\mathbf{SU(3)}$  Coulomb gas as a system of coupled lattice $\mathbf{XY}$ spins }
\label{su3lattice}

In this section, we map the Coulomb gas of bions and $W$-bosons for the $SU(3)$ theory to the 
$\Z_3\times \Z_3$-preserving ``vector" spin model discussed in the Introduction, see eqn.~(\ref{z3sqrdmodel}). We provide two alternative, but equivalent, descriptions of the spin-system. 

\subsubsection{First description}

The first description uses a  two-component vector $\vec\theta_x =  (\theta_x^1, \theta_x^2)$, which is a periodic variable:
\begin{equation}
\label{periodicroot}
   \vec\theta_x \equiv    \vec\theta_x  + 2 \pi \vec\alpha_1 \equiv \vec\theta_x  + 2 \pi \vec\alpha_2  \qquad {\rm or} \qquad 
      \vec\theta_x \equiv    \vec\theta_x  + 2 \pi  \Gamma_r~,
            \end{equation}
   whose periodicity is determined  by  ($2 \pi$ times)  
   the root  lattice $ \vec\alpha_i \in \Gamma_r$ of $SU(3)$. The vector $\vec\theta_x$ is
associated with a site $x$ on a two-dimensional lattice.   We shall take the partition function of the model to be defined as a path integral over the compact $\vec\theta$, $   Z = \int {\cal{D}} \theta  e^{ - \beta H}$ (in the path-integration,  $\theta$ needs to be integrated over the unit cell of the root lattice $\Gamma_r$), where: 
\begin{equation}
\label{z3sqrdmodel1}
- \beta H =   \sum_{x; \hat\mu = 1,2}  \sum_{i=1}^3 {\kappa \over 4 \pi} \cos  2 \vec\nu_i \cdot ( \vec\theta_{x  + \hat\mu} -\vec\theta_{x}) + \sum_{x} \sum_{i=1}^3 \tilde{y}   \cos 2 (\vec\alpha_i - \vec\alpha_{i-1}) \cdot  \vec\theta_{x} .
\end{equation}
Here,  $\vec\nu_i$ are the three two-component weights of the defining representation of $SU(3)$, $\vec\alpha_i - \vec\alpha_{i-1} \equiv \vec{Q}_i   $ is the magnetic bion charge defined in (\ref{bioncharge}).
The factor of  two in  both terms  ensures the periodicity (\ref{periodicroot}) of $\theta$.\footnote{\label{conventions}In this footnote, we  give an explicit basis for calculations  and various properties of weights and roots of $SU(3)$.  As per our trace normalization, ${\rm Tr}(t_a t_b)=\half \delta_{ab}$, the weights of the defining representation are
$  \vec\nu_1= (\half, \frac{1}{2 \sqrt 3}), \; \vec\nu_2= (-\half, \frac{1}{2 \sqrt 3}),   \; 
\vec\nu_3= (0,  -\frac{1}{ \sqrt 3})$. The roots are differences of weights and are given by  
 $\vec\alpha_1 =  \vec\nu_1 -  \vec\nu_2=(1,0)$,   $\vec\alpha_2=  \vec\nu_2 -  \vec\nu_3=(-\half, \frac{\sqrt 3}{2})$, and affine root  $\vec\alpha_3=  \vec\nu_3 -  \vec\nu_1= (-\half, -\frac{\sqrt 3}{2})$. Their length is, following normalization of the trace, normalized to one.   
 The following relations are valid for general $SU(N)$ and are useful in practical calculations. 
 The weights, now represented by $N$ ($N$$-$$1$)-dimensional vectors $\vec{\nu}_i$, obey:
 \begin{eqnarray}
 &&\sum_{i=1}^{N} \vec\nu_i=0,   \qquad 
\vec\nu_i \cdot \vec\nu_j = \sum_{a=1}^{N-1} (\nu_i)^a (\nu_j)^a =   \frac{\delta_{ij}}{2} - \frac{1}{2N},  \qquad i, j=1, \ldots, N~, \cr 
 && \sum_{i=1}^{N} (\nu_i)^a (\nu_i)^b =   \frac{\delta^{ab}}{2},   \qquad a,b=1, \ldots, N-1 ~,
\end{eqnarray} 
and they generate the weight lattice $\Gamma_w$ of $SU(N)$. 
The roots obey:
\begin{eqnarray}
&& \vec\alpha_i  \equiv   \vec\nu_i - \vec\nu_{i+1}  ,  
 \vec\nu_i \cdot  \vec\alpha_j   = \vec\nu_i \cdot  (  \vec\nu_j -  \vec\nu_{j+1} )= \half (\delta_{ij} - \delta_{i, j+1})~,\cr 
&& \vec\alpha_i\cdot \vec\alpha_j= (  \vec\nu_i-  \vec\nu_{i+1} )\cdot(  \vec\nu_j-  \vec\nu_{j+1} )= \delta_{ij}- \half \delta_{i, j+1} 
 -  \half \delta_{i, j-1} ~,
\end{eqnarray} 
and they generate the root lattice  $\Gamma_r$. The root lattice is a sublattice of the weight lattice, and it is coarser. The quotient is   $\Gamma_w/ \Gamma_r= \Z_N$.   }

The continuum description of the lattice model (\ref{z3sqrdmodel1}) can be found by taking 
the naive continuum limit and keeping in mind that the two scalars are periodic according to  (\ref{periodicroot}). Expanding the first term in (\ref{z3sqrdmodel1}) to quadratic order around the origin,\footnote{Note that this is the  unique minimum of the  ``kinetic" term in the fundamental domain of (\ref{periodicroot}). } we obtain a canonical (up to an overall prefactor) kinetic term for the $\vec\theta(x)$ field.  
Explicitly, denoting $\nabla_\mu \theta_x = \theta_{x +\hat{\mu}} - \theta_x$, we have:
 \begin{eqnarray}
 (-\beta H)^{\rm quad.}_{\tilde{y}=0} && = -   \frac{ \kappa}{2 \pi}  \sum_{x, \mu}   \sum_{i=1}^{3}   (\vec\nu_i \cdot \nabla_\mu \vec\theta_{x})^2 = 
-   \frac{ \kappa}{2 \pi}  \sum_{x, \mu}  \sum\limits_{a,b=1}^{2} \sum_{i=1}^{3}   (\nu_i)^a  (\nu_i)^b  \nabla_\mu  \theta_{x }^a \nabla_\mu \theta_{x}^b  \cr
   && =    -  \frac{ \kappa}{4\pi}    \sum_{x, \mu}      (\nabla_\mu  \vec\theta_{x}) \cdot (\nabla_\mu \vec\theta_{x})   \longrightarrow   - \frac{\kappa}{4\pi}  \int d^2 x  \;    (\partial_{\mu} \vec\theta)^2~.
   \label{cont}
 \end{eqnarray}
  We shall take the partition function of the continuum version of the model to be defined as a path integral over the compact $\vec\theta$, $   Z = \int {\cal{D}} \theta  e^{ - \beta H}
$: 
\begin{equation}
\label{su310}
-\beta H = - {\kappa \over 4 \pi} \; \int_{\R^2} (\partial_\mu  \vec\theta)^2  + 
2 \zeta \sum_{i=1}^3    \int_{\R^2}  \cos 2 \vec{Q}_i \cdot \vec\theta \;, \qquad \qquad 
\vec\theta \equiv \vec\theta+ 2 \pi  \Gamma_r~,
\end{equation} 
where $\vec{Q}_i = \vec\alpha_i - \vec\alpha_{i-1}$ is the bion charge.

Expanding the interaction potential, exactly as in (\ref{XYZ431}), we observe that the partition function of (\ref{su310}) precisely reproduces 
 the magnetic bion interactions and their contribution to the partition function    (\ref{sunpartition11})  with $\zeta \sim \xi_{bion} \sim y_m$.  Eqn.~(\ref{su310}) also incorporates 
 the $W$-boson interactions as well as the Aharonov-Bohm type  $e$-$m$ interactions, as in the discussion of the $SU(2)$ case. The $W$-bosons, once again, arise as vortices of the spin model. 
To see this,  recall that in the continuum version  $\vec\theta$ is periodic by elements of the root lattice (\ref{periodicroot}). Thus, a vortex at the origin can be written  as:
 \begin{equation}
 \label{vortexsu3}
 \vec\theta(\varphi) = \left( n^1 \vec\alpha_1 +  n^2 \vec\alpha_2 \right) \varphi\,, 
\end{equation}
where $n^1, n^2  \in \Z$ and $\varphi$ is the polar angle. Following Nelson \cite{Nelson},  we  take the fugacity of a vortex 
 labeled by $(n^1, n^2)$  at $x$  to be given by:\footnote{We note that the lattice model produces a fugacity exactly of the form (\ref{su34}), with $y$ determined by the details of the lattice cutoff; that this is so can be seen also from the continuum description, where the self-energy of a vortex is  proportional to $(n^i  \vec\alpha_i)^2$. In the Villain representation of the spin model, a  fugacity term   (\ref{su34}) can be added with arbitrary $y$.}

 \begin{equation}
 \label{su34}
 y^{ (n^i(x) \vec\alpha_i)^2 } = y^{(n_1^2 + n_2^2 - n_1n_2)}~.
 \end{equation}
Thus, for small $y$, the leading-order contribution to the vortex partition function is due to vortices with winding number vectors  denoted by $\vec{n}_A$, $A=1,2,3$:
\begin{equation}
\label{su35}
\vec{n}_1 = (1,0),~ \vec{n}_2 = (0,1), ~ {\rm and} ~ \vec{n}_3 = (-1, -1),
\end{equation}
and their corresponding  anti-vortices.   
Note that the vortex with $ (n_1,n_2)=(-1,-1)$ corresponds to the affine root 
$\vec\alpha_3 = -\vec\alpha_1  - \vec\alpha_2$ of the $SU(3)$ algebra. It is pleasing to see that  a $W$-boson 
associated with the affine root---which is only present in the compactified gauge theory, and can be referred to as a Kaluza-Klein $W$-boson---arises naturally, with fugacity equal to that of the other two $W$-bosons, from the spin-system as the vortex with charge assignment  $(-1,-1)$. 
Each of the three vortices (\ref{su35}) gives the same contribution, proportional to $y$, to the vortex partition function. Any vortex with winding numbers different from (\ref{su35}) (and the corresponding antivortex)  gives a contribution to the partition function proportional to a higher power of $y$ and its effect is thus  suppressed with respect to that of (\ref{su35}). 

Next, we can also easily see that the interactions of two vortices  (\ref{vortexsu3}),  separated by a distance $r$, with vorticities by $n(0)^i \vec\alpha_i$ and $n(r)^j \vec\alpha_j$,  is given by:
\begin{equation}
\label{vortexintsu3}
- \kappa  \; n(0)^i \vec\alpha_i \cdot \vec\alpha_j n(r)^j  \; \ln (rT)~,
\end{equation} which, when restricted to the leading vortices (\ref{su35}), reproduces the interactions between the three types of $W$-bosons in (\ref{sunpartition11}). 
After also taking into account the $e$-$m$ interactions, exactly as in the $SU(2)$ discussion, we find that
the affine spin-system  (\ref{z3sqrdmodel1}) is equivalent to the   gauge  theory with partition function (\ref{sunpartition11}).

{\flushleft{\bf  The $\bm{ SU(3) / \Z_3}$ theory:}}
The description (\ref{z3sqrdmodel1}) and its continuum form (\ref{su310}) are appropriate to the 
$SU(3)/\Z_3$ theory. They allow the introduction of dynamical monopoles of smaller charge than the bions, but do not allow the inclusion of particles of charge smaller than that of $W$-bosons. 
This means that the theory does not possess a center symmetry, but a topological symmetry.
The symmetry is  $\Z_3^{\rm d\chi}  \times \Z_3^{\rm t}$  where the first factor  is 
 identified with $\Z_3$ discrete chiral symmetry  and the latter is  the $\Z_3$ topological symmetry associated with $\pi_1(SU(3)/\Z_3)$. 
 As (\ref{z3sqrdmodel1}) shows, both $\Z_3$ symmetries are broken at low $T$, where the spin model is semi-classical. In the high-temperature phase, following direct analogy with the $SU(2)$ discussion, none of the discrete symmetries is spontaneously broken. 

{\flushleft{\bf  The $\bm {SU(3)}$ theory:}}
In analogy with the $SU(2)$ theory, we may construct a sigma model Lagrangian to describe the $SU(3)$ theory as well. In that case, the $SU(3)$ theory allows the introduction of dynamical matter fields whose charge is  smaller  than the $W$-bosons', but  does not allow the inclusion of monopoles whose charge is   smaller than that of 't Hooft-Polyakov monopole. 
Consequently, the theory  does not possess a topological symmetry, but just the 
$\Z_3^{\rm c}$ center symmetry. The discrete symmetry of this construction is 
 $\Z_3^{\rm d\chi}  \times \Z_3^{\rm c}$. 
At low temperature,   $\Z_3^{\rm d\chi}$  chiral   symmetry is broken and  $\Z_3^{\rm c}$ is unbroken. At high temperature, this is reversed:  $\Z_3^{\rm c}$ is broken, and 
 $\Z_3^{\rm d\chi}$  chiral   symmetry is unbroken.

Electric-magnetic duals of  (\ref{su310},\ref{z3sqrdmodel1}) as well as a more detailed study of  the phase diagram of the $SU(3)$ QCD(adj) theory will be considered elsewhere.

\subsubsection{Second  description}

Here, we mention briefly a second lattice action which may be 
easier to simulate and which is equivalent to (\ref{z3sqrdmodel1}).  
Recall that the fields in (\ref{z3sqrdmodel1})  have  a non-cartesian 
periodicity, determined by  $2 \pi \Gamma_r$,  the root lattice, and a canonical diagonal kinetic term for the associated continuum sigma model. It is also useful to provide a description in terms of two compact scalar fields on $\R^2$ that we denote  $\phi^i(x) \equiv \phi^i(x) + 2 \pi$ ($i=1,2$), which are both periodic by $2 \pi$. However, now the metric of the sigma model is non-diagonal.  
The lattice model is:
\begin{eqnarray}
\label{z3sqrdmodel1alt}
- \beta H &&=   \sum_{x; \hat\mu = 1,2} {\kappa  \over 2 \pi} 
\Big[ \cos ( \phi^1_{x  + \hat\mu} -\phi^1_{x}) +  \cos ( \phi^2_{x  + \hat\mu} -\phi^2_{x}) +
\half  \sin ( \phi^1_{x  + \hat\mu} -\phi^1_{x})  \sin ( \phi^2_{x  + \hat\mu} -\phi^2_{x})  \Big] \qquad  \cr
&& +  \; \; 
\sum_{x} \;  \tilde{y} \left( \cos 3 \phi^1_{x}+ \cos 3 \phi^2_{x}+ \cos 3 (\phi^1_{x}-\phi^2_x)\right)~,
\end{eqnarray}
where\footnote{As promised in the Introduction, we note that   the above action, with $\tilde{y}=0$, is the one relevant in the study of melting of the two-dimensional triangular-lattice crystals \cite{Nelson}.}
 $\tilde{y} \sim \xi_{bion} \sim y_m$.
The continuum version of (\ref{z3sqrdmodel1alt}) does not have a diagonal kinetic term. Instead, there is a metric $g_{ij}$ ($i,j=1,2$) on the   target space,   encoding the interactions between vortices (i.e., the $W$-bosons), 
  $g_{ij} \equiv \vec\alpha_i \cdot \vec\alpha_j $. 
The  continuum limit of   (\ref{z3sqrdmodel1alt}) is:
\begin{eqnarray}
\label{su310alt}
-\beta H = - {\kappa \over 4 \pi}  \int_{\R^2}  g_{ij}   \partial_\mu \phi^i \partial^\mu \phi^j +  2 \xi_{bion} \int_{\R^2}\left( \cos 3 \phi^1 + \cos 3 \phi^2 + \cos 3 (\phi^1 - \phi^2) \right)~.
\end{eqnarray} 
 The $\Z_3 \times \Z_3$ symmetry-breaking nature of the perturbation is also more manifest in this formulation.  The charges $(3, 0), (0,3)$, and $(3,-3)$ correspond precisely to magnetic  bion charges in 
 the $\phi$-frame.  

The relation of (\ref{z3sqrdmodel1alt}) to (\ref{z3sqrdmodel1}) (and of (\ref{su310}) to (\ref{su310alt})) can be most easily elucidated by the  change of variables:
\begin{equation}
\label{rectification}
\phi^1 = - \theta^1 +   \theta^2/\sqrt{3}~,~~   
\phi^2 = - \theta^1 -  \theta^2/\sqrt{3}~.
\end{equation}
The transformation (\ref{rectification})  rectifies the fundamental domain (\ref{periodicroot}) of $\vec{\theta}$ 
to that of $\phi^{1,2}$ ($\phi^i \equiv \phi^i + 2 \pi$) and can be easily seen to map (\ref{su310alt}) to (\ref{su310}).
The kinetic term  of the lattice model  (\ref{z3sqrdmodel1}) when expressed in terms of $\phi^{1,2}$, reads:
\begin{equation}
\label{z3sqrdmodel12}
- \beta H^{kin}  =   \sum_{x; \hat\mu = 1,2} {\kappa  \over 4 \pi} 
\Big[ \cos ( \phi^1_{x  + \hat\mu} -\phi^1_{x}) +  \cos ( \phi^2_{x  + \hat\mu} -\phi^2_{x}) +
  \cos( \phi^1_{x  + \hat\mu} -\phi^1_{x} -  \phi^2_{x  + \hat\mu} +\phi^2_{x})  \Big]  ~,
  \end{equation}
and differs from the kinetic term in (\ref{z3sqrdmodel1alt}) by an overall factor of $1/2$ and the third term. However,
we believe this difference to be inessential.  First, it is easy to see that both (\ref{z3sqrdmodel12}) and the kinetic term in (\ref{z3sqrdmodel1alt}) have a unique minimum in the fundamental domain  at $\nabla_\mu \phi^1 = \nabla_\mu\phi^2 = 0$ and, hence, there are no ``doublers" in either description. Second, the expansions of both (\ref{z3sqrdmodel12}) and the kinetic term in (\ref{z3sqrdmodel1alt}) around the minimum coincide to quadratic order in $\nabla_\mu \phi^{1,2}$, so that the models are identical in the Villain approximation; we expect that the difference (a four-derivative term) will not be essential in the continuum limit.

\section{Possible  future studies}

\label{future}

Some possible directions related to the  affine XY-spin model/gauge theory equivalence that are   worthy of pursuit and would directly extend the results of this paper are listed below:
\begin{enumerate}
\item
One can study the renormalization group equations for higher-rank QCD(adj) to next order in fugacities and search for fixed points, similar to  studies of  extended dual sine-Gordon theories \cite{Boyanovsky:1988ge,Boyanovsky:1989mc,Boyanovsky:1990iw}. 
\item
In the $SU(3)$ case,  using 2d CFT methods, as in \cite{Lecheminant:2002va}, one can focus on the self-dual point and attempt to infer the existence and nature of the fixed-point (critical) theory. 
\item
A mean-field analysis could be useful to gain some idea of the behavior of the system in the region of large fugacities (to which the leading order RGEs appear to drive the theory). Recall that in the BKT transition, there is a tricritical point separating the small-fugacity continuous transition   from a line of first-order transitions at large fugacity, revealed by a mean-field analysis, and observed in experiments and simulations,  see, e.g.,\cite{Herbutbook, Minnhagen1,Minnhagen2,Diehl}.
\item Numerical studies of the phase transition in the   2d lattice models  with short-range interactions,  dual to the long-range interaction Coulomb gas, should be possible. These would be useful to determine  the order of the deconfinement transition.
\item The study of non-equilibrium properties of thermal gauge theories is  a subject of current interest, e.g., for RHIC physics. It would be of interest to extend the   $\R^{ 2} \times \S^1_\beta \times \S^1_L$ framework of this paper to allow the study of the near-critical behavior of quantities such as the conductivity and viscosity.
\end{enumerate}

{\flushleft T}here are also other interesting directions, broadly related to the technology proposed here, which we intend to study more systematically:

\begin{enumerate}
\item {\it  Orbifold/orientifold  equivalences and spin systems for general  QCD-like theories:}
In the large-$N$ limit of gauge theories, there are exact  orbifold  and orientifold equivalences relating 
 gauge theories with different microscopic matter content
\cite{Armoni:2003gp, Kovtun:2003hr}. These equivalences 
have remnants in the small-$N$  semi-classical regimes, as emphasized in  \cite{Shifman:2008ja, Unsal:2008ch}.  Therefore, by using double-trace deformations, our study should generalize in interesting ways to QCD with fundamental, anti-symmetric, and symmetric representation fermions. In particular, the semi-classical domains are also  suited to check the universality of the phase diagrams of   QCD-like theories related through valid  equivalences
\cite{Hanada:2011ju} satisfying the necessary and sufficient conditions  \cite{Kovtun:2003hr}.
\item{\it  Chiral gauge theory/spin system mapping:}
We certainly hope that the numerical cross-checks of the spin system/gauge theory equivalence will be successful and useful  in vector-like gauge theories. If so, then 
  it is worth   generalizing  the equivalence to chiral gauge theories, by 
  using the techniques of \cite{Shifman:2008cx,Poppitz:2009uq}.
\item  {\it Zero-mode  Hamiltonian with fermions:}
In order to understand the meaning of the topological symmetry better, we have constructed 
the zero-mode Hamiltonian for the theory on  $\T^2 \times \S^1_L$, where 
$\S^1_L$ is parametrically smaller than  $\T^2$.  We aim to study the bosonic and fermionic zero-mode quantum mechanics for the quantum field theories  systematically. 
 It seems to us that this framework may provide opportunities to relate  the instantons with  fractional topological charge proposed by 't Hooft \cite{thooft1} on toroidal compactification 
 and the monopole-instantons or bions  which appear semi-classically  on $\R^3 \times \S^1_L$. 
The magnetic bions and monopoles, depending on the theory, provide reliable (i.e., semi-classically calculable) confinement mechanisms on    $\R^3 \times \S^1_L$. On the 
other hand, ref.~\cite{GonzalezArroyo:1995zy} provided numerical evidence for the relevance 
of objects with fractional topological charge on the torus, for which analytic solutions are not known. We aim to gain a better understanding of their interconnection. This project is ongoing. 
\item  {\it Zero-mode  Hamiltonian and $\theta$-angle dependence:}
We expect that our framework  on  $\R \times \T^2 \times \S^1_L$  will provide new insights 
into the  topological $\theta$-angle dependence of observables in  gauge theories. This project is ongoing. 
\end{enumerate}

   \acknowledgments 
  We thank  Philip Argyres,  Ben Burrington, Rajamani Narayanan,   and Arun Paramekanti 
  for useful discussions on various topics relevant to this paper. 
 The work of M.A. and E.P. was supported in part by the  National Science and Engineering Council of Canada (NSERC).
     
 \appendix 

\section{RGEs for the $\mathbf{SU(2)}$(adj) magnetic-bion/W-boson Coulomb plasma}
\label{rgeappendix}

Consider the partition function of a form more general than (\ref{SU2gas}), namely that of a neutral gas of unit ``electric" and charge-$p$ ``magnetic" charges, where the interaction strengths are given by $\kappa$ for the electric charges Coulomb interaction and $1 \over \kappa$ for the magnetic charges; in addition, there is the appropriate Aharonov-Bohm phase (\ref{ABinteraction2}). The dimensionless fugacities are $y_e$ and $y_m$, respectively, and $a$ is the ``lattice spacing". The partition function reads:
\begin{eqnarray}
\label{SU2gas11}
&Z&_{EM, p} \nonumber \\
&=& \sum_{N_m \ge 0, q_a=\pm 1 } \sum_{{N_e \ge 0}, q_A = \pm 1}  { \left( {y_m \over  a^{2}} \right)^{2N_m} \over \left( N_m! \right)^2  }
{ \left({{y_e} \over a^{2}} \right)^{ 2 {N_e} }  \over \left({N_e}!\right)^2 } \prod_j^{2 N_m} {d^2 R_j}   \prod_A^{2 N_e} {d^2 R_A}  \\
&\times& \exp \left[
 {p^2 \over \kappa} \sum_{a >  b} {q_a q_b} \ln {|\vec R_a-\vec R_b|\over a} + \kappa \sum_{A >  B} {q_A q_B} \ln {|\vec R_A-\vec R_B|  \over a}
+ i p \sum\limits_{a,B} q_B q_a \Theta(\vec{R}_B - \vec{R_a})
 \right] ~.  \nonumber
\end{eqnarray}

The map of (\ref{SU2gas11}) to the bion/W-gas partition function (\ref{SU2gas}) is straightforward, using the change of variables (\ref{variables}). Our goal is to derive the RGEs (\ref{rges}) to leading order in fugacities $y_e$, $y_m$, describing the behavior of the three parameters $\kappa, y_e, y_m$ under course graining of the partition function. These equations are well-known and we only present our derivation to make our exposition self-contained (we believe ours is the simplest derivation to leading order in fugacities; for higher-order calculations, see  \cite{Boyanovsky:1988ge}).

Consider first the renormalization of the fugacities. They can be most easily obtained by considering the lowest-order contributions to $Z_{EM, p}$. For example, the  $(N_m, N_e) =(1, 0)$ term reads:
\begin{equation}
\label{fugacity}
\left(y_m(a) \over a^2\right)^2 \int d^2 r_1 d^2 r_2 \left|r_1 - r_2 \over a\right|^{- {p^2 \over \kappa}} \sim y_m^2(a) \left( L \over a \right)^{4 - {p^2 \over \kappa}}~,
\end{equation}
where we have indicated that the fugacity on the l.h.s. is the one appropriate for lattice spacing $a$, the $r_{1,2}$ integrals were performed from  $a$ to the size of the system $L$.  Demanding RG invariance of the contribution (\ref{fugacity}) to the partition function:
\begin{equation}
\label{fugacityrge}
 y_m^2(a) \left( L \over a \right)^{4 - {p^2 \over \kappa}} = y_m^2(e^b a)  \left( L \over e^{b} a \right)^{4 - {p^2 \over \kappa}}\,, 
\end{equation}
and differentiating (\ref{fugacityrge}) w.r.t. $b$ at $b=0$, we obtain:
\begin{equation}
\label{ymagneticrge}
{d y_m \over d b} = (2 - {p^2 \over 2 \kappa}) y_m~,
\end{equation}
the third of eqns.~(\ref{rges}) given in the text (with $y_m = \tilde{y}$; note also that to leading order in the fugacity we can neglect the change of $\kappa$ in (\ref{fugacityrge})). The RGE for $y_e$, the second of eqns.~(\ref{rges}), follows by duality or from similarly considering the $(N_m,N_e) = (0,1)$ contribution to the partition function:
\begin{equation}
\label{yelectricrge}
{d y_e \over d b} = (2 - {\kappa \over 2}) y_m\,.
\end{equation}
It is easy to check that the scaling of the fugacities (\ref{ymagneticrge}) and (\ref{yelectricrge}) also ensures that the entire partition function is invariant under course-graining. 

To find the effect of screening and anti-screening on the interaction between magnetic or electric charges, consider for definiteness the interaction between two external magnetic charges with $q=+1$ at $\vec{x}$ and $q = -1$ at $\vec{y}$. To leading order, ignoring any fluctuation contribution, the interaction between  external magnetic charges is given by:
\begin{equation}
\label{magneticinteraction}
e^{- {p^2 \over \kappa(a)} \ln |x-y|}~,
\end{equation}
where  we have indicated that the strength $\kappa$ is the tree-level value appropriate for a cutoff $a$. Our goal now is to compute the effective strength of the interaction appropriate to a course grained system (i.e., find  $\kappa(e^b a)$). In other words, we want to find the change of the interaction (\ref{magneticinteraction}) due to fluctuating electric and magnetic dipoles with sizes between $a$ and $e^b a$. 
The expression for their effect takes the following form (where we use the shorthand notation $r_{1y} \equiv |\vec{r_1} - \vec{y}|$ and $\Theta_{x2} = \Theta(\vec{x} - \vec{r_2})$, etc.):
\begin{eqnarray}
\label{kappaeff1}
e^{- {p^2 \over \kappa(e^b a)} \ln r_{xy}} &=& \left[ e^{- {p^2 \over \kappa(a)} \ln r_{xy}} + \left(y_m \over a^2\right)^2  \int d^2 r_1 d^2 r_2 e^{- {p^2 \over \kappa(a)}(\ln r_{xy} + \ln r_{12} + \ln r_{x1} - \ln r_{x2} - \ln r_{1y} + \ln r_{2y} ) } \right. \nonumber \\
&&~~~~~ \left. + \left({y_e \over a^2}\right)^2  \int d^2 r_1 d^2 r_2 e^{- \kappa(a) \ln r_{12} - {p^2 \over \kappa(a)} \ln r_{xy} + i p (\Theta_{x2} - \Theta_{x1} + \Theta_{y1}-\Theta_{y2})} + \ldots 
\right] \\
&\times & \left[1 + \left(y_m \over a^2\right)^2  \int d^2 r_1 d^2 r_2 e^{- {p^2 \over \kappa(a)} \ln r_{12} } 
+ \left({y_e \over a^2}\right)^2  \int d^2 r_1 d^2 r_2 e^{- \kappa(a) \ln r_{12} }+ \ldots  \right]^{-1}~.\nonumber
\end{eqnarray}
The meaning of the various terms above is as follows: the term $\sim y_m^2$ represents the contributions of fluctuating magnetic dipoles (a positive charge at $r_2$ and a negative charge at $r_1$) to the interaction between the external magnetic  charges and the  term  $\sim y_e^2$ similarly represents the contribution of an electric dipole (a positive charge at $r_2$ and a negative charge at $r_1$). In each case the energy is equal to the sum of the energies of interaction of all charges in the configuration, and to the order of the calculation involves six pairwise interactions in each case, as shown explicitly in (\ref{kappaeff1}). The last term  in (\ref{kappaeff1}) is the $1/Z$ normalization factor of the expectation value of the potential between the external charges, also written to leading order in fugacities. 
When this term is expanded to order $y_{e,m}^2$, the expression for the effective interaction takes the simpler form:
\begin{eqnarray}
\label{kappaeff2}
e^{- {p^2 \over \kappa(e^b a)} \ln r_{xy}} &=& e^{- {p^2 \over \kappa} \ln r_{xy}}  \left[ 1+ \left(y_m \over a^2\right)^2  \int d^2 r_1 d^2 r_2 e^{- {p^2 \over \kappa}\ln r_{12}} \left(e^{- {p^2 \over \kappa}( \ln r_{x1} - \ln r_{x2} - \ln r_{1y} + \ln r_{2y} ) } - 1\right) \right. \nonumber \\
&&~~~~~~~~~~~~~ \left. + \left({y_e \over a^2}\right)^2  \int d^2 r_1 d^2 r_2 e^{- \kappa \ln r_{12} } \left( e^{- i p (\Theta_{x1} - \Theta_{x2} - \Theta_{1y} + \Theta_{2y})} -1\right)  
\right] ~,
\end{eqnarray}
which, as appropriate, only takes into account the interactions between the dipoles and the external charges (the coupling $\kappa$ on the r.h.s. is the one appropriate at the scale $a$, but beginning with (\ref{kappaeff2}), we do not explicitly indicate this for brevity; also for the time being we do not explicitly show the limits on the integration  over the positions of the fluctuating dipoles $r_{1,2}$). 

Recall now that we are interested in computing the effect of magnetic and electric fluctuating dipoles of small sizes on the effective interaction between the external charges. To this end, introduce centre of mass ($\vec{R}$) and relative ($\vec{r}$)  coordinates: 
\begin{eqnarray}
\label{kappaeff3}
e^{- {p^2 \over \kappa(e^b a)} \ln r_{xy}} &=& e^{- {p^2 \over \kappa} \ln r_{xy}}  \left[ 1+ \right.\\
&+&\left. \left(y_m \over a^2\right)^2  \int d^2 r e^{- {p^2 \over \kappa} \ln r} \int d^2 R  \left(e^{- {p^2 \over \kappa}( \ln r_{R -x, {r \over 2}} - \ln r_{R -x, -{r \over 2}}  - \ln r_{R -y, {r \over 2}} + \ln r_{R -y, -{r \over 2}} ) } - 1\right) \right. \nonumber \\
&+&\left. \left({y_e \over a^2}\right)^2  \int d^2 r e^{- \kappa \ln r} \int d^2 R \left( e^{- i p (\Theta_{R -x, {r \over 2}} - \Theta_{R -x, -{r \over 2}}  - \Theta_{R -y, {r \over 2}} + \Theta_{R -y, -{r \over 2}})} -1\right)  \nonumber
\right] ~, 
\end{eqnarray}
and note that for small $r$, $f(r_{R -x, {r \over 2}}) - f(r_{R -x, -{r \over 2}})  - f(r_{R -y, {r \over 2}}) + f(r_{R -y, -{r \over 2}}) \simeq \vec{r} \cdot \vec{\nabla}_R(f( r_{R, x}) - f (r_{R, y}))$ (we have in mind that the function $f(r_{R,x})$ stands for either $\ln(r_{R,x})$ or $\Theta_{R,x}$). Next, we note that expanding the exponents and averaging over the direction of $\vec{r}$, the contribution linear in $\vec{r}$ vanishes, while the second order contribution becomes:
\begin{eqnarray}
\label{kappaeff4}
e^{- {p^2 \over \kappa(e^b a)} \ln r_{xy}} &=& e^{- {p^2 \over \kappa} \ln r_{xy}}  \left[ 1+ \right.\\
&+& \left. \left(y_m \over a^2\right)^2 {\pi p^4 \over 2 \kappa^2} \int\limits_{a}^{e^b a} d r r^3 e^{- {p^2 \over \kappa} \ln r} \int d^2 R  \left( \vec{\nabla}_R (\ln r_{R, x} - \ln r_{R, y} )\right)^2 \right. \nonumber \\
&-&\left. \left({y_e \over a^2}\right)^2   {\pi p^2 \over 2 }\int\limits_{a}^{e^b a} d r r^3 e^{- \kappa \ln r} \int d^2 R \left(    \vec{\nabla}_R(\Theta_{R, x} - \Theta_{R,y}\right)^2 \nonumber
\right] ~.
\end{eqnarray}
Above, we explicitly indicated that we are only taking into account  the (anti-)screening contribution of electric and magnetic dipoles with sizes between the old and new cutoffs $a$ and $e^b a$. Now we use the identity $(\vec{\nabla}_R \ln r_{R,x})^2 = (\vec{\nabla}_R \Theta_{R,x})^2$---an expression of the ``electric-magnetic" vortex-charge duality in 2d---to recast (\ref{kappaeff4}) as:
\begin{eqnarray}
\label{kappaeff14}
e^{- {p^2 \over \kappa(e^b a)} \ln r_{xy}} &=& e^{- {p^2 \over \kappa} \ln r_{xy}}  \left[ 1+ \right.\\
&+& \left. \left[ \left(y_m \over a^2\right)^2 {\pi p^4 \over 2 \kappa^2} \int\limits_{a}^{e^b a} d r r^3 e^{- {p^2 \over \kappa} \ln r} -  \left({y_e \over a^2}\right)^2   {\pi p^2 \over 2 }\int\limits_{a}^{e^b a} d r r^3 e^{- \kappa \ln r} \right] \times \right. \nonumber \\
& & \left. \times \int d^2 R  \left( \vec{\nabla}_R (\ln r_{R, x} - \ln r_{R, y} )\right)^2 \right]. \nonumber
\end{eqnarray}
Finally, we  integrate $I(x,y)=  \int d^2 R  \left( \vec{\nabla}_R (\ln r_{R, x} - \ln r_{R, y} )\right)^2$ by parts, using the neutrality of the system to discard the boundary term, to obtain $I(x,y) = 4 \pi \ln r_{x,y} - 4 \pi \ln r_{0,0}$, and ignoring the constant term, which does not contribute to the potential between the external charges,
 substitute back into (\ref{kappaeff14}) to obtain:
\begin{eqnarray}
\label{kappaeff5}
e^{- {p^2 \over \kappa(e^b a)} \ln r_{xy}} &=& e^{- {p^2 \over \kappa} \ln r_{xy}}  \left[ 1- {p^2} \ln r_{x,y} \times \right.\\
&\times& \left. \left(  \left({y_e \over a^2}\right)^2   {2 \pi^2  }\int\limits_{a}^{e^b a} d r r^3 e^{- \kappa \ln r} - \left(y_m \over a^2\right)^2 {2 \pi^2 p^2 \over  \kappa^2} \int\limits_{a}^{e^b a} d r r^3 e^{- {p^2 \over \kappa} \ln r}  \right)  \right]~\nonumber . 
\end{eqnarray}
Re-exponentiating, we find that the effect of electric and magnetic fluctuating dipoles of sizes between $a$ and $e^b a$ renormalize the strength of the interaction between external magnetic charges as follows:
\begin{eqnarray}
\label{kappaeff6}
{1 \over \kappa(e^b a) } &=& {1 \over \kappa(a)} +   \left({y_e \over a^2}\right)^2   {2 \pi^2  }\int\limits_{a}^{e^b a} d r r^3 e^{- \kappa \ln r} - \left(y_m \over a^2\right)^2 {2 \pi^2 p^2 \over  \kappa^2(a)} \int\limits_{a}^{e^b a} d r r^3 e^{- {p^2\over \kappa} \ln r} ~.
\end{eqnarray}
Finally, we differentiate w.r.t. $b$ and take $b=1$ and, remembering that the interaction is really $\ln{ r\over a}$, obtain: 
\begin{equation}
\label{kappaeff7}
{d \over d b}\left( {1 \over \kappa } \right)= 2 \pi^2 \left( y_e^2 - y_m^2 {p^2 \over \kappa^2}\right)~,
\end{equation}
or equivalently, 
\begin{equation}
\label{kappaeff8}
\dot  \kappa = 2 \pi^2 \left( p^2 y_m^2 -   \kappa^2 y_e^2 \right)\,.
\end{equation}
For $p=4$,  this is exactly the first of eqns.~(\ref{rges}).

\section{The critical point of $\mathbf{SU(2)}$(adj) and the approach to ${\mathbf T_c}$}
\label{z4rgflow}

We begin by rewriting the RGEs (\ref{rges}) of the $\Z_4$ model in terms of the variables:
\begin{equation}
\label{variables2}
X \equiv {\kappa - 4 \over 2}, ~~ Y\equiv 4 \pi y, ~~ Z \equiv  4\pi \tilde{y}~,
\end{equation}
 keeping up to quadratic terms in $X,Y,Z$, which is sufficient for studying the approach to the fixed point:
 \begin{eqnarray}
 \label{rges2}
 \dot{X} &=& Z^2 - Y^2\,, \nonumber \\
 \dot{Y} &=& - X Y\,, \\
 \dot{Z} &=& X Z \nonumber~. 
 \end{eqnarray}
These equations have two integrals of motion: 
\begin{eqnarray}
\label{rges3}
Y Z &=& c^\prime ~,\\
Y^2 + Z^2 - X^2 &=& c \nonumber~,
\end{eqnarray}
hence the RGEs can be solved in quadratures.  

See Figure \ref{fig:RGFLOW} (Section \ref{rgessection}) for an illustration of the RG flow (\ref{rges2}),
 which we will now analyze in some detail.

\subsection{The critical RG trajectory and determination of $\mathbf{T_c}$}
We begin by considering trajectories that end on the critical line. To this end note that $c^\prime$ is determined by the initial (UV cutoff scale) values of the fugacities, $c^\prime = Y(0) Z(0) = 16 \pi^2 y_0 \tilde{y_0}$, with $y_0, \tilde{y}_0$ given in (\ref{variables}). The fixed point of (\ref{rges2}), whose location on the fixed line $\kappa =4, y = \tilde{y}$ is determined by the initial conditions,   is at $\bar{X} = 0$, $\bar{Y} = \bar{Z} = \sqrt{c^\prime}$. The last equation in (\ref{rges3}) then implies that the fixed point trajectory is the one for which $c = 2 c^\prime$. 
We can now  solve (\ref{rges3}) for $X$, using this value  for $c$:
\begin{equation}
\label{xsoltn}
X = \pm {\sqrt{ (Y^2- c^\prime)^2} \over Y}~.
\end{equation} We can now use this to integrate the equation for $Y(t)$ from (\ref{rges2}), taking (without loss of generality) $Y(0) > \sqrt{c^\prime}$ and a positive sign for $X$:
\begin{equation}
\label{ysoltn}
Y(t) = \sqrt{c^\prime} \tanh\left( \sqrt{c^\prime} t + \tanh^{-1} {Y(0) \over \sqrt{c^\prime}} \right)~,
\end{equation}
showing clearly that this trajectory approaches $\bar{Y} = \sqrt{c^\prime}$ as $t \rightarrow \infty$. 
The corresponding solution for $Z$ is:
\begin{equation}
\label{zsoltn}
Z(t) = {c^\prime \over Y(t)} = { \sqrt{c^\prime} \over \tanh\left( \sqrt{c^\prime} t + \tanh^{-1} {Y(0) \over \sqrt{c^\prime}} \right)}~.
\end{equation}
The solution for $X(t)$ is given by (\ref{ysoltn}) substituted in (\ref{xsoltn}). As $t \rightarrow \infty$ this clearly approaches the $\bar{X}=0$ fixed point, while as $t \rightarrow 0$, we have:
\begin{equation}
\label{xsolution1}
X(0) = {Y(0)^2 - c^\prime \over Y(0)} = {Y(0)^2 - Z(0) Y(0) \over Y(0)} = Y(0) - Z(0)~.
\end{equation}
Note that $Y(t)$ and $Z(t)$ monotonically change from $Y(0)$, $Z(0)$, respectively, to $\sqrt{c^\prime} = \bar{Y}=\bar{Z} $ as $t$ changes from $0$ to $\infty$. If the initial values $Z(0)$ and $Y(0)$ are small enough, and close to the fixed-point value $\sqrt{c^\prime}$, $X(t)$ is also small for all $t \in (0, \infty)$, so that the approximations made in arriving at (\ref{rges2}) hold. Thus,  the entire RG trajectory to the critical point lies within the validity of the perturbative RGEs (\ref{rges2}).  

We conclude that the above analysis of the approach to the critical point is reliable, provided the initial values of the fugacities and $\kappa -4$  are small enough---as they are for our small-$L$ $SU(2)$(adj) theory. The solutions to   the RGEs (\ref{ysoltn}, \ref{xsolution1}) ending at the critical point are then reliable all along the flow---in contrast with the flow in the $\Z_2$ model (and more  generally, in the $\Z_{p < 4}$ models) relevant, e.g., in the deformed-Yang-Mills study of \cite{Simic:2010sv}. 

Equation (\ref{xsolution1}) is a constraint on the initial conditions for a trajectory to end at the critical point. Thus, it can be used to determine $T_c$. Recalling the definitions (\ref{variables2}), (\ref{variables}), as well as (\ref{wfugacity}) and (\ref{bionfugacity}) for the $W$-boson and bion fugacities, using $m_W = {\pi\over L}$, and taking the lattice spacing to be the bion size $r_*$ (\ref{bionpersistencecondition}), we have:
\begin{equation}
\label{tcequation}
 {g^2 \over 2 \pi L T_c} - 4   =  4 \pi r_*^2 \left( (2 n_f + 1) {T_c \over L} \; e^{- {\pi \over L T_c}} - {A  \over L^3 \; T_c \; g^{14 - 8 n_f}}   \; e^{- {8 \pi^2 \over g^2} (1 + c g)} \right), ~
\end{equation}
where $g \equiv g(L)$ is the four-dimensional running coupling at the scale $L$, $c= \sqrt{n_f - 1 \over 3}$, and $A$ is the unknown numerical coefficient of the bion fugacity (only the dependence on the coupling is known, see  \cite{Anber:2011de} for details). 

The point of equation (\ref{tcequation}) is that the critical temperature can  in principle be precisely calculated by careful matching of the effective 2d theory to the 4d UV theory in the  small-$L$ regime. Clearly, one can  iteratively solve (\ref{tcequation}) to find the deviation of $T_c$ from $g^2 \over 8 \pi L$ (equivalently, the deviation of $\kappa_0$ from $4$). However,  since the fugacities are exponentially small, it is clear that $T_c = {g^2 \over 8 \pi L}$ is an exponentially accurate estimate---the shift of $T_c$ is easily seen to be  of order ${f(g) \over L} e^{- {8 \pi^2 \over g^2}}$, where $f(g)$ has a known leading-order dependence on $g$ but is itself only known up to a numerical coefficient.

\subsection{The correlation length as $\mathbf{ T \rightarrow T_c}$}
\label{corrappx}

Let us now analyze the behavior of almost critical trajectories.   
For the critical trajectory, we have from (\ref{xsolution1}) that $X(0)_c = Y(0)_c - Z(0)_c$. Now, consider a trajectory with $X(0) = X(0)_c - \Delta$ and use the definitions  $X = {\kappa - 4 \over 2}$, $\kappa = {g^2 \over 2 \pi L T}$, and $T_c \simeq {g^2 \over 8 \pi L}$ to find:
\begin{equation}
\label{delta}
\Delta \equiv - 4 \; { T - T_c\over T_c}.  
\end{equation}
A trajectory with the chosen value of $T$ has to also obey (\ref{rges3}), in particular, at $t=0$:
\begin{equation}
c = Y(0)^2 + Z(0)^2 - X(0)^2 \simeq Y(0)^2_c + Z(0)^2_c - X(0)_c^2 + G \Delta = 2 c^\prime  + G \Delta,
\label{noncriticalc}
\end{equation} 
where we used $c = 2 c^\prime$ for the critical trajectory, expanded for $T$ near $T_c$ (small $\Delta$), and denoted by $G$ the coefficient of the first term in the expansion of $Y^2 + Z^2 - X^2$ near $T_c$. The value of $G$ will not play a role in determining the critical properties (it can be calculated using (\ref{tcequation})).
Let us now study the non-critical RG trajectory of (\ref{rges2}), for which  $Y(t)$ is given by:
\begin{equation}
\label{corr1}
\int\limits_{Y(0)}^{Y(t)} { d Y \over \sqrt{ Y^4 - c\; Y^2 +  (c^\prime)^2}} = - t. 
\end{equation}
 For $c$ given by (\ref{noncriticalc})
we rewrite (\ref{corr1}) as:
\begin{equation}
\label{corr2}
\int\limits_{Y(0)}^{Y(t)} { d Y \over \sqrt{  (Y^2 - \ c^\prime)^2 -     G \Delta\; Y^2 }} = - t~.
\end{equation} 
It is possible to integrate (\ref{corr2}) explicitly, but for the purposes of studying the divergence of the correlation length as $\Delta \rightarrow 0$, the argument below is sufficient. Clearly, for $\Delta = 0$, the trajectory is given by the critical solution (\ref{ysoltn}). It is clear that for small $\Delta$, the trajectory will be essentially the same as the critical one up to values of RG ``time" $t_*$ when $|Y(t_*)^2 - c^\prime|^2 \sim |G \Delta|
Y(t_*)^2$. Thus, we can use the critical solution up to times of order $t_*$, which we can estimate by requiring that the two quantities below are of the same order of magnitude:
\begin{eqnarray}
\label{corr3}
|G \Delta| Y(t_*)^2 &\sim& |\Delta|  \tanh^2 (\sqrt{c^\prime} t_*) \sim |\Delta| (1 - 2 e^{- 2 \sqrt{c^\prime} t_*})~,\nonumber \\
 |Y(t_*)^2 - c^\prime|^2 &\sim& (1 - \tanh^2 (\sqrt{c^\prime} t_*))^2   \sim  e^{- 4 \sqrt{c^\prime} t_*}~, 
\end{eqnarray}
where we assumed large $t_*$ and neglected numerical coefficients. 
Equating the two lines in (\ref{corr3}), we have  $t_* \sim - \ln (|\Delta|^{1\over 4 \sqrt{c^\prime}})$, hence the correlation length $\zeta \sim e^{t_*}$ scales  as $\Delta \rightarrow 0$ as:
 \begin{equation}
 \label{corr4}
  \zeta \sim   |\Delta|^{-{1 \over 4 \sqrt{c^\prime}}}~,
 \end{equation}
i.e., with a non-universal exponent, whose value  depends on the fugacities of the bions and $W$-bosons, $y$ and $\tilde{y}$. The approach to $T_c$ is governed by their  values  at the fixed line (equal to $\sqrt{c^\prime}$), for which, according to (\ref{rges3}) we have $c^\prime = 16 \pi^2 y_0 \tilde{y}_0$ (where the subscript denotes the UV values of the fugacities). Thus, we can  write, using the map (\ref{variables2}):
\begin{equation}
\label{corr5}
\zeta \sim |T - T_c|^{ -  {1\over 16 \pi \sqrt{y_0 \tilde{y}_0}} }~, 
\end{equation} 
thus we have $\nu = - {1 \over 16 \pi \sqrt{y_0 \tilde{y}_0}}$, in the usual convention $\zeta \sim |T - T_c|^{ - \nu}$. 
Recall  that the theory is in a gapped phase  on both sides of $T_c$ (related by the electric-magnetic duality) and hence the scaling holds upon approaching $T_c$ from both sides.
The singular part of the free energy is:
\begin{equation}
\label{freeenergy}
 F \sim \zeta^{-2} \sim |T - T_c|^{1 \over 8 \pi \sqrt{y_0 \tilde{y}_0}}~.
 \end{equation}
  Thus, for small fugacities ($y_0 \sim e^{ - {8 \pi^2 \over g^2}}$  in the small-$L$ $SU(2)$(adj) case), the transition in the $\Z_4$ gas is of quite high (but finite, unlike the Kosterlitz-Thouless transition) order determined by the integer part of $2 \nu$.

\section{RGEs for the $\mathbf{SU(N_c)}$ case}
\label{sunrges1}

In this section we consider the RGEs in the $SU(N_c)$ case. At temperatures much smaller than $M_W$ and larger than the photon mass, the system can be describes as a $2D$ neutral gas of $N_c$ distinguishable copies of "electric" and "magnetic" charges. 
To derive the RGEs to leading order in fugacities $y_e$ and $y_m$, we now closely follow Appendix \ref{rgeappendix}. The renormalization of the fugacities can be obtained from the lowest-order contributions of the partition function. 
Taking the contribution  to (\ref{sunpartition11})  with   $N^p_{e+}=  N^p_{e-}=1$ (for some fixed $p$), we obtain: 
\begin{eqnarray}
\label{rgefory}
\nonumber
&&\frac{dy_e}{db}=\left(2-\frac{\vec\alpha_p\cdot\vec\alpha_p}{2}\kappa_e \right)y_e = \left(2 - \frac{\kappa_e}{2} \right) y_e\,,
\end{eqnarray}
while taking  $N^p_{m+}=  N^p_{m-} =1$, we obtain:
\begin{eqnarray}
\label{fugacities in SUn}
\qquad \qquad \qquad  \; &&\frac{dy_m}{db}=\left(2-\frac{2 \vec Q_p\cdot\vec Q_p}{\kappa_m} \right)y_m = \left(2 - {6 \over \kappa_m}\right) y_m, \; \; {\rm for} \; N_c \ge 3,
\end{eqnarray}
where we indicated that the last equality in the second equation holds for $N_c \ge 3$ only.

To derive the RGE for $\kappa_e$, we consider  two external electric charges of the type $p$, where $1\leq p \leq N$ with charges $q=+1$ at $\vec x$ and $q=-1$ at $\vec y$. Ignoring the fluctuating contribution, we find that the interaction between these two charges is given by:
\begin{eqnarray}
e^{-\vec \alpha_p\cdot \vec \alpha_p \kappa_e(a)\ln|x-y|}\,.
\end{eqnarray}
Now, we take into account the contribution from the fluctuations of the electric and magnetic dipoles of the various types. The expression for the net effect takes the form:
\begin{eqnarray}
\nonumber
&&e^{-\vec\alpha_p\cdot\vec\alpha_p\kappa_{e}(e^{b}a)\ln r_{xy}}=e^{-\vec\alpha_p\cdot\vec\alpha_p\kappa_e\ln r_{xy}}\\
\nonumber
&&\times\left[1+\left(\frac{y_e}{a^2}\right)^2\int d^2 r_1 d^2 r_2\sum_{i=\{0,\pm1\}} e^{\kappa_e(a)\vec\alpha_p\cdot\vec\alpha_{p+i}\left(\ln r_{x1}-\ln r_{x2}-\ln r_{y1}+\ln r_{y_2}\right)-\kappa_e(a)\vec\alpha_{p+i}\cdot\vec\alpha_{p+i}\ln r_{12}}   \right.\\
\nonumber
&&\left.\quad +\left(\frac{y_m}{a^2}\right)^2\int d^2 r_1 d^2 r_2\sum_{i=\{-1,0,1,2\}} e^{2 i\vec\alpha_p\cdot\vec Q_{p+i}\left(\Theta_{x1}-\Theta_{x2}-\Theta_{y1}+\Theta_{y2}\right)- 
\frac{4}{\kappa_m(a)} \vec Q_{p+i}\cdot\vec Q_{p+i}\ln r_{12}}\right]/Z\, ,
\end{eqnarray}
where we used the fact that while every $W$ boson interacts with itself and its two nearest-neighbors in the Dynkin diagram, a $W$-boson interacts with four kinds of bions, as per (\ref{qproducts}) and (\ref{alphaproducts}).
The partition function $Z$ to leading order in fugacities reads (omitting the contributions of particles that do not interact with the external ones as they do not contribute to the (anti-)screening):
\begin{eqnarray}
\nonumber
Z&=&1+\left(\frac{y_e}{a^2}\right)^2 \int d^2 r_1 d^2 r_2\sum_{i=\{0,\pm1\}} e^{-\kappa_e(a)\vec\alpha_{p+i}\cdot\vec\alpha_{p+i}\ln r_{12}}\\
&&+\left(\frac{y_m}{a^2}\right)^2 \int d^2 r_1 d^2 r_2\sum_{i=\{-1,0,1,2\}} e^{-\frac{4}{\kappa_m(a)}\vec Q_{p+i}\cdot\vec Q_{p+i}\ln r_{12}}\,.
\end{eqnarray}
Expanding the $Z^{-1}$ terms to order $y_{e,m}^2$ we find:
\begin{eqnarray}
\nonumber
&&e^{-\vec\alpha_p\cdot\vec\alpha_p\kappa_{e}(e^{b}a)\ln r_{xy}}=e^{-\vec\alpha_p\cdot\vec\alpha_p\kappa_e\ln r_{xy}}\\
\nonumber
&&\left[1+\left(\frac{y_e}{a^2}\right)^2\int d^2 r_1 d^2 r_2\sum_{i=\{0,\pm 1\}} e^{-\kappa_e(a)\vec\alpha_{p+i}\cdot\vec\alpha_{p+i}\ln r_{12}}\left(e^{\kappa_e(a)\vec\alpha_p\cdot\vec\alpha_{p+i}\left(\ln r_{x1}-\ln r_{x2}-\ln r_{y1}+\ln r_{y2}\right)}-1\right)   \right.\\
\nonumber
&&\left.\quad +\left(\frac{y_m}{a^2}\right)^2\int d^2 r_1 d^2 r_2\sum_{i=\{-1,0,1,2\}}e^{- \frac{4}{\kappa_m(a)}\vec Q_{p+i}\cdot\vec Q_{p+i}\ln r_{12}}\left( e^{2 i\vec\alpha_p\cdot\vec Q_{p+i}\left(\Theta_{x1}-\Theta_{x2}-\Theta_{y1}+\Theta_{y2}\right)}-1\right)\right]~.
\end{eqnarray}
Using the center of mass $\vec R$ and relative  $\vec r$ coordinates and proceeding as we did in the RGEs of SU(2) we obtain:
\begin{eqnarray}
\nonumber
&&e^{-\vec\alpha_p\cdot\vec\alpha_p\kappa_{e}(e^{b}a)\ln r_{xy}}=e^{-\vec\alpha_p\cdot\vec\alpha_p\kappa_e\ln r_{xy}}\left[1+ 4\pi\ln r_{xy}\right.\\
\nonumber
&&\left.\times\left(\frac{\pi \kappa_e^2(a)}{2}\left(\frac{y_e}{a^2}\right)^2\sum_{i=0,\pm 1}(\vec\alpha_{p}\cdot\vec\alpha_{p+i})^2\int_{a}^{e^{b}a} dr r^3 e^{-\kappa_e(a)\vec\alpha_{p}\cdot\vec\alpha_{p}\ln r} \right.\right.\\
&&\left.\left.\quad-\frac{\pi}{2} \left(\frac{y_m}{a^2}\right)^2\sum_{i=-1,0,1,2} 4 (\vec\alpha_{p}\cdot\vec Q_{p+i})^2\int_{a}^{e^{b}a} dr r^3 e^{- \frac{4}{\kappa_m(a)}\vec Q_{p}\cdot\vec Q_{p}\ln r}\right)\right]\,.
\end{eqnarray}
Re-exponentiating we find:
\begin{eqnarray}
\nonumber
\kappa_e(e^ba)&=&\kappa_e+\frac{2\pi^2}{\vec\alpha_{p}\cdot\vec\alpha_{p}}\left(- \kappa_e^2(a)\left(\frac{y_e}{a^2}\right)^2\sum_{i=0,\pm1}(\vec\alpha_{p}\cdot\vec\alpha_{p+i})^2\int_{a}^{e^{b}a} dr r^3 e^{-\kappa_e(a)\vec\alpha_{p}\cdot\vec\alpha_{p}\ln r}\right.\\
&&\left.+8\pi^2 \left(\frac{y_m}{a^2}\right)^2\sum_{i=-1,0,1,2}(\vec\alpha_{p}\cdot\vec Q_{p+i})^2\int_{a}^{e^{b}a} dr r^3 e^{-\frac{4} {\kappa_m(a)}\vec Q_{p}\cdot\vec Q_{p}\ln r}\right)\,.
\end{eqnarray}
Differentiating w.r.t. $b$ we obtain:
\begin{eqnarray}
\label{RG for kappae}
\frac{d\kappa_e}{db}= 2\pi^2 \left(- \kappa^2_ey_e^2\sum_{i=0,\pm1}(\vec\alpha_{p}\cdot\vec\alpha_{p+i})^2+4 y_m^2\sum_{i=-1,0,1,2}(\vec\alpha_{p}\cdot\vec Q_{p+i})^2  \right)\,.
\end{eqnarray}
One can redo the same analysis for two external magnetic charges of strengths $\vec Q_p$ and $-\vec Q_p$ located at $x$ and $y$  to find:
\begin{eqnarray}
\label{RG for kappam}
\frac{d}{db}\left(\frac{1}{\kappa_m}\right)=\frac{2\pi^2}{\vec Q_{p}\cdot \vec Q_{p}}\left(-\frac{4 y_m^2}{\kappa_m^2}\sum_{i=0,\pm1,\pm2}(\vec Q_{p}\cdot\vec Q_{p+i})^2+y_e^2\sum_{i=-2,-1,0,1}(\vec Q_{p}\cdot\vec \alpha_{p+i})^2  \right)\,.
\end{eqnarray}
Equations (\ref{fugacities in SUn}), (\ref{RG for kappae}), and (\ref{RG for kappam}) constitute the RGEs for the $SU(N_c)$ case with $N_c \ge 5$ (the restriction comes from having to have five bions that interact with each other; for $N_c = 3,4$ the equations will require a minor modification).

\end{document}